\documentclass{elsart}  
\usepackage{epsfig}

\newcommand{\pxdir}{p_{x}^{\mathrm{dir}}}
\newcommand{\fdo}{F_{\mathrm{DO}}}
\newcommand{\drnd}{D_{\mathrm{rnd}}}
\newcommand{\zpoi}{Z_{\mathrm{POI}}}
\newcommand{\zsum}{Z_{\mathrm{sum}}}
\newcommand{\ekin}{E_{\mathrm{kin}}}
\newcommand{\ecm}{E_{\mathrm{cm}}}
\newcommand{\ecoll}{\mathcal{E}_{\mathrm{coll}}}
\newcommand{\eth}{\mathcal{E}_{\mathrm{th}}}
\newcommand{\qval}{Q_{\mathrm{val}}}
\newcommand{\tlab}{\theta _{\mathrm{lab}}}
\newcommand{\tcm}{\theta _{\mathrm{cm}}}
\newcommand{\tcmew}{\theta _{\mathrm{cm}}^{\mathrm{EW}}}
\newcommand{\ylabiw}{y_{\mathrm{lab}}^{\mathrm{IW}}}
\newcommand{\ylab}{y_{\mathrm{lab}}}
\newcommand{\ecut}{\epsilon _{\mathrm{cut}}}
\newcommand{\ccut}{C_{\mathrm{cut}}}
\begin{document}

\begin{frontmatter}
\title{Central collisions of Au on Au at 150, 250 and 400 A MeV}  

\author[gsi]{W.~Reisdorf},
\author[gsi]{D.~Best},
\author[gsi]{A.~Gobbi},
\author[gsi,hd]{N.~Herrmann},
\author[gsi]{K.D.~Hildenbrand},
\author[gsi]{B.~Hong}, 
\author[gsi]{S.C.~Jeong},
\author[gsi]{Y.~Leifels},
\author[gsi]{C.~Pinkenburg},
\author[gsi]{J.L.~Ritman},
\author[gsi]{D.~Sch\"ull},
\author[gsi]{U.~Sodan},
\author[gsi]{K.~Teh}, 
\author[gsi]{G.S.~Wang},
\author[gsi]{J.P.~Wessels}, 
\author[gsi]{T.~Wienold},
\author[clermont]{J.P.~Alard},
\author[clermont]{V.~Amouroux},
\author[rudjer]{Z.~Basrak},     
\author[clermont]{N.~Bastid},
\author[moscow]{I.~Belyaev},
\author[clermont]{L.~Berger},
\author[rossendorf]{J.~Biegansky},
\author[florence]{M.~Bini},     
\author[clermont]{S.~Boussange},
\author[romania]{A.~Buta},
\author[rudjer]{R.~\v{C}aplar},
\author[rudjer]{N.~Cindro},
\author[strasbourg]{J.P.~Coffin},
\author[strasbourg]{P.~Crochet},
\author[strasbourg]{R.~Dona},
\author[clermont]{P.~Dupieux},
\author[rudjer]{M.~D\v{z}elalija},
\author[hungary]{J.~Er\"o},
\author[hd]{M.~Eskef},  
\author[strasbourg]{P.~Fintz},
\author[hungary]{Z.~Fodor},
\author[clermont]{L.~Fraysse},      
\author[clermont]{A.~Genoux-Lubain},
\author[hd]{G.~Goebels},
\author[strasbourg]{G.~Guillaume},
\author[kurchatov]{Y.~Grigorian},
\author[hd]{E.~H\"afele},
\author[rudjer]{S.~H\"olbling},
\author[strasbourg]{A.~Houari},  
\author[clermont]{M.~Ibnouzahir},
\author[clermont]{M.~Joriot},   
\author[strasbourg]{F.~Jundt},
\author[hungary]{J.~Kecskemeti},
\author[warsaw]{M.~Kirejczyk},
\author[hungary]{P.~Koncz},     
\author[moscow]{Y.~Korchagin},
\author[rudjer]{M.~Korolija},   
\author[rossendorf]{R.~Kotte},
\author[strasbourg]{C.~Kuhn},
\author[clermont]{D.~Lambrecht},
\author[moscow]{A.~Lebedev},
\author[kurchatov]{A.~Lebedev},
\author[romania]{I.~Legrand},
\author[strasbourg]{C.~Maazouzi},
\author[kurchatov]{V.~Manko},
\author[warsaw]{T.~Matulewicz},
\author[florence]{P.R.~Maurenzig},
\author[hd]{H.~Merlitz},
\author[kurchatov]{G.~Mgebrishvili},
\author[rossendorf]{J.~M\"osner},
\author[hd]{S.~Mohren},
\author[romania]{D.~Moisa},
\author[clermont]{G.~Montarou},
\author[clermont]{I.~Montbel},  
\author[clermont]{P.~Morel},      
\author[rossendorf]{W.~Neubert},
\author[florence]{A.~Olmi},       
\author[florence]{G.~Pasquali},   
\author[hd]{D.~Pelte},
\author[romania]{M.~Petrovici},
\author[florence]{G.~Poggi},      
\author[clermont]{P.~Pras},       
\author[strasbourg]{F.~Rami},
\author[clermont]{V.~Ramillien},
\author[strasbourg]{C.~Roy},
\author[kurchatov]{A.~Sadchikov},    
\author[hungary]{Z.~Seres},
\author[warsaw]{B.~Sikora},
\author[romania]{V.~Simion},
\author[warsaw]{K.~Siwek-Wilczy\'{n}ska},
\author[moscow]{V.~Smolyankin},
\author[florence]{N.~Taccetti},  
\author[strasbourg]{R.~Tezkratt},
\author[strasbourg]{L.~Tizniti},
\author[hd]{M.~Trzaska},
\author[kurchatov]{M.A.~Vasiliev},
\author[strasbourg]{P.~Wagner},
\author[warsaw]{K.~Wisniewski}                  
\author[rossendorf]{D.~Wohlfarth},
\author[moscow]{A.~Zhilin}
 
\bigskip
\begin{center}
{FOPI Collaboration}
\end{center}

\address[gsi]{Gesellschaft f\"ur Schwerionenforschung, Darmstadt, Germany}
\address[clermont]{Laboratoire de Physique Corpusculaire, IN2P3/CNRS, 
                  and Universit\'{e} Blaise Pascal, Clermont-Ferrand, France}
\address[rudjer]{Rudjer Boskovic Institute, Zagreb, Croatia}
\address[moscow]{Institute for Theoretical and Experimental Physics, 
                 Moscow, Russia}
\address[rossendorf]{Forschungszentrum Rossendorf, Dresden, Germany}
\address[florence]{I.N.F.N. and University of Florence, Italy}       
\address[romania]{Institute for Physics and Nuclear Engineering, 
                  Bucharest, Romania}
\address[strasbourg]{Centre de Recherches Nucl\'{e}aires and 
               Universit\'{e} Louis Pasteur, Strasbourg, France}
\address[hungary]{Central Research Institute for Physics, Budapest, Hungary}
\address[hd]{Physikalisches Institut der Universit\"at Heidelberg, 
             Heidelberg, Germany}
\address[kurchatov]{Kurchatov Institute, Moscow, Russia}
\address[warsaw]{Institute of Experimental Physics, Warsaw University, Poland}

\begin{abstract}
Collisions of Au on Au at incident energies of 150, 250 and 400 A MeV were
studied with the FOPI-facility at GSI Darmstadt.
Nuclear charge (Z $\le$ 15) and velocity of the products were
detected with full azimuthal acceptance at laboratory angles
1$^{\circ}$ $\le$ $\theta _{lab}$ $\le$ 30$^{\circ}$.
Isotope separated light charged particles were measured with movable
multiple telescopes in an angular range of $6-90^{\circ}$.
Central collisions representing about $1\%$ of the reaction cross section
were selected by requiring high total transverse energy, but vanishing
sideflow.
The velocity space distributions and yields of the emitted fragments are
reported.
The data are analysed in terms of a thermal model including radial flow.
A comparison with predictions of the Quantum Molecular Model is presented.
\bigskip


\end{abstract}

\end{frontmatter}

\section{Introduction}
\label{introduction}
The properties of hadronic matter under various conditions of density and
temperature are of high interest both in their own right and owing to their
astrophysical connection: evolution of the early universe and of neutron
stars.
In earthly laboratories hadronic matter in a state far off its ground state
can only be studied in heavy ion collisions at incident energies that are
well above the Coulomb barrier.
Such studies in the energy regime from about 100 A MeV to 2 A GeV were
initiated in the mid 1970's at the Bevalac in Berkeley.
It became soon clear that the high multiplicity and diversity of particles
emerging from such collisions required sophisticated detection systems
covering large angular ranges while having high particle identification
power and high granularity and/or multiple hit capabilities.
Due to the large variety of possible event topologies, there was also a
need to register and sort a very large number of collisions in order to
obtain statistically significant information.

With the advent  of 
 $4\pi$ particle detectors like the Streamer Chamber \cite{sandoval83}, 
the Plastic Ball \cite{baden82} and, somewhat later,  
Diogene \cite{alard87}, first dynamic signals of hot
expanding nuclear matter were seen in the form of collective flow, i.e.
correlated emission patterns involving many particles.
Together with studies of the chemical composition of the emitted fragments
(single nucleons, clusters, produced mesons) and its possible connection to
the entropy of the created hot and compressed matter,
and further complemented by space-time evolution studies inferred from
two-particle interferometry,
 great hopes arose that the
properties of hadronic matter, namely its equation of state, and possible
phase transitions, could be inferred from the data.

The experimental and theoretical situation in the second half of the
eighties was summarized in a series of excellent review papers
\cite{csernai86,stoecker86,stock86,clare86,gutbrod89}.
Despite the tremendous progress achieved, it remained clear however that
considerable efforts were still needed both experimentally and
theoretically to achieve a sufficiently detailed understanding of the
highly complex dynamics of these reactions in order to be able to extract
the more fundamental issues.

At the beginning of the nineties a new generation of $4\pi$ detectors
emerged both at the Bevalac and at SIS/ESR, Darmstadt,
 that had increased particle
identification power as well as higher granularity.
One of the tasks of these new detector systems is to provide complete
triple-differential coverage of the populated momentum space while
achieving good particle identification and minimal apparatus distortions.

The detector FOPI, installed at SIS/ESR, is such a system.
It was built in two phases: Phase I, covering the forward hemisphere
($1-30^{\circ}$ in the laboratory) and based on time-of-flight plus
energy-loss methods to identify and characterize the emerging particles,
and Phase II, which extended the detector to $4\pi$ geometry and added
magnetic analysis.
The first major experiment was concerned with Au on Au collisions in the
energy range extending from 100 to 800 A MeV.
It was performed with the Phase I detectors \cite{gobbi93} complemented by
movable multiple telescopes \cite{poggi93} that covered also backward angles.

In the present work we show and analyse data on very central collisions
limiting ourselves to incident energies of 150, 250 and 400 A MeV.
The corresponding center-of-mass (c.o.m.) energies of 37-95 A MeV extend
from energies comparable to the nucleonic Fermi energy in ground state
nuclei to well above it.
As we shall see, cluster formation in this regime is still the rule, rather
than the exception.
While 'hard' nucleon-nucleon collisions play an increasing role, in addition
to the mean field that seems to dominate reactions at lower energies, by
far most of these collisions on a 'microscopic' level are expected to be
elastic:
The internal nucleon degrees of freedom are still essentially dormant and
in particular pion production, although well above measuring sensitivity,
does not play an important part in the general balance of energy.

Under these conditions the fact that Phase I was still 'pion-blind' does
not represent a serious problem in a study aimed at obtaining a general
overview of the triple-differential momentum-space population and the
composition of the most abundant particles.
The lower energy limitation was governed by the necessity to have
sufficient coverage of the mid-rapidity part of phase-space in view of
threshold limitations affecting heavier clusters.

Indeed, one of the perhaps surprising outcomes of this investigation of
{\em central} collisions was the still copious production of
intermediate-mass fragments that we shall define to be fragments with
nuclear charge $Z\geq 3$ and abbreviate as IMF in accordance to established
practice in the literature.
Processes involving the 'simultaneous' emission of many IMF's have been
termed 'multifragmentation' and have found widespread interest, see for
instance \cite{gross90,moretto93,hirschegg94,bondorf95}.

Older exclusive information on heavy cluster formation from Bevalac times
was limited to an experiment at 200 A MeV \cite{doss87} concentrating on
sideflow: a remarkable result of this study was the demonstrated higher
sensitivity of the IMF's to flow, a fact that had been anticipated in the
framework of hydrodynamical model calculations \cite{csernai83}.
Since then, new information from central or semicentral collisions at
energies beyond 100 A MeV has emerged from work done both at Berkeley and
Darmstadt
\cite{alard92,kuhn93,tsang93,jeong94,hsi94,petrovici95,dzelalija95,kotte95,ramillien95,poggi95,kunde95,lisa95,partlan95,wang95,wang96}.

One of the important new observations, favoured by the sensitivity of IMF's
to flow, was the overwhelming and unambiguous evidence of a new kind of
axially {\em symmetric} or 'radial' flow 
\cite{jeong94,hsi94,petrovici95,poggi95,lisa95} as a
signature of highly exclusive central collisions.
This flow was far more important in magnitude than the sideflow predicted
\cite{stoecker80}
and discovered \cite{gustafsson84,renfordt84} earlier and was characteristic of
{\em participant} matter, in contrast to the sideflow phenomenon which was
primarily interpreted as a spectator phenomenon.

The development of an isotropic 'blast' in energetic collisions had been
explicitly proposed by Bondorf {\em et al.} \cite{bondorf78} in the
framework of an    adiabatic expansion of a hot ideal-gas .
This provided one possible mechanism of local cooling that seemed necessary
to account for the formation of clusters.
A year later, Siemens and Rasmussen \cite{siemens79} first proposed that
the data then available gave evidence for a 'blast wave'.
Much later, evidence came from emulsion studies \cite{barz91,bauer93} in
which it was first shown that in events with a high degree of sphericity,
fragments were emitted that were characterized by an approximately {\em
linear rise of the average kinetic energy with size}.
This phenomenon clearly went beyond the expectations of a thermal model
\cite{mekjian78} based on an equilibrated system {\em enclosed in a box}. 

Our collaboration, taking advantage of combined high multiplicity and
low directivity event selection, first showed that the heavy clusters
indeed were emitted from essentially one source located at midrapidity
\cite{alard92}.
Soon after, we came to the conclusion, with use of statistical models, that
the occurrence of so many IMF's had to be associated with a surprisingly
low entropy \cite{kuhn93,dzelalija95}.
The new radial flow phenomenon was then quantized \cite{jeong94} and a
first attempt using the model of ref.~\cite{bondorf78} was made
\cite{petrovici95} to trace back the initial conditions of compression and
temperature from the measured momentum distributions and an approximate
account for the chemical composition of the fireball could be done.
It also became clear that radial flow was going to have an essential
influence on the quantitative interpretation of small relative-momentum
correlations between IMF's \cite{kotte95} aimed at determining the size of
the fireball at freeze out.

Further work concentrated on the isotope-separated light charged particles
($Z=1,2$) \cite{poggi95} and it became evident that the simultaneous
understanding of both the chemical composition (light {\em and} heavy
clusters) of the fireball {\em and} the momentum space distributions of all
the observed particles was going to be a special challenge.

The present work concentrates on both of these aspects, presenting the data
in more complete detail than before.
After a brief reminder of the major features of the Phase I apparatus
\cite{gobbi93} in section~\ref{experimental} we outline and discuss in
section~\ref{centrality} our method for obtaining a highly central sample
of events.
The aim is to isolate a maximal-size piece of hot nuclear matter with a
minimum of contamination by spectator matter.
The isolation of these highly central events will then allow us to take a ,
still qualitative, look at the general topology in momentum space
concentrating, among others, on the gradual evolution with incident energy.
 
Then we explore in sections~\ref{velocity} and \ref{chemical}
 the possibility of accounting for
the data (momentum space distribution and chemical composition)
 in the framework
of a thermal model, which however, accounts for the presence of radial
flow.
A preliminary extraction of the radial flow values was reported in refs.
\cite{reisdorf94,reisdorf94a}.
This flow considerably diminishes the available chaotic part of the energy.
The presence of IMF's allows us to determine the flow profile with higher
accuracy than before. 
An important point will be that the analysis keeps track of the energy and
charge balance.
After subtraction of the collective, or flow, energy, the remaining
chaotic energy is interpreted as thermal energy that is then used in
section~\ref{chemical} as input into statistical models to see if chemical
composition can be understood in terms of a quasihomogenous population of
the locally available phase space.
Statistical model descriptions of the data
 that are {\em not} constrained by a fixed available
thermal energy have been remarkably successful \cite{kuhn93} even 
 at much higher energies
\cite{pbm95}.
   
The results of these efforts will then bring us to consider the dynamical
alternative (section~\ref{qmd}) in the framework of the Quantum Molecular
Dynamics (QMD) model \cite{aichelin86}.
In particular we will explore with this model to what degree the generation
of transverse momentum results in quasi-isotropic topologies of central
events, and we will consider the amount of scaling of radial flow and of
sideflow and their possible connection.
We will also take a look at the capability of such microscopic models to
predict the probability of clusterization and the possible interconnection
of the latter with the time scale of the explosion.
  
A summary of the data and conclusions, together with a brief outlook
(section~\ref{summary}) will close this work.

\section{Experimental procedure}
\label{experimental}
Au on Au collisions at incident energies ranging from 100 to 800 A MeV
were studied with the Phase I of the FOPI detector \cite{gobbi93}
at the SIS/ESR accelerator facility, GSI Darmstadt. 
Only the data obtained at 150, 250 and 400 A MeV will be presented here.
The beam intensities were typically $10^5$ ions/s.
Targets of $0.5\%$ (for lower energies) up to $2\%$ (for higher energies)
interaction length (i.e. 200 mg/cm$^2$) were used. 
Most events were registered under 'central trigger' conditions, some under
'mimimum bias' conditions.
The central trigger consisted in a high multiplicity threshold on the
External Wall (PM$>$16, 17, 21, respectively, at 150, 250 and 400 A MeV)
which selected an impact parameter range up to
$8.5\pm 0.5$ fm in a sharp cutoff model.
An average of $10^{6}$ events was collected at each energy under central
trigger conditions. Unless otherwise stated only these events will
be analysed in the following.

Phase I of the detector covered the laboratory polar angles from $1.2^{\circ}$
to $30^{\circ}$.
It consists of a Forward Wall of scintillators with a granularity exceeding
that of the earlier Plastic Ball experiments \cite{baden82,gutbrod89}
by roughly a factor three.
The Inner Wall consisting of 252 trapezoidal scintilator paddles extends
from $1.2^{\circ}$ to $7.5^{\circ}$, and the External Wall (512 scintillator
bars) covers the angles from $7^{\circ}$ to $30^{\circ}$.
Each scintillator provides an energy loss and a time-of-flight signal
allowing to determine the element number $Z$ and the velocity vector of the
charged particles.
In order to achieve lower detection thresholds a shell of 188 thin energy
loss detectors (cluster detectors) is mounted in front of the Wall.
It consists of an ensemble of gas-filled ionization chambers (Parabola) in
front of the External Wall and thin plastic scintillator paddles (Rosace)
in front of the Inner Wall.
A helium bag is placed between the target and the Forward Wall.
With this setup charge identification up to Z=15 is obtained, with
thresholds in the external part of the detector increasing from 14 A Mev
for Z=1 fragments to 50 A MeV for Z=15 fragments.
The energy thresholds are slightly higher in the inner part due to larger
flight path and larger thickness of the Rosace paddles in comparison to the
Parabola ionization chambers.

In some of the runs the FOPI detector was complemented by 8 multiple
Si-CsI(Tl)  telescopes \cite{poggi93} which could be moved in the
horizontal plane over an angular range from $6^{\circ}$ up to $90^{\circ}$.
These detectors, operated in coincidence with the FOPI detector,
 allowed isotope separation for Z=1 to 2 and had lower
thresholds of 8 A MeV or less.
The telescope data have been discussed in an earlier publication
\cite{poggi95}, but will be included here again in a combined analysis.

The nominal beam energies indicated are understood to be
mid-target energies taking into account all energy losses upstream.
Likewise, in the following, all momentum space distributions of the
observed fragments are corrected for their path and charge dependent energy
losses on the way to the detectors.
The systematic errors of the measured velocities are estimated to be 1.5 \%
or $3 \beta$ \%, whichever is larger ($\beta$ is the velocity in units of
the light velocity). 

\section{Centrality selection and event topologies}
\label{centrality}
\subsection{Choosing central events}
\label{choosing}
Using effective sharp radii \cite{blocki77} for Au one can estimate a
'geometrical' cross section of 5850 mb corresponding to a maximal impact
parameter b=13.6 fm.
Ideally, in a central collision, one would like to minimize the
contribution of non-overlapping matter. 
In a clean-cut straight-trajectory model the 'spectator' fraction can be
estimated to reach the level of 10\% at an impact parameter of 1.4 fm or a
cross section of 60 mb which is just about 1\% of the total cross section.
Because of the very high 'background' of non-central collisions it is clear
that global event-observables that are used for the selection of central
collisions should not only have a unique {\em average} correlation with the
impact parameter, but also a small {\em fluctuation} around this average
value.
  
In finite systems such fluctuations are unavoidable.
Also, in the present energy regime Fermi velocities remain comparable to
the c.o.m. velocity in the incident channel making a clean separation of
'spectators' and 'participants' difficult as we shall see.

\begin{figure}[p]
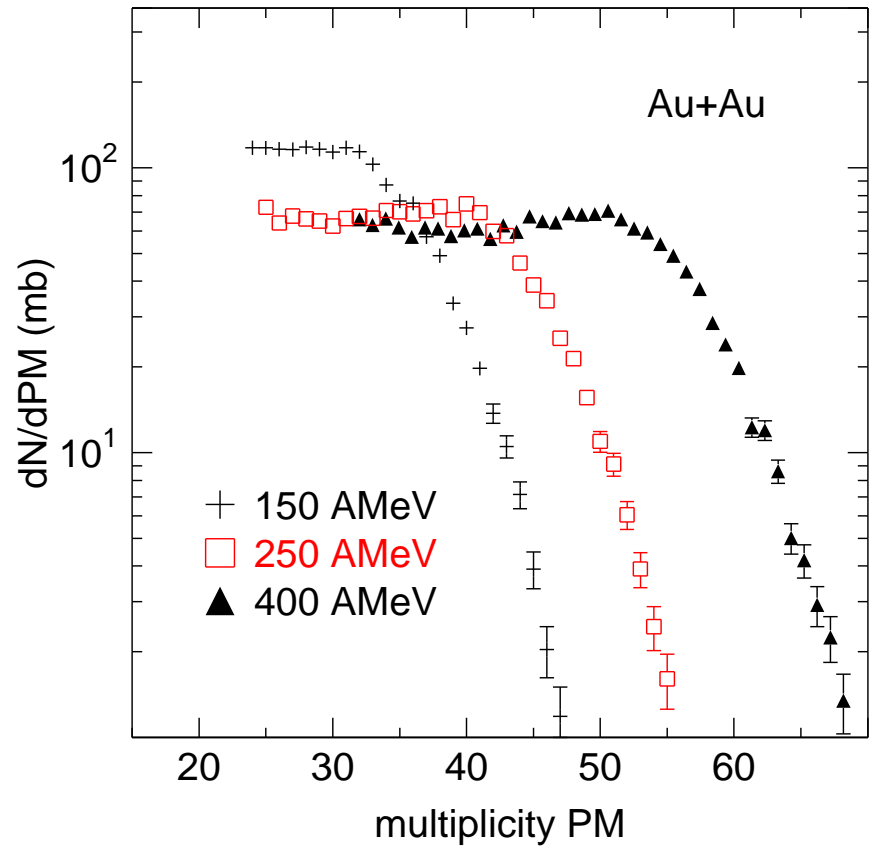

\caption{\small Measured External Wall multiplicity (PM) distributions for
Au on Au at three indicated energies.}
\label{fig:pm}
\end{figure}

Traditionally, high particle multiplicities have been associated with
central collisions.
Fig.~\ref{fig:pm} shows charged-particle multiplicity ('PM') distributions
measured with the External Wall.
In the figure the
distributions are cut off at a total cross section level of about 1500
mb (7 fm). 
With the depicted multiplicity range   contamination from non-target
background is estimated (from target-free runs) to be below the 10\% level
in the worst case and altogether negligible around and beyond the plateau
edge (see Fig.~\ref{fig:pm}).
Future detailed comparisons with event-simulating models should of course
take into account the granularity (see section~\ref{experimental})
 of the External Wall and the resulting
multiple-hit probabilities .
  
The presence of a plateau (on a logarithmic scale) in these PM distributions
allows to define a limiting value at the upper edge of the distribution
that is characterized by a yield equal to half the plateau value
\cite{gutbrod89}. 
In our case the summed cross section beyond this limiting value turns out
to be about 250 mb (it varies somewhat with energy) corresponding to
'geometric' cuts of about 3 fm.
This still relatively large cross section makes one suspect that the
multiplicity fluctuation seen in the tailend of the distribution is rather
large limiting the 'resolution' of the associated impact parameter.
The other drawback of the selection of events with large multiplicities is
the bias this implies concerning the chemical composition of the hot object
one wants to study: by requiring high average multiplicities $ <M>  $
one obviously fixes the average size of the emitted clusters which is just
$A/<M>$ ($A$ is the total mass).
  
Another alternative to select the most central collisions is to look for
events with a maximal transverse energy of the emitted fragments.
If one assumes that
transverse energy is created predominantly via nucleon-nucleon
collisions, the picture is that the number of such collisions is maximal
when there is optimal overlap between target and projectile.
Since in the present energy regime particle production is modest, the sum
of total longitudinal, $E_{\ell}$, and transversal, $E_{t}$,
kinetic energies is constant by energy conservation and hence in a perfect
$4\pi$ apparatus any function of $E_{t}$ and/or $E_{\ell}$ is
essentially equivalent.
Simulations with the Quantum Molecular Dynamics (QMD) model
\cite{aichelin86} model showed that in our Phase I geometry the ratio,
dubbed ERAT, defined by
\begin{equation}
\label{eq:erat}
ERAT \equiv \frac{E_{t}}{E_{\ell}}
\equiv \frac{\sum_{i} p_{ti}^{2}/(m_{i}+E_{i})}
            {\sum_{i} p_{\ell i}^{2}/(m_{i}+E_{i})} \; ; \; \; \;
\end{equation}
is particularly suitable for event sorting \cite{reisdorf92}.
In this definition \cite{daniel95}
$p_{t}$, $p_{\ell}$ are transverse and longitudinal momenta,
$m$ is the rest mass, $E$ the total energy (including the rest mass) of a
fragment.
As we are not measuring masses in Phase I, we replace $m_{i}$ by
$2Z_{i}m_{N}$ ($Z$ the nuclear charge,$m_{N}$ the nucleon mass) when
calculating ERAT.
The sums are limited to the forward hemisphere in the
c.o.m., as the backward hemisphere was not well covered by the Phase I
setup.
One other advantage of this choice of a dimensionless observable is that it
is 'scale invariant' in the sense that it will  not vary with incident
energy if the event shapes are        scale invariant.
In a naive physics interpretation and for central collisions one may   
expect three possible outcomes for ERAT: ERAT$\approx 2$, may        be
associated with an approximately thermalized fireball.
Values significantly below two  tend to indicate transparency, while
values well above two signal squeeze out of nuclear matter at
$90^{\circ}$ due perhaps to shock-like effects as predicted long ago in the
framework of hydrodynamical calculations \cite{scheid74,sobel75}.
  
\begin{figure}[p]
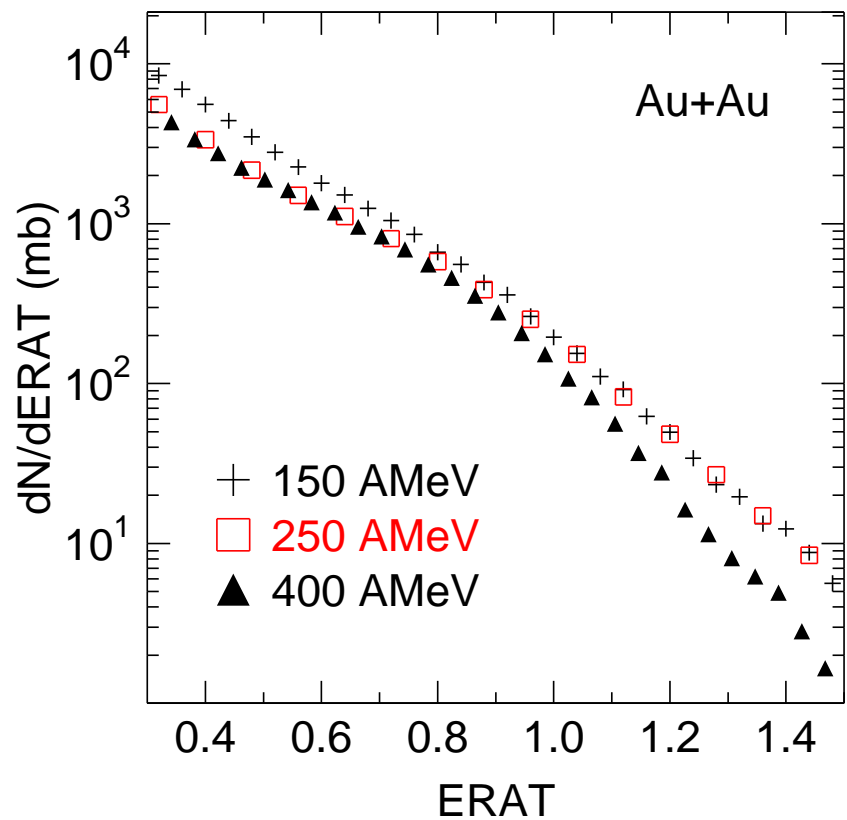

\caption{\small Measured distributions of the ratio of transverse to 
longitudinal kinetic energies (ERAT) for
Au on Au at three indicated energies.}
\label{fig:erat}
\end{figure}

Measured distributions of ERAT (for values larger than 0.3) are shown in
Fig.~\ref{fig:erat}.
Apparatus limits (see section \ref{experimental})
 roughly cut in half the true ERAT value 
(see also section~\ref{qmd}).
For these distributions an additional cut on the External Wall multiplicity
PM ($> 16, 17, 21$, respectively for $E/A=$ 150,250,400 MeV) was done.
With these thresholds on multiplicity background contributions are
virtually eliminated provided ERAT$> 0.3$.
We found that the applied multiplicity cut had no consequence for ERAT$>
0.5$. 

The approximate scaling properties for the ERAT distributions can be judged
from the Figure.
The statistical errors are smaller than the symbol sizes.
We are reluctant to
attach physics meaning to the small variations with incident energy
as the systematic errors (0.07 units along the ERAT axis beyond the 100 mb
level)
are estimated to be only slightly smaller than the
differences visible in the Figure.
We shall come back to ERAT distributions in section~\ref{qmd}.

Since our prime aim was to select the most central events, it was desirable
to obtain model independent information on how well centrality is achieved
when applying various cuts on global observables such as PM or ERAT.
In the limit of zero impact parameter the event shapes ought to be axially
symmetric on the average.

\begin{figure}[p]
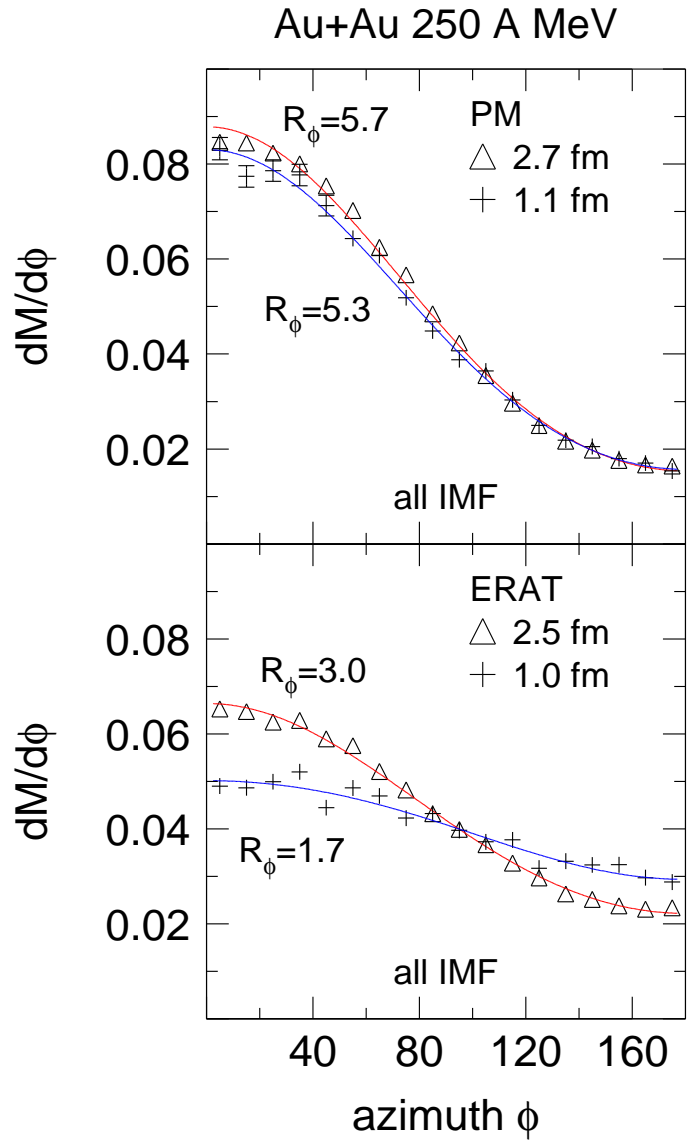

\caption{\small Azimuthal angle distributions of intermediate mass
fragments for multiplicity (PM) and transverse energy (ERAT) selections
corresponding to the indicated geometrical cuts. The intensity ratios
$R_{\phi}(0^{\circ}/180^{\circ})$ obtained by least square fits (see text)
are also indicated.}
\label{fig:azimuth1}
\end{figure}

\begin{figure}[p]
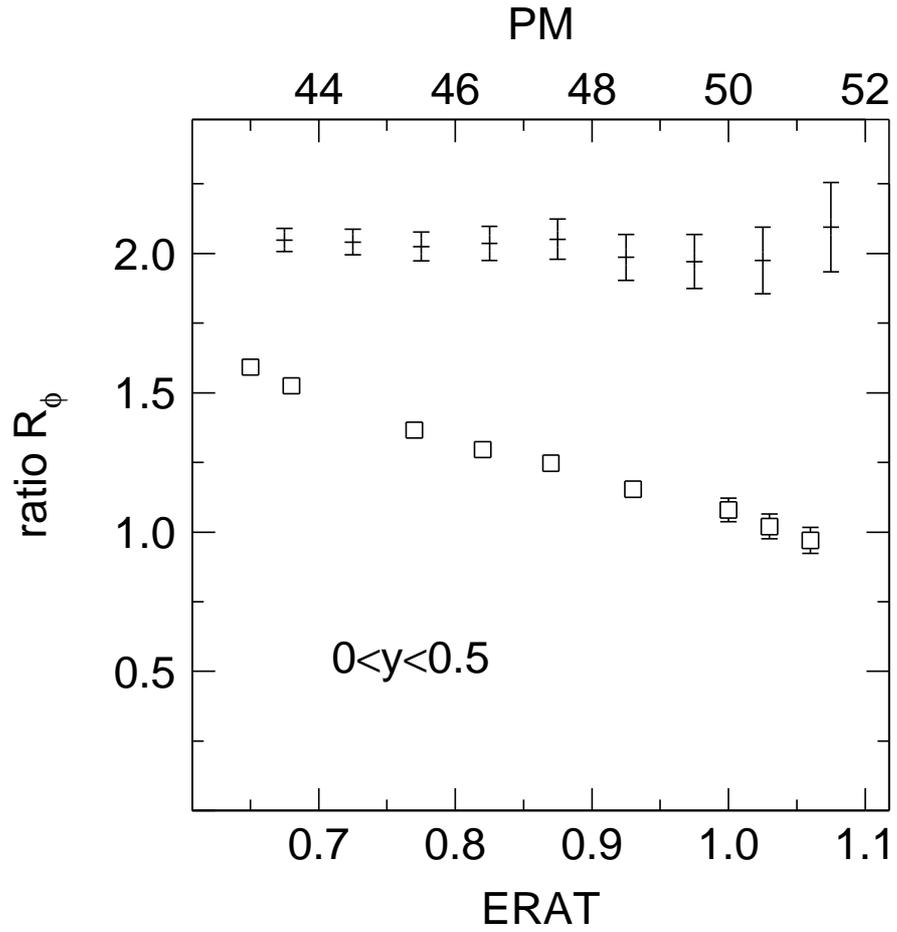

\caption{\small Azimuthal asymmetry variation of intermediate mass
fragments with multiplicity (upper abscissa) and with ERAT (lower abscissa,
open squares). An additional constraint on the scaled rapidity, indicated
in the figure, was applied.}
\label{fig:azimuth2}
\end{figure}

Using the method of Danielewicz and Odyniec \cite{daniel85} we have
determined azimuthal distributions of the IMF.
Due to side-flow \cite{gustafsson84,renfordt84} these distributions in
general are peaked at an azimuth $\phi=0$ in the reaction plane if only
fragments in the forward hemisphere are considered.
The particular sensitivity of heavier fragments to flow phenomena has been
anticipated theoretically, see for instance ref.~\cite{csernai83,stoecker82},
 and was first observed experimentally in ref.~\cite{doss87}.
IMF's are therefore a rather sensitive probe of deviations from centrality.
Figure~\ref{fig:azimuth1} shows azimuthal distributions (limited to forward
rapidities and $E/A=250$ MeV) under various cut thresholds on PM and    
ERAT.
Comparable integrated cross sections (corresponding to effective impact
parameters b=1 or 2.5-2.7 fm), were taken as indicated in the Figure.
The data were least-squares fitted using
\begin{equation}
\label{eq:dmdphi}
dM/d\phi = a_{0} (1 + a_{1} \cos \phi + a_{2} \cos 2\phi )
\end{equation}
where the parameters are $a_{0}$, $a_{1}$ and $a_{2}$ and the azimuthal
angle is $\phi$.
With a PM selection one finds quite sizeable azimuthal asymmetries with
ratios $R_{\phi}(0^{\circ}/180^{\circ})$ in excess of five, almost
independent on the PM threshold (as long as it is sufficiently high).
The $R_{\phi}$ ratios are considerably smaller when an ERAT selection is
done and keep decreasing as the ERAT threshold is increased to the
statistical limit (see also \cite{roy96}).
These trends remain even when a midrapidity cut is
applied in addition, as can be seen in Fig.~\ref{fig:azimuth2}:
$R_{\phi}$ is saturated at a value of two for all PM$>43$ while it reaches
gradually the nominal value for isotropy when ERAT$\geq 1.0$.
In the range shown the integrated cross section
varies from 350 mb (3.3 fm) down to 20 mb (0.8 fm).
Thus, 'perfect' isotropy is reached only for ERAT selections and at the
cost of very severe cuts.

As it is our aim to study the momentum space distribution as well as the
chemical composition of the central fireball, a compromise has to be made
between the need for good centrality and the minimum number of events
required for adequate statistics.
Because of finite number fluctuations there is also the worry about
statistical noise causing one to look at 'atypical' events.
We shall come back later to the severe autocorrelation effects that occur
in extreme-cut situations.
From the above study we concluded that while ERAT was a better choice than
PM to achieve centrality, there was a need for another selection criterion
that would allow to 'open up' the severe ERAT cut necessary to achieve
azimuthal isotropy.
The idea followed was to define global observables that would be a measure of
sideflow and to request {\em small sideflows while maintaining high
transverse energies}.

\begin{figure}[p]
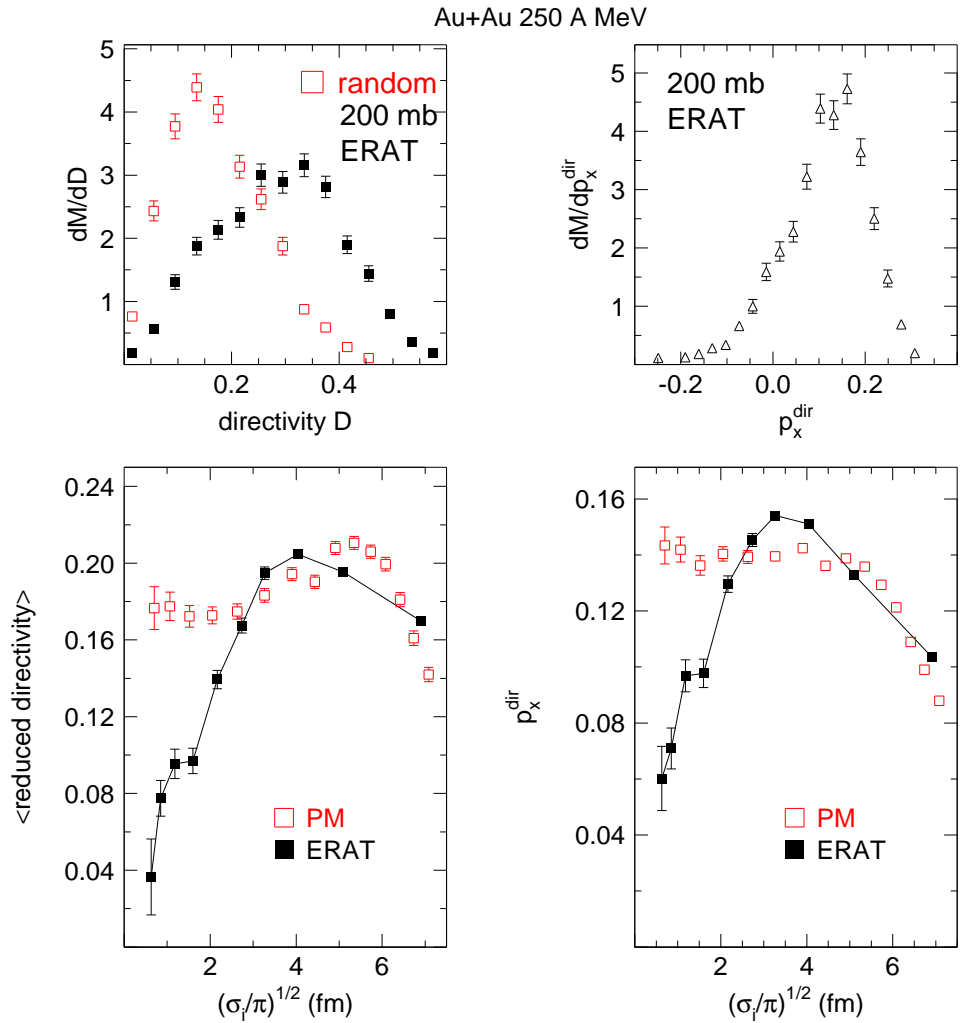

\caption{\small Directivity and in-plane flow for Au on Au at 250 A MeV.
The upper panels show the directivity distribution (left) and the
distribution of average in-plane transverse momenta for a cross section
sample of 200 mb cut off from the tail of the ERAT distribution.
An azimuthally randomized directivity distribution (open symbols) is also
shown for comparison in the left panel.
The lower panels show the corresponding variation of the first moments with
the cross section interval (reduced to an effective impact
parameter) obtained by binning with the multiplicity PM (open symbols) or
with the ERAT (full symbols).}
\label{fig:dirpx}
\end{figure}

Introducing the scaled four-velocity $u \equiv (\vec{u},u_{4})$ with
$\vec{u} = \vec{\beta}\gamma/\beta_{p}\gamma_{p}$,
 where $\vec{\beta}$ is the velocity in units
of $c$, $\gamma = ( 1 - \beta ^{2})^{-1/2}$, $u_{4}=\gamma/\gamma^{p}$
and the index $p$ refers to the incident projectile in the c.o.m.,
 we define the directivity
\cite{alard92,beckmann87}
\begin{equation}
\label{eq:directivity}
D = |\sum_{i} Z_{i} \vec{u}_{ti}| / \sum_{i} Z_{i} |\vec{u}_{ti}|
\end{equation}
and \cite{daniel85,wienold93}
\begin{equation}
\label{eq:fdo}         
\fdo   = \sum_{j,i\neq j} Z_{i} \vec{u}_{ti} Z_{j} \vec{u}_{tj}          
         / \sum_{j,i\neq j} Z_{i} Z_{j}
\end{equation}
\begin{equation}
\label{eq:pxdir}       
\pxdir = \sum_{i} Z_{i} u_{xi} / \sum_{i} Z_{i}                      
\end{equation}
where $\vec{u}_t$ is the transverse component of $\vec{u}$ and $u_{xi}$ is
given by 
\begin{equation}
\label{eq:uxi}         
u_{xi} = \vec{u}_{ti} \cdot \vec{Q}_{i} / | Q_{i} |                      
\end{equation}
with
\begin{equation}
\label{eq:qi}         
\vec{Q}_{i} = \sum_{j\neq i} Z_{j} \vec{u}_{tj}                          
\end{equation}
Owing to the Phase I limitations all sums in eqs. \ref{eq:directivity} to
\ref{eq:qi} are limited to fragments i emitted into the forward hemisphere
in the c.o.m. and we use nuclear charges instead of the more commonly used
masses.
Note that $j\neq i$ in the sums for $\fdo $ and $\vec{Q}_{i}$.
The observable $\pxdir$      is called {\em mean transverse momentum per
nucleon in the reaction plane}, although it is dimensionless in the scaled
units used 
here and, due to the condition $j\neq i$, the vector $\vec{Q}_{i}$ varies
with $i$.
The variables $D$ and $\fdo $ do not refer to a reaction plane
explicitly.
In the absence of side-flow $\fdo $ and $\pxdir$      tend to be small
and negative due to momentum conservation constraints.
The same is true for the {\em reduced directivity} 
$D^{(0)}= D-\drnd $ where the 'random' directivity $\drnd $ is the
value of $D$ averaged for the same subset of events but after randomization
of the azimuthal emission angles.
Typical distributions of $D$, $\drnd $ and $\pxdir$  for an ERAT-selected
event sample corresponding to the most central 200 mb (ERAT200) 
are shown in the upper panels of Fig.~\ref{fig:dirpx}.
The evolution of the first moments of these distributions with the average
value of ERAT or PM within narrow intervals is shown in the lower panels
where the abscissa has been converted to the effective impact parameters.
One sees that both measures of axial asymmetry 
(as well as $\fdo $ \cite{wienold93}, which is not shown) have completely analogue
trends: 
for PM binning they never reach the value of zero,
 while for ERAT binning a clean maximum
is resolved showing that below an effective impact parameter of 4 fm
collisions are reached that have increasing axial symmetry as the ERAT
value is {\em raised} and hence are interpreted as being more central.

\begin{figure}[p]
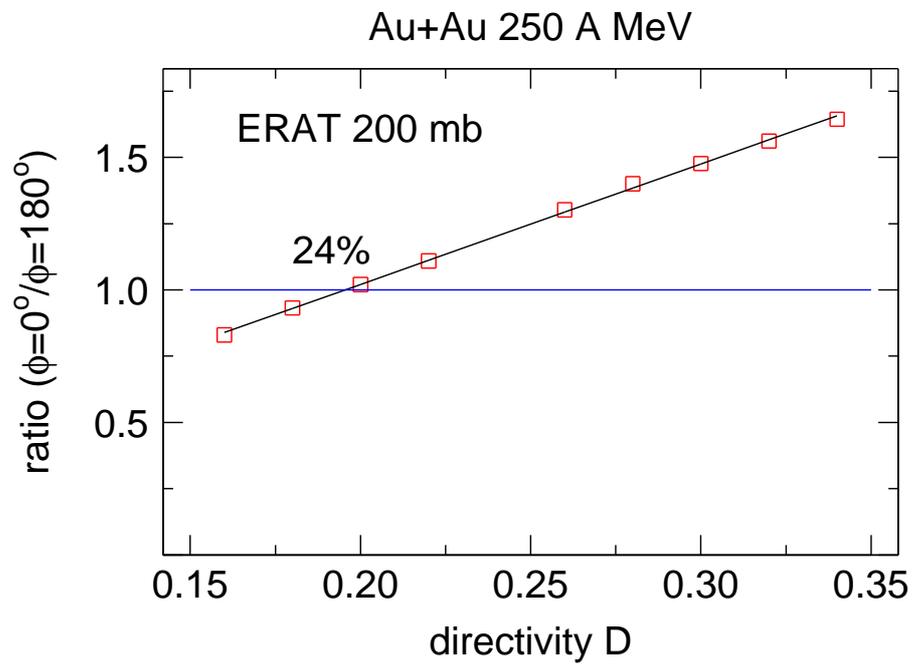

\caption{\small Azimuthal asymmetry as function of the directivity cut
for Au+Au events (250 A MeV) from the ERAT 200 mb sample.}
\label{fig:dircut}
\end{figure}

As all measures of sideflow      were seen to be largely equivalent for our
purpose, we opted to combine ERAT with the $D$ cut.
Note that $D$, again, is a scale invariant quantity in particular when the
multiplicity dependent average random value is subtracted.
The threshold value of $D$ was determined empirically, see
Fig.~\ref{fig:dircut}, by taking the ERAT200 sample and gradually lowering
the upper limit of accepted events while controlling the azimuthal ratio
$R_{\phi}$ of the IMF.
These event samples will be dubbed ERAT200D in the following.
In order to assess the dependence of the results on the selection scheme an
alternate sample ERAT50 with approximately 50 mb, but without directivity
selection, was also considered in the analyses to be described in the
sequel.
Occasionally we shall compare also with PM selected samples.
A summary of the specifications of all these event selections is given in
table~\ref{tab:eventsample}.
While the indicated cross sections are quite accurate on a relative basis,
they share an estimated common systematic uncertainty of about 10\%.

\begin{figure}[p]
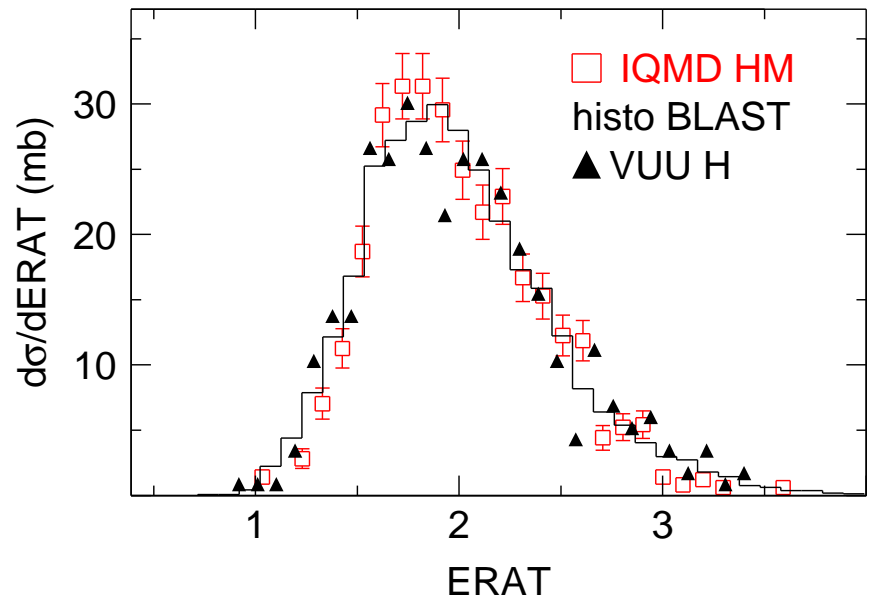

\caption{\small ERAT distributions (charged particles only) for central Au
on Au collisions in $4\pi$ geometry at 250 A MeV.
Histogram: isotropic blast model.
Open square symbols: IQMD \protect \cite{hartnack93,bass95}
 prediction for the hard
momentum-dependent equation of state.
Full triangles: VUU calculation \protect \cite{daniel92}
 with finite-number
fluctuations added and with abscissa rescaled to give an average value of two.}
\label{fig:eratfluct}
\end{figure}

Before presenting data under specific selection criteria it is useful to
make a few remarks on the possibility of bias when choosing events in the
tails of distributions of global observables.
Even such a heavy system as Au on Au is subjected to trivial finite-number
fluctuations that sit on top of potentially interesting 'physics'
fluctuations.
An idea of the magnitude of such fluctuations for the observable ERAT,
obtained only from charged particles, but in $4\pi$ geometry, is given in
fig.~\ref{fig:eratfluct} which shows the result of central event
simulations with three different models. 
Only two of the models will be discussed here, the third one will be picked
up in section~\ref{qmd}.
The histogram in the figure is an isotropic blast model simulation (that
will be described in detail in section~\ref{blast}), which by definition
predicts an average ERAT value of two.
In each event the model conserves energy, momentum, mass and charge.
Except for this constraint it is assumed that the multiplicities of the
different species (taken from experiment) fluctuate in an uncorrelated way.
Similar assumptions have been used with success in ref.~\cite{delzoppo95}.

The resulting, {\em purely statistical}, fluctuation of ERAT around the
value of two is considerable.
As this fluctuation is unavoidable in finite systems, it should not come as
a surprise that a simulation with  a Quantum Molecular Dynamics 
\cite{aichelin86} code IQMD, \cite{hartnack93,bass95}, to be
described in section~\ref{qmd}, comes out to be indistinguishable (see
fig.~\ref{fig:eratfluct}). 
Here we choose an energy (250 A MeV) and parameterization that happens to
give also an average ERAT value of two.
In principle QMD also allows for non-trivial fluctuations, but the
finite-number fluctuations are dominating.
  
If one were to put an event selection cut in figure~\ref{fig:eratfluct} at
very low (or very high) ERAT values, that are significantly different from
the average value, one would sample events that are {\em not} isotropic, but
fluctuate between prolate and oblate shapes.
High ERAT values for example indicate a surplus of transverse momenta and
hence preferred emission perpendicular to the beam axis.
By far the simplest way to distinguish if the high ERAT values represent a
{\em collective}, i.e. correlated fluctuation, or a trivial, uncorrelated
one, is to always exclude the socalled Particle Of Interest (POI) from the
trigger condition.
Loosely speaking: if out of 50 particles, 49 are going left, then if the
move is {\em collective}, the last particle, our POI, will do so also with
a large probability.
If the move was an uncorrelated fluctuation, the POI will be uninfluenced
by the random fluctuation of the other particles.
   
From an extensive series of test simulations we learned that
failure to use this simple recipee will, in the tails of global
distributions, lead to severe autocorrelation effects, that distort the
true distributions. 
We found, as intuitively obvious, that such autocorrelation effects were
rising with the deviation of the sampled events from the average event and
with $\zpoi /\zsum $, i.e. the ratio of the charge of the POI to the sum
of charges contributing to the global observable.
Therefore, in the sequel, we have always excluded the POI from the cut
condition.
  
In many ways the procedure is the same as the well accepted procedure of
eliminating the POI from the reaction plane determination, see our
eq.~\ref{eq:qi}.
The removal of {\em one} particle from a global observable (which in
practice is not measured with {\em all} particles anyhow) does not 
significantly change the binning property of this observable.
If out of $n$ particles of interest in a given event $m\leq n$ have been
selected, the event is accepted with a probability $m/n$. 
This still allows to determine cross sections for certain cuts.
The deviation from the yes/no logic of sharp cuts is not considered a
serious problem. 

\subsection{Topologies of central events}
\label{topologies}
Resuming now our investigation of highly central events,
it is  revealing to take a look at the variation of the event topology
in the two-dimensional space of transverse 4-velocity and
rapidity.
In the following we shall make use of scaled quantities: in particular
rapidities, $y$, are defined in the c.o.m. system and divided by the beam
rapidity, and we recall that the transverse four-velocities, $u_{t}$ 
are scaled with the $\beta\gamma$
value of the projectile in the c.o.m..
In this system of units target (projectile) rapidity is at
$-1$ (+1).
Scaling properties expected in the hydrodynamics context will be discussed
briefly in section~\ref{sideflow}.

\begin{figure}[p]
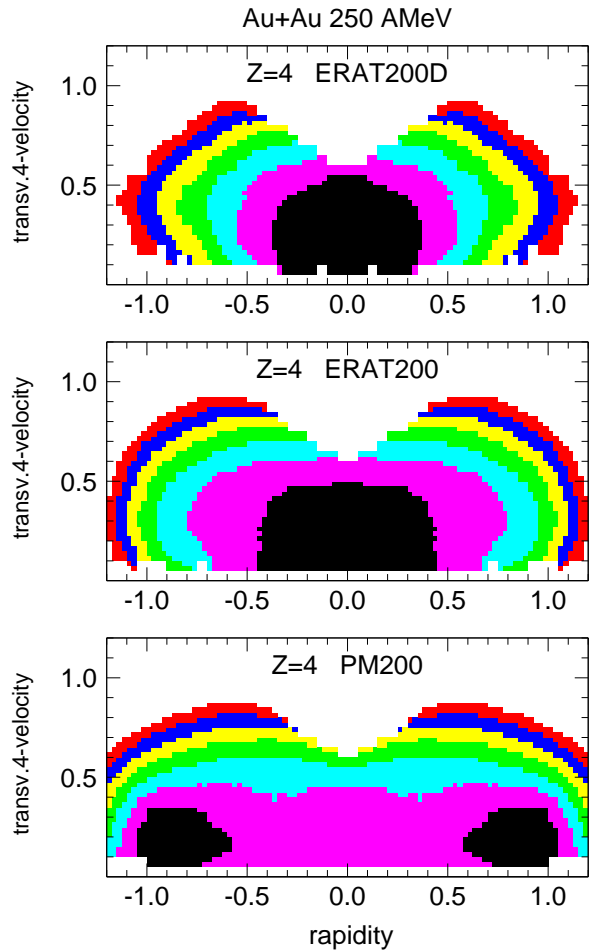

\caption{\small Invariant cross section $d^{2}\sigma/u_{t}du_{t}dy$ contour
plots (scaled units) for Be (Z=4) fragments observed in the system Au+Au at
250 A MeV. Only the forward hemisphere was measured and the data were then
reflected at midrapidity. The contours are separated by factors 1.5.
The three panels from bottom to top are for the selections PM200, ERAT200
and ERAT200D.}
\label{fig:ptyc}
\end{figure}

We start with Fig.~\ref{fig:ptyc} which shows contour plots for the
invariant cross sections $d^{2}\sigma/u_{t}du_{t}dy$
 of Z=4 fragments under the cuts PM200, ERAT200 and
ERAT200D.
In the non-relativistic limit, and with the scaling just mentioned, events
which are isotropic on the average, should lead to circular contours.
A remarkable variation of the topology is revealed:
The sizeable 'spectator' contributions in the PM200 sample have disappeared
when ERAT selections are made, the directivity cut leads to manifestly more
compact configurations on the rapidity axis \cite{alard92}.
For the selection ERAT200D
there is no question that these fragments are predominantly emitted from a
source centered at midrapidity confirming our earlier findings
\cite{alard92}.
Although near-isotropy in the {\em polar} angles is reached (see also
section~\ref{velocity}, the
ERAT200D {\em  azimuthal} distributions are isotropic by
selection), the contours reveal remarkable structural deviations that one
could associate to 'bounced-off' particles \cite{stoecker82}.
It is clear that finite number fluctuations, corona effects, and the need
for a finite sample to achieve statistical significance, are preventing the
isolation of a 'pure' fireball even if it exists.

Is the near-isotropy reached for all energies?
For the cuts, ERAT200D, we compare the topologies for Z=3
fragments at the three studied energies in Fig.~\ref{fig:ptye}: 
for this particular charge
one sees an evolution towards more prolate distributions as
the incident energy is lowered, indicating either more 'transparency' at
these lower energies or a higher difficulty, despite the same selection
procedure, to isolate 'truly central' collisions.
This sort of ambiguity is difficult to remove in a convincing way.
Naively, one expects that Pauli blocking might severely limit the
stopping power at energies increasingly close to nucleonic Fermi energies.
On the other hand,
one might also expect a highly increased 'cross-talk' between
initially overlapping parts and 'spectator' parts of the two nuclei
since the Fermi velocity is comparable to the incident-beam velocity.
The increasing 'spectator'-'participant' separation is shown     
in Fig.~\ref{fig:ptypm}, where
the topologies  are compared
using the selection PM200.
At the lower energy, besides the obvious lack of separation of 'sources'
the reader should also notice more subtle effects, such as the shift of the
'spectator' parts away from their 'nominal' positions at $y=\pm 1$.
This could be a faint remnant of dissipative effects known from many
studies of deep-inelastic heavy-ion collisions 
\cite{schroeder84} at near barrier energies all
the way up to 30 A MeV \cite{bougault95,quednau93}. 
In this case important fractions of the dinuclear system are {\em
collectively} decelerated.
Special one-body dynamics \cite{swiatecki81,feldmeier87}, valid when
two-body scatterings are Pauli-blocked, has been made responsible for such
phenomena.

\section{Velocity distributions and the blast model}              
\label{velocity}
It is our stated aim to explore to what degree a thermal model is able to
account for the data, taking into account the presence of flow.
We shall proceed in two steps:\\
a) extract the collective part of the available energy by a flow-analysis
of the measured momentum-space distributions and interprete the remaining
energy as thermal energy;\\
b) use this thermal energy as input to statistical model assumptions.
Task a) will be accomplished in the present section, task b) is reserved to
section~\ref{chemical}.
In accomplishing task a) we shall make three important simplifying
assumptions:
\begin{enumerate}
\item  we shall assume the flow is isotropic,
\item we shall allow for only a single local temperature $T$
\item we shall assume the whole system (i.e. two Au nuclei) can be
described in this way
\end{enumerate}
Let us briefly comment on these 'naive' assumptions:\\
1) clearly the ERAT200D topologies shown in fig.~\ref{fig:ptye} do not
represent a perfectly isotropic fireball. 
The introduction of more complex ideas, namely the introduction of
anisotropic 
flow patterns, would introduce additional
parameters.
As we shall see the deviations of the model fits to the data are typically
on the $20\%$ level, a modest inaccuracy in the framework of a phase-space model
and our present ambitions.\\
2) The introduction of a single local temperature is of 
importance if one wants to extract flow {\em phenomenologically} from the
data.
This assumption leads to the key equation for isotropic blasts (neglecting
relativistic corrections)
\begin{equation}
\label{eq:avekin}
<\ekin > = a + b \cdot A_{f}
\end{equation}
which states that the average kinetic energies of fragments with mass
$A_{f}$ grows linearly with mass, in contrast to purely thermal models
where it would be constant.
Equation (\ref{eq:avekin}) holds for {\em any} flow profile, as long as this
profile is common to all fragments.
The constant $b$ is the flow per nucleon, which is thus seen to be
determined unambiguously (within the model, of course) provided $<\ekin >$ 
has been measured for a sufficiently large range of fragment masses with
sufficient accuracy.
The mass lever is the only lever that can determine radial flow.
Of course flow profiles can influence the {\em shapes} of the spectra which
therefore can be used as testing ground providing additional information on
the flow mechanism.
The limitation to one local temperature makes sense only if the freeze-out
of multinucleon clusters, which are our prime interest here, is fast on the
typical expansion time scale.
As we shall see, in the central event samples considered here there is no
compelling reason to introduce complex temperature profiles. 
Also, our analysis in the framework of a self-similar adiabatic expansion
of a hot and initially compressed ideal gas, with parameters constrained by
our data, ended up with modest local variations of the final temperature of
the clusterized matter in the bulk region \cite{petrovici95}.
If the basic one-temperature assumption should prove untenable from
{\em experimental} observations, then we share the caveats expressed in
\cite{konopka95} on the possibility to determine flow and temperature
in a model-independent way just from the particle spectra.\\
3) The assumption that the whole system is participant removes any excuses
to neglect energy and mass balance and is therefore a powerful constraint.
Such an assumption is meaningful only if surface losses on the way to
'equilibrium freeze-out' are sufficiently small, say $20\%$ or less.
Experimentally, one can investigate this possibility by varying the
volume/surface ratio, i.e. by varying the size of the heavy ion system.
Such experiments are on the way.
The thermal model would be inadequate if 'preequilibrium' losses were to be
a large fraction of the system.

Here we shall mention that an alternative treatment of our data in
ref.~\cite{heide95,heide95a} in the framework of a hybrid model, combining
a transport theoretical model with a statistical model came to the
conclusion that two thirds of the nucleons (in a collision at 250 A MeV)
had escaped prior to equilibrium removing more than their share of the
available energy and leaving a relatively cold remnant.
The decay of this small remnant was then treated with the statistical
model.
The authors were able to reproduce the {\em slope} of the measured nuclear
charge distribution, but were off by a factor of three in the overall
absolute yields (our data were misrepresented by a factor three in
ref.~\cite{heide95}, see erratum \cite{heide95a}).
The fate of the two thirds of preequilibrium nucleons was left open. 

Tractable ways to handle preequilibrium are being explored
\cite{neubert96}. In section~\ref{qmd} we will use a model
\cite{aichelin86} that does not require (even local) equilibrium. 
%
%
\subsection{The blast model implementation}
\label{blast}
As outlined before,
we shall assume that at times close to freeze-out (i.e. free streaming) the
dynamics has evolved to a situation which can be described approximately by
an isotropically expanding system with a
local  temperature $T$ .
Following Cooper and Frye \cite{cooper74} we write for the invariant single
particle distribution in momentum space
\begin{equation}
\label{eq:cooper}
E \frac{d^{3}N}{d^{3}p} =
\int_{\sigma} f(x,p) p^{\mu} d\sigma _{\mu}
\end{equation}
where $p^{\mu}$ is the four-momentum, $E$ the energy including the rest
mass, $d\sigma _{\mu}$ is an element of the 3-surface $\sigma$ and $f(x,p)$
is a Lorentz-invariant distribution-function.
For a fixed time $t_{l}$ ($dt_{l} = 0$) in the local, comoving frame we can
write $d\sigma _{\mu} \equiv (d^{3}x_{l},\vec 0)$ and
%
\begin{equation}
\label{eq:cooper1}
E \frac{d^{3}N}{d^{3}p} =
\int_{\sigma} f_{l}(x_{l},p_{l}) E_{l} d^{3} x_{l}
\end{equation}
where the index $l$ holds for the local frame and where we can switch to
the center-of-mass (c.o.m.) frame by Lorentz-boost
\begin{equation}
\label{eq:lorentz}
E_{l} = \gamma _{f} (E - \vec{\beta _{f}} \vec{p}) 
\end{equation}
\begin{equation}
\label{eq:lorentz1}
d^{3}x_{l} = d^{3}x/\gamma_{f}                
\end{equation}
$\beta_{f}$ is the flow velocity which in general will depend on the
location, $\gamma^{-2} = 1 - \beta^{2}$, and $\vec{p}$ is the 3-momentum.
Equation (\ref{eq:lorentz1}) follows from the fixed-time condition
\cite{cooper74}.

In the thermal limit an invariant particle number density $n_{l}(x_{l}) =
n(x)$ can be defined.
Using the approximation of classical statistics we write
\begin{equation}
\label{eq:fxp}
f_{l}(x_{l},p_{l}) = N_{o} \frac{g}{(2\pi )^{3}} e^{\mu /T} n(r)
e^{-E_{l}/T}
\end{equation}
where $N_{o}$ is a normalization constant,
$g$ the spin degeneracy and $\mu$ the chemical potential. 
We allow the particle number density to vary with the radial distance $r$
from the center-of-mass of the total system.
If we assume that there is a unique flow-velocity versus position
correlation $\beta_{f}(r)$ at any fixed time, then we can take $r$ to be
the inverted function $r(\beta_{f})$ of $\beta_{f}(r)$
and hence define a flow
density $n_{f}(\beta_{f})$ by the condition
\begin{equation}
\label{eq:flowprofile}
n(r) r^{2} dr \equiv n_{f}(\beta _{f}) \beta _{f}^{2} d\beta _{f}
\end{equation}
Assuming that freeze out is fast on the typical expansion time scale,
 then passing from the local
frames to the c.o.m. frame and integrating over angles one obtains for the
velocity distribution ($u \equiv \beta \gamma$)
\begin{eqnarray}
\label{eq:flow}
\frac{dN}{d\beta} =
   \frac{4\pi m}{TK_{2}(m/T)} \gamma^{3} u^{2}  \nonumber \\
   \int 
          \exp ^{-\frac{\gamma_{f}E}{T}}
          \left [
            \left ( \frac{T}{E}+\gamma _{f} \right )
            \frac{\sinh \alpha }{\alpha }
            -\frac{T}{E} \cosh \alpha 
          \right ]
   n_{f}(\beta_{f}) \beta_{f}^{2} d \beta_{f}/\gamma_{f}
\end{eqnarray}
$K_{2}$ is the modified Bessel function of order 2
and $\alpha = muu_{f}/T$, ($u_{f}=\gamma_{f}\beta _{f}$).
As we will be using $dN/d\beta$ as a probability distribution for the {\em
shapes} of spectra (see later) we have also given the proper normalization
constants, which imply that the integration of eq.~(\ref{eq:flowprofile})
over $r$, resp. $\beta _{f}$, gives unity as well, i.e. $n(r)$ and
$n(\beta_{f})$ are also redefined as probability densities.
(We recall that {\em absolute} yields are analysed in
section~\ref{chemical}.)

Eq.~(\ref{eq:flow}) can be converted to $dN/dE_{k}$ by considering that the
kinetic energy $E_{k}$ is given by $E_{k}=m (\gamma - 1)$ and $dE_{k} =
dE = m\beta \gamma ^{3}d\beta $.
For the limit case where all particles are on a shell sharing a single flow
velocity one finds
\begin{equation}
\label{eq:S}
dN/dE_{k} = \frac{pE}{m^{2} T K_{2}(m/T)}  
            \exp \left (-\frac{\gamma_{f} E}{T} \right )
          \left [
            \left ( \frac{T}{E}+\gamma _{f} \right )
            \frac{\sinh \alpha }{\alpha }
            -\frac{T}{E} \cosh \alpha 
          \right ]
\end{equation}
This Ansatz was used in ref.~\cite{siemens79} and also by the EOS
Collaboration \cite{lisa95} to analyse their data.
We shall call this  'Shell' or 'Bubble' scenario in the following.

In our attempts to reproduce the data we have used besides this
single-velocity formula the
following options:
\begin{itemize}
\item $n_{f}(\beta_{f})=n_{o}$ for $\beta_f \leq \beta_{fo}$ and zero
otherwise, giving a
box-shaped profile and hence called 'Box' scenario               
(this Ansatz was used in \cite{poggi95,daniel95}).
\item $n_{f}(\beta_{f})$ is a Woods-Saxon distribution with size parameter
$\beta_{fo}$, 
relative diffuseness $a_{f}=0.1$ and a cutoff at
$\beta_{f}=\beta_{fo}(1+2a_{f})$ (called scenario WS1)
\item as before but with double the diffuseness $a_{f}=0.2$ (scenario WS2).
\end{itemize}
 
Although we want to put constraints on the flow density profile from our
data, no attempt was made to find the 'best possible' profile, as
the neglection of evaporative
influences on the spectral shapes (see later) make such an attempt
premature at the current stage of our analysis. As we shall see, there is
also a fundamental limitation to the 'resolution' of such profile shapes
due to the thermal convolution.

In general there is no simple relation between the particle number density
and the flow profile densities as this must depend on the details of the
dynamic evolution. 
The formulation eq.(\ref{eq:flow}) has the advantage of expressing final
velocity distributions in a form that is amenable for analysis of
experimental distributions.
It is tempting however to interprete the shape of the
flow profile {\em at freeze out} with the help of the scaling relation
\begin{equation}
\label{eq:hubble}
\beta _{f} = r  H
\end{equation}
where $H$ is a 'nuclear Hubble constant'. The analogy (and also the
differences) of this 'red-shift' scenario with the expansion of the
universe in the Big Bang model has been discussed by Mekjian
\cite{mekjian78}.
It is interesting to note that microscopic transport model calculations
in the present energy domain
\cite{daniel95,konopka95} also lead to almost linear dependence of the
flow velocity on the distance towards the end of the evolution.
Trivially, for sufficiently long times of 'free-streaming',
relation~(\ref{eq:hubble}) will always be valid asymptotically as the
expanding system has moved to distances large compared to radii at which
frequent hard collisions are still taking place.
A linear relation between velocity and position {\em at all times} in the
dynamic evolution (not just at or after freeze-out) is the characteristic
of a class of 'self-similar' solutions to the ideal-gas hydrodynamics
\cite{sedov57}
which were introduced into the
nuclear context in  \cite{bondorf78} and  applied to some
of our data in an earlier publication \cite{petrovici95}.
Finally, it should be mentioned that such scaling can also be justified 
in {\em collisionless} (repulsive) mean-field expansions.

It should be clear, however, that the validity of equation~(\ref{eq:hubble})
is not essential for our present analysis.
In particular, we are not trying here to backtrace from the freeze-out
configuration to earlier times of maximum compression and minimal or zero
flow.
We refer the reader to \cite{petrovici95} for instructive results obtained
from an analysis of our data at 250 A MeV in terms of self-similar expansion
dynamics. 

Our object was to determine the collective (flow) energy $\ecoll $
from the measured charge-separated spectra.
Energy conservation requires that
\begin{equation}
\label{eq:energy}
\ecoll  + \eth = \ecm   + \qval  
\end{equation}
where $\ecm $ is the center-of-mass energy in the incident channel,
$\qval $ is the Q-value of the reaction and $\eth $ is the
thermal energy.
  
Once the velocity profile (Shell, Box, WS1, WS2) is fixed, the collective part of
the spectra is determined by a single scaling parameter $\beta _{fo}$.
Because of Eq.~(\ref{eq:energy}) the thermal part is then fixed as well. For
the calculation of the spectral shapes using Eq.~(\ref{eq:flow}) we then
still need the temperature T.
However, T is fixed if we use classical statistics: T is
determined by the multiplicity M of all emitted particles (finite-number
fluctuations are discussed later).
In the non-relativistic limit we have
\begin{equation}
\label{eq:T}
\eth   = \frac{3}{2} M T
\end{equation}
Here we have ignored the internal excitation energies of the final observed
fragments.
The temperature from eq.~(\ref{eq:T}) is 'effective' in the sense that it
is not the temperature at freeze-out time, since the
observed multiplicity has been raised due to late particle decays.
Eq.~(\ref{eq:T}) has however the virtue of being compatible with energy
conservation and will only be used for the 'chaotic' part of the spectral
shapes.
Later in section~\ref{chemical} we fix the thermal {\em energy} not the
temperature to calculate fragment yields.
Statistical evaporation does not change on the average the local flow
velocity as it must obey momentum conservation in the c.o.m. of the
decaying particle, hence the separation into a flow term and a specific
local term is not affected.
Or expressed more simply: {\em statistical particle emission does not create
flow}.
However the resulting spectra, although still 'chaotic' 
(in the sense that the details of the evaporative processes are not
resolved) are no longer
thermal in the strict sense. 
They could have some structural mass dependences, in contrast to true
thermal spectra.
It is assumed here that this dependence is unsystematic and hence can be
ignored in a first approximation.
Close inspection of the actual particle spectra (section \ref{kinetic})
will help to assess this simplification.

The multiplicity M in eq.~(\ref{eq:T}) is obtained as follows:
Neglecting pion production the neutron multiplicity $M_{n}$ is calculated
using
\begin{equation}
\label{eq:Mn}
M_{n} = 236 - (M_{d}+M_{h}) - 2 (M_{t}+M_{\alpha}) - \sum _{Z=3}^{12} ZM_{z}
\end{equation}
where 236 is the total number of neutrons available in the double-gold
system and where one uses the measured multiplicities $M_{z}$ (for
fragments with nuclear charge Z), $M_{t}$ (for tritons), $M_{d}$ (for
deuterons), $M_{\alpha}$ (for $^4$He) and $M_{h}$ (for $^3$He).
Eq.~\ref{eq:Mn} implies that each IMF (Z=3-12, yields with $Z>12$ are
negligible for central collisions at these energies), carries as many
neutrons as protons. 
The yields of H and He isotopes were measured \cite{poggi95} only at 150
and 250 A MeV.        
For the 400 A MeV data, where the charges but not the isotopes were
separated, we assumed that the H isotope ratios were the same as at 250 A
MeV , while the $^3$He/$^4$He ratio was extrapolated to be 0.78 from a
linear fit of the measured energy dependence at 100-250 A MeV. 

We have determined the ratio $\ecoll /\eth $ with the
constraints of eqs.~(\ref{eq:energy}) and (\ref{eq:T})
  by use of a maximum-likelihood
method which was applied either to single events or to groups of $N$ events
assembled together to form one 'macro'event.
$N$ could be any number, usually between 1 and 400. 
Geometrical and threshold limits, which could cause a bias, were taken into
account in the following way:
we maximize the logarithm of the likelihood $\mathcal{L}(\beta _{f0})$
defined by
\begin{equation}
\label{eq:likeli}
\mathcal{L} = \prod _{i} f_{i}/N_{i}
\end{equation}
where the product is over all, or alternatively a subset of, the registered
particles $i$ in a single, resp. a macroevent and
\begin{tabbing}
$f_{i}$ \= $=$ \= $(dN/d\beta )_{i}$ \= \hspace{1cm} within detector
acceptance \\
\> $=$ \> $0$ \> \hspace{1cm} outside detector acceptance
\end{tabbing} 
$N_{i}$ is given by the normalization condition
\begin{equation}
\label{eq:norm}
\int f_{i}/N_{i} d\beta = 1
\end{equation}
$(dN/d\beta )_{i}$ is given by eq.~(\ref{eq:flow}) and is different for each
mass, but averages over dependencies on polar angle.
 Since we resolved the charges, rather than the masses, we assumed
$A=2Z$.

As the detector acceptance depends on nuclear charge (thresholds) and
laboratory angle $\tlab $ (angular cuts, shadows) we have the
dependence $N_{i}=N_{i}(Z,\tlab )$.
In order to ease the calculation of the normalization constants, we have
applied to the data a 'clean-cut' filter consisting of three types of cuts:
\begin{enumerate}
\item
angular cuts: only fragments between $\tlab  = 1.2^{\circ}-30^{\circ}$,
but not within the partially shadowed angles $5^{\circ}-7.5^{\circ}$ and
$19^{\circ}-21.5^{\circ}$ were accepted for the fit.
\item
For the External Wall ($\tlab  > 7.5^{\circ}$) the thresholds
and other efficiency cuts were replaced by (conservative) c.o.m. angle
cuts, $\tcm > \tcmew$, given in Table~\ref{tab:limits}
\item
For the Internal Wall Z-dependent rapidity cuts were done, $\ylab (Z) >
\ylabiw $, that are also given in the table.
\end{enumerate}
This clean-cut filter is more restrictive than the usual PHASE I filter
\cite{hoelbling92} adopted by our Collaboration for filtering theoretical
data prior to comparison with our results.

Technically, the collectivity is varied in steps of $10\%$ using pre-stored
functions and calculating $N_{i}(Z_{i},\theta _{i})$ numerically. For the
final result a quadratic interpolation is done.
To check the procedure, 'theoretical' blast events, after filtering, were
also fed into the analysis instead of the experimental events and it was
found that the original blast parameters were recovered despite the
apparatus influences.
Under the condition that most events have a sufficient number of particles
with different mass, the fitting procedure converges very fast: event-wise
fluctuations of apparent collectivity were about $25\%$ and the fluctuation
in a 'macro'event diminished roughly with the square-root of the assembled
number of events.
The method therefore allows in principle to be extremely selective in the
choice of the event sample.
As a curiosity let us add: in a perfect $4\pi$ detector that would measure
and correctly identify all particles, including neutrons, just ten events
would be sufficient to fit     the radial flow with an accuracy of $5\%$ in the
energy regime considered here.

Instead of equation (\ref{eq:T}) which implies a mass-independent thermal
energy of $3/2T$ for each emitted fragment, we actually use the
relativistically correct (but still classical) formula
\begin{equation}
\label{eq:Trel}
\eth  = \sum_{i} 3T+ m_{i}(K_{1}(x_i)-K_{2}(x_i))/K_{2}(x_i)
\end{equation}
where $K_{n}$ are modified Bessel functions of argument $x_i=m_{i}/T$.
For a fixed value of the total thermal energy $\eth $ one has to
determine T selfconsistently by using the mass-dependent multiplicities
extrapolated to $4\pi$.
As the extrapolation depends in turn on the fitted scenario, the procedure
requires iterations.
Only one iteration was necessary as the relativistic corrections on the
thermal energies remain modest in the present energy range. 

As the velocity profile is 'folded' with the temperature its precise shape
is not resolved, however the 'resolving power' increases with the mass of
the observed fragment. 
This resolving power is demonstrated in Fig.~\ref{fig:blast}.
The panels show 'blast' spectra for a scenario at 250 A MeV with an assumed
50/50 ratio of $\ecoll /\eth  $ in scaled units. The
mass distribution was taken from experiment. The panels demonstrate three
major points:
\begin{itemize}
\item while the fully integrated rapidity spectra for mass $A=8$ shows
noticeably flatter tops for the single-velocity scenario 'Shell' than the more
complex velocity profiles 'Box' and 'WS2', the difference turns dramatic when
one applies in addition a cut on the scaled transverse four-velocity
 ($u_{t} < 0.6$) :
the upper left panel shows how a 'bubble' ('Shell' scenario )
configuration is influencing             the rapidity spectrum;
\item the same drawing for mass $A=1$ shows practically no difference
between the various scenarios: extracting details of the blast profile is
therefore close to impossible with single nucleons (and a fortiori kaons or
pions);
\item a very large sensitivity is also found in the transverse
four-velocity spectra if a rapidity cut ($|y| < 0.5$) is
performed; our Phase I apparatus limit ($u_{t} < 0.6$ at $90^{\circ}$ in 
the c.o.m.)
 is high enough to judge if the 'Bubble' hole of the 'Shell' scenario exists.
Since the blast scenario is isotropic the approximate equivalence between
panels a and d in Fig.~\ref{fig:blast} is expected, but real data may not
have this symmetry.
\end{itemize} 
\subsection{Experimental data and the blast model}
\label{expdata}
A summary of the flow characteristics deduced from the blast model analysis
is given in Table~\ref{tab:results}.
One of the main results is that $(62\pm 8)\%$ of the available energy is
stored into radial flow.
This value is significantly larger than the values recently published by the EOS
collaboration \cite{lisa95} , but not too far from our earlier, more
preliminary analyses \cite{reisdorf94,reisdorf94a} and in line with,
although somewhat larger than values from analyses reported in
\cite{jeong94,hsi94,petrovici95}.
The associated average flow velocity $<\beta _{f}>$ is also given in the
Table.

In the following we shall present a sample of our data and use the blast
model fits as reference.
We recall that once the velocity profile is fixed the fit has only one free
parameter for the description of the spectral shapes of all the fragments.
The absolute yields will be treated in section~\ref{chemical}.
It is clear that we cannot present the full richness of the data and we
point out that other sectors of the data can be made available on request.
If not otherwise specified, we compare with the data obtained using the ERAT
plus directivity selection, ERAT200D, defined in section~\ref{centrality}
and use scaled units.
Since FOPI was not a full $4\pi$ detector at the time of this experiment,
we have chosen some cuts in phase space (outlined in the following
sections)
 which we believe to be minimally
distorted by apparatus effects, aside of course for the well defined 
geometrical limits.   
Further theoretical efforts to understand these data in more fundamental
detail as we claim here should therefore be minimally plagued by obscure
'apparatus effects'.
We remind that the 'third dimension' is averaged out in these data, but as
discussed earlier, due to the high centrality selection, and in particular
the directivity cut, azimuthal variations are reasonably small.

As the (unmeasured) mass fluctuation of the hydrogen fragments is,
relatively speaking, large, these fragments are not very useful in our 
special fitting context. 
Excluding the hydrogen fragments from the fit generally lead to a better
description of the other spectra and increased the deduced collectivity at
150 and 250 A MeV by about 10 \%. This is due to the fact that the   
average mass of Z=1 isotopes is somewhat lower \cite{poggi95} than
the value $A=2$ assumed to estimate the likelihood.
This effect is stronger at 400 A MeV where the average mass of hydrogen
particles is expected to be significantly less than 2.
In order to treat the data for the three incident energies on the same
basis we opted to exclude the hydrogen isotopes from the fit at all three
energies.
We shall however compare the fit prediction also to the
measured hydrogen spectra with and without isotope separation in order to
assess the deviations from the blast model reference.

We shall start with rapidity spectra,
then switch to average kinetic energies as function of mass,
kinetic energy spectra, transverse-four-velocity spectra and, 
finally, polar angle
distributions.

\subsubsection{Rapidity spectra}         
\label{rapidity}

In the following we shall show rapidity distributions with a cut,
$u_{t}$ on the transverse four-velocity.
From the previous section we have learned that the rapidity spectrum with a
cut on the transverse four-velocity could discriminate between various
blast profiles if sufficiently heavy clusters are available.
In a simultaneous fit to all data (both in terms of various cluster sizes
and in terms of the full measured phase space) it is however not obvious 
{\em a priori} on what the fit will eventually compromise.
In Fig.~\ref{fig:rapidity250} we show from the data at 250 A MeV the rapidity
distributions for clusters of nuclear charge $Z=4$ and $Z=5,6$ together
with the blast model fits using the 'Shell','Box', and 'WS2' velocity profiles
defined in section~\ref{blast}.
Apart from the obvious peaking of the data
 at midrapidity, which confirms once more the
central-source origin of these clusters, one can also see that the 'Bubble'
or 'Shell' scenario is evidently in conflict with the data (lower panels).
Although the 'Box' scenario (middle panels) avoids the two-humped
feature of the 'Bubble' scenario, it is still seen to be too flat.
Only the Woods-Saxon profile 'WS2' gives a reasonable account (note the
{\em linear} scale) of the data.
While there is a clear need for a finite diffuseness, we found the
difference to a scenario with half the diffuseness, 'WS1', (not shown), to
be too small to make a definite choice.
We finally opted for the large diffuseness because it yields a somewhat
smaller thermal energy $\eth $.
The reasons why we want to minimize the latter (while preserving a good
description of the velocity space distribution) will become clear in
section~\ref{chemical}.
So in the following we shall show only fits with the scenario 'WS2'.
With this scenario fixed, the blast fits are a {\em one-parameter}
optimization: the ratio $\ecoll /\eth $.
Quantitative results are summarized in Table~\ref{tab:results}.
The leading error estimate (in the frame of the model, of course)
can be found in  row two, $\ecoll $, and stems from the uncertainty of the
details of the flow density profile and small differences between the
ERAT200D and ERAT50 cuts. 

In Figures \ref{fig:rapidity150} and \ref{fig:rapidity400} we show rapidity
distributions for various fragments observed at 150, resp. 400 A MeV
and compare with scenario 'WS2'.
Several remarks can be made.  
The blast model describes the rapidity distributions of the IMF ($Z>2$) with
an accuracy of about $10 \%$ or better.
The width of the distributions of the lighter particles tend to be larger,
the deviation from the overall fit is most pronounced for the hydrogen
fragments, especially at the lower energy.

\subsubsection{Average kinetic energies}
\label{average}
Fig.~\ref{fig:avekin} shows the measured average kinetic energies as a
function of mass in the c.o.m. polar angle range $25-45^{\circ}$.
In this angular range apparatus cuts are negligible at 400 A MeV and small
($< 2\%$ for $Z<6$, $5\%$ for $Z=8$) at 150 A MeV.
It should be noted that we measure velocities and nuclear charges.
In the figure we have primed both the mass and the kinetic energy because
the actually plotted quantities are $2Z<\ekin /A>$ versus $2Z$.
This representation evidences the approximately linear dependence on mass
($\approx 2Z$) and
stresses once more the fundamental deviation from a
thermodynamically equilibrated system that is enclosed in a box rather than
exploding into the vacuum.
In principle, as stressed before,
in a blast model description of the data with {\em one}
global temperature, the slope $d<\ekin >/dA$ fixes the collective energy
{\em independently on the profile}.
However, as we shall see, this is not the case for fitting strategies that
require a good reproduction of the full momentum space population, rather
than reproducing the more limited information contained in the slopes
$d<\ekin>/dA$.
The results of the blast model fits with our preferred scenario, 'WS2', are
shown as solid lines for all three incident energies.
On the panel for the 250 A MeV data we have also shown the curves obtained
with use of the scenario's 'Shell' (dotted) and 'Box' (dashed).
As for the rapidity distributions we see a gradual improvement of the data
reproduction when going to increasingly complex profiles with, however, a
tendency to 'saturate': eventually the slope does not change much any more
when going from 'Box' to 'WS2' (the 'WS1' profile which is not shown is of
course intermediate between the latter two).
This saturation in the mass dependence and hence the collectivity is also
shown in Fig.~\ref{fig:collectivity} for the 150 A MeV data and turned out
to be very similar for all three studied energies.
A priori it may seem surprising that a fit such as 'Shell' would completely
miss the measured mass dependence.
We recall that the fit has to make a compromise in order to reproduce 
the data in the
full measured phase space  with the additional constraint of energy 
conservation: had the 'Shell' slope in panel (a) of Fig.~\ref{fig:avekin} come
out steeper, and hence closer to the
average kinetic energy data, then the double-hump in the
rapidity distributions for the 'Shell' scenario,
 see Fig.~\ref{fig:rapidity250} would
have come out even more dramatic, in total disagreement with this sector of
phase space.

We have explored how robust the deduced collectivity was against ignoring
all fragments emitted forward of $45^{\circ}$.
We found only insignificant variations of the deduced
collectivities, if at all flow was {\em increasing} by a few $\%$:
$62.7\pm 2.2$, $63.3\pm 1.8$, $67.6\pm 2.3 \%$ for the three energies,
respectively and for the WS2 scenario. These values can be compared with
row 2 of table~\ref{tab:results}.
The main difference of the data between $25^{\circ}$ and $45^{\circ}$ and
the full set of data is visible in Fig.~\ref{fig:avekin} as an overall
offset of the curves representing the globally fitted energies which is
most pronounced for the 400 A MeV data and seems to be limited to nuclear
charges $Z \leq 5$ (mass' $\leq$ 10).
  
In Fig.~\ref{fig:avekinlp} we show on a blown-up scale the average kinetic
energies of the hydrogen and helium isotopes at
$\tcm =(60-90)^{\circ}$ which have been discussed in an earlier
publication \cite{poggi95} and compare with the global trend of the blast
model fit.
These data, obtained with separate telescopes, but during the same runs,
were triggered by a selection on ERAT corresponding to approximately 250 mb
and were not part of the fit procedure which used only data from the
Plastic Wall.
The global trend of the blast, which is identical to the one shown in
Fig.~\ref{fig:avekin}, represents a fair compromise in view of the fact that
the model used here is confined to a linear mass dependence and that
hydrogen fragments were excluded altogether from the fit.
It is obvious however that there is an important scatter of the data points
away from the smooth trend.
This scatter is larger at the lower energy, where in particular the 
alpha energy is quite low and the proton energy rather high.
From the data \cite{poggi95} at 100, 150 and 250 A MeV one finds that the
average kinetic energies of the protons emitted close to $90^{\circ}$ is
equal to the incident center of mass energy per nucleon plus an additional
energy of $(26.5\pm 2)$ MeV which is slightly higher than the average
kinetic Fermi-energy of ground-state nuclei.
This is remarkable as this is an {\em average} energy of single nucleons
emitted at rather large angles in {\em central} collisions.
Relatively speaking Fig.~\ref{fig:avekinlp} shows that the 'surplus' proton
energy over the blast model prediction increases with decreasing incident
energy if one scales the energies. 

Another striking observation from these light-particle data is the fact
that $^{3}$He fragments have higher kinetic energies than the $^4$He
fragments, a trend opposite to flow considerations.
This '$^{3}$He puzzle' has now been seen by three groups
\cite{poggi95,lisa95,doss88}.

It is not obvious from looking at the five data points in both panels
of fig.~\ref{fig:avekinlp} which of the points is really 'anomalous', but it
is clear that analyses that use only these light particles to extract
radial flow values \cite{poggi95,lisa95} can come to different conclusions
concerning flow.

The analysis of the EOS group \cite{lisa95} has overlap with our present
analysis at 250 and 400 A MeV.
Since the error bar for the 400 A MeV data given in ref.~\cite{lisa95} is
large we shall discuss only the 250 A MeV results here.
Using the single-velocity formula \cite{siemens79} and concentrating on
{\em all light charged particles} the EOS group extracts a radial flow of
$17\pm 5$ A MeV at this energy which is only $50\%$ of ours.
Their blast-model 'flow-line' (i.e. $\ekin $ versus $A$) runs through
their p,d,t and $\alpha$ data and is well below the $^{3}$He point.
A more detailed comparison
of \cite{lisa95} with \cite{poggi95} is difficult as there are
some discrepancies: the EOS average energies for the H isotopes are about
20 MeV lower, the $^{4}$He energy is about 30 Mev lower, but the $^{3}$He
values agree with \cite{poggi95}. These differences are disturbing.

The flow analysis presented in \cite{poggi95} used {\em only the hydrogen
isotopes}, i.e. the slope in fig.~\ref{fig:avekinlp} given by the average
kinetic energies of H, $^2$H and $^3$H which came out to be 20 A MeV at an
incident energy of 250 A MeV.
This procedure {\em does not include the Coulomb part} of the flow 
(which essentially acts as an energy offset when the mass is changed but
not the charge $Z$),
 which
we estimate to be 5 MeV per nucleon (see shortly) giving a total of 25 A MeV
or $74\%$ of our present value deduced from all fragments, {\em excluding
the hydrogen isotopes}.

When passing from $^3$H to $^3$He, in a naive picture, only the Coulomb
part of the flow should change, since the charge, but not the mass was
changed.
In ref.~\cite{poggi95} the energy of $^3$He was found to exceed that of
$^3$H by $21.5\pm 3$ MeV for incident energies in the range 100-250 A MeV.
Several independent estimates concur to say that Coulomb effects at best
could explain half this energy difference.
A simulation, done in ref.\cite{poggi95} using the statistical model code
WIX \cite{randrup93} and starting from a rather condensed spherical
configuration with nucleonic density $\rho=0.8\rho _{0}$ ($\rho _{0}=0.17$
fm$^{-3}$) taking into account Coulomb fields, predicts a 'Coulomb flow' of
about 12 MeV per unit charge.
A transport model simulation \cite{daniel95,daniel92},
 also shown in \cite{poggi95},
came to a similar energy difference of about 10 MeV and, finally, a
measurement for Au on Au at 35 A MeV showed \cite{dagostino96}, for central
collisions, kinetic energies slightly below 5 MeV {\em per nucleon},
 i.e. roughly
again 10 MeV {\em per charge}. 
 
There are several possible reasons for the 'irregularities' in
light-particle energies that make them subtle to use for flow energy
determinations in the present energy range.
The most obvious one is the occurrence of late evaporative decay, a more
subtle one are reorganisation processes during clusterization \cite{bondorf95a}.The former would involve alpha particles much more than the $^3$He
fragments, which would tend to look more 'primordial'\cite{gutbrod82}.
Indeed, the low alpha energies could be accounted for in a simulation
including evaporative contributions \cite{poggi95}.
One could also speculate that tritons would in general have higher
contributions from decays than the Coulomb-inhibited $^3$He's.
However, the t/$^3$He difference could not be explained in
ref.~\cite{poggi95}.
As the statistical model has other difficulties, as we shall see in section
\ref{chemical}, a firm conclusion is not yet possible.

The reorganization processes, or self-organization alluded to in
ref.~\cite{bondorf95a}, concern the processes of amalgamation of nucleons
to clusters. Energy and momentum conservation in such local processes
demand that the condensation energy be taken up by the surroundings, that
do not clusterize, i.e. by single nucleons. This could explain the high
energies of single protons even at $90^{\circ}$ in the c.o.m..  

Characteristic structural influences of the mean kinetic energies are also
visible from Fig.~\ref{fig:avekin}, for example the S-like shape
particularly visible with the 400 A MeV data. 

\subsubsection{Kinetic energy spectra}
\label{kinetic}
How well does the blast model reproduce the {\em shape} of the kinetic
energy spectra?
This is shown in Fig.~\ref{fig:ekin250}
for charge separated fragments at $(25-45)^{\circ}$ and Fig.~\ref{fig:ekin250lp}
for the light-particle data at $(60-90)^\circ$.
The reproduction of the data by the WS2 blast is rather good. 
The Z=1 theoretical spectrum around $35^{\circ}$ was constructed assuming that the d/p
and t/p ratios are the same than at $90^{\circ}$
\cite{poggi95}, but then substituting for
the conversion from velocities to energies the assumption $A=2Z$ used in
the experimental evaluation.
The resulting rather steeply sloped spectrum represents the most serious
deviation of the model from the data.

In Fig.~\ref{fig:ekin250lp} the model calculation used again the known
isotopic ratios, but not the {\em absolute} values which were fixed from
the $4\pi$ extrapolated Z-distribution resulting from the fit to the
Plastic Wall data. 
In view of this, the reproduction of the data also on an absolute scale is
satisfactory and shows the internal consistency of the analysis.   
There are some subtle deviations of the model from the data.
The most striking ones concern protons and $^{4}$He.
Relative to the blast model the data show a lack of low-momentum yields for
the protons and a lack of high-momentum yields for the $^{4}$He fragments.
Both of these observations tend to support our speculations made in the
previous section.

\subsubsection{Transverse four-velocity spectra}
\label{transverse}
Another aspect of the data, the transverse momentum distributions, is shown
in Fig.~\ref{fig:pt} for the 150 and 400 A MeV data for two cuts on
rapidity.
The $30^{\circ}$ laboratory cut starts influencing the data at
$u_{t}>0.6$ ($y<0.5$) and $u_{t}>1.0$ ($y>0.5$).
There is an excellent description of the midrapidity part, even for Z=1.
None of the spectra
shows a trace of a bubble (or hollow cylinder) depression.
The high rapidity part is generally underestimated, revealing the  
deviations from the assumed isotropy (more prolate event shapes)
 at larger rapidities.
This trend is larger for the Z=1 and 2 fragments than for the heavier
fragments and for the latter it is larger at 150 A MeV than at 400 A MeV
showing once more the higher conversion to isotropy at the higher energy
that was visible already in the topology plots in section \ref{centrality}. 

\subsubsection{Polar angular distributions}
\label{polar}
In the limit of very central collisions flow effects with directional
preference can no longer be quantized with respect to a 'reaction plane'
\cite{daniel85} since the high degree of axial symmetry achieved in such
collisions, together with finite-number fluctuations, precludes an accurate
reconstruction of this plane.
In that case one expects {\em polar} angular distributions to exhibit
possible signatures of anisotropic flow.

In particular the 'squeeze-out' effect
 \cite{dlhote88,gutbrod89a,gutbrod90,wang96}, if
dominated by compressional, possibly shock-like effects, as implied by the
name \cite{stoecker82}, should then become maximal and lead to a pronounced
peaking around $90^{\circ}$ in the c.o.m..
Such predictions were made long ago \cite{scheid74} in the framework of
classical hydrodynamics and were also made  recently by 
microscopic calculations \cite{daniel95,daniel92,konopka94a}.
The other important  cause of the preferential 'out-of-plane'
emission of high-momentum ($u > 1$, in scaled units)
 fragments (if done in the proper
flow frame that avoids 'spectator' contaminations \cite{gutbrod90}), namely
shadowing effects caused by absorption or at least 'cooling down' in
relatively cold non-participant matter, should be minimal in the case of
truly central events in symmetric systems.

We expect that data obtained with the completed FOPI detector will
elucidate this important question in the near future.
With our Phase I limitation momentum spectra at $90^{\circ}$ are cut off at
scaled momenta $p>0.6$ .
Nevertheless it is useful to take this exploratory look at polar angle
distributions in highly central collisions for at least two reasons:
a) since we have used an isotropic blast model as reference for comparison
with the data, it is important to assess how well isotropy is fulfilled at
least to the degree to which this is possible at present;   
b) the high sensitivity of the heaviest observed fragments to flow in
general, demonstrated earlier, might lower the limit of $p$ above
which directional flow effects become measurable as compared to observing
single nucleons \cite{lambrecht94}.
Indeed, for less central events we have seen clear squeeze-out signals even
in this limited geometry when looking at IMF's \cite{bastid96}.
 
In the following we shall concentrate on the 400 AMeV data as compressional
effects are likely to be stronger at this energy than at 150 or 250 AMeV
and because threshold effects are less important. 
A more detailed investigation of the 150 AMeV data
will be published separately \cite{roy96}.  
We recall (see section \ref{choosing}) that here we have attempted to
minimize autocorrelation (which distorts the distributions severely, in
particular for heavy fragments) by excluding the particle of interest from
the calculation of the selecting global quantity (PM, ERAT, Directivity).
 
Figure \ref{fig:polar1} shows for fragments of different size the degree of
compatibility of the data (cut ERAT200D) with the isotropic scenario.
Deviations from isotropy (taking into account the geometric cuts) are on
the $20\%$ level or less.
In particular the yields of charge-one fragments show a surplus at angles
forward of $50^{\circ}$ which increases with decreasing angle.
Due to the $1/\sin \theta$ weighting this surplus is actually less than
$10\%$ in terms of count rates and general balances in charge and energy.
The overshoot at angles below $45^{\circ}$ (necessarily balanced by a slight
undershoot at higher angles) is higher at 150 AMeV where it reaches $30\%$
and extends to all fragment sizes.
The more prolate topology at this lower energy, when using the present
selection methods, has been displayed already in section \ref{centrality}. 

If one makes a cut $u \leq 0.5$ on the scaled four-velocities, one can
check, at 400 AMeV, the isotropy from $10^{\circ}$ to $160^{\circ}$.
This is shown in Fig.~\ref{fig:polar2} for different global event selections.
Isotropy holds within $20\%$ or better for all these selections.
A few finer points can be noted:
for the 200 mb selection on PM a faint remnant of side peaking near
$25^{\circ}$ is visible; after application of the directivity cut, PM200D,
the cross section drops to 20 mb and statistical limitations prevent one
from observing deviations from isotropy below the $20\%$ level.
The selections ERAT200 ($\approx 200$ mb) and ERAT200D ($\approx 80$ mb)
are very similar, but tend to show a forward/backward asymmetry on the
order of $10\%$ despite the removal of the autocorrelation.
Although we cannot exclude completely double-hit distortions, there are at
least two other 'true physics' reasons why one would see asymmetries of
this order of magnitude when using our selection methods.
First, we define ERAT, Eq.~(\ref{eq:erat}), using only particles emitted in the
forward hemisphere,i.e. our selection criterion is not symmetric with
respect to the c.o.m.. 
Applying the same technique (and filtering) to our
blast model simulation, we were able to reproduce such asymmeties and
interprete them as forward-backward fluctuations due to finite number
effects. 
A second effect, that was not incorporated in our simulation (at the
present stage) is the fact that the autocorrelation might not be completely
removed, if many IMF's were to be accompanied by a splitting partner at the
late evaporative stage (so that the 'primary' correlation is not completely
removed as the splitting partner is not identified).

When using again the $u$ cut, the data at 250 A MeV and 150 A MeV
look very similar, even in detail, and are therefore not shown here.

To assess the importance of the isotropy in this 'inner' part (momentum
space) of the fireball, one should be aware that the $u<0.5$
cut is rather severe and represents only $20-25\%$ of the full spectra at
$45^{\circ}$ (c.o.m.).
It should also be remembered that only the ERAT200D and PM200D selections
are {\em azimuthally} symmetric (by construction), while the ERAT200 and
the PM200 selections, even with a mid-rapidity cut $0<y<0.5$, are not
(see Fig.~\ref{fig:azimuth2}).
  
Our general conclusion concerning polar angular distributions is that they
do not show convincing evidence for {\em dramatic} shock effects (i.e.
$90^{\circ}$ {\em polar} squeeze out) as predicted semi-quantitatively in
the early days \cite{scheid74,stoecker82}.
Since out-of-plane yield enhancement {\em has} been seen \cite{bastid96}
under similarly restrictive cut conditions, especially for collisions with
rather large impact parameters (7 fm), one is also tempted to conclude that
the presence of non-participant matter is required for the mechanism of
'squeeze-out'.
Full phase space coverage in future data should further qualify these
statements.


\section{Chemical composition}
\label{chemical}
Besides the velocity-space distributions the second topic of high interest
in these central collisions is the distribution of the yields for the
various fragment species.
In Fig.~\ref{fig:zrobust} we show as example the measured fragment charge
distributions at 250 A MeV incident energy.
Various centrality cuts were applied (see Table \ref{tab:eventsample}):         
PM200, ERAT200, ERAT200D, ERAT50.
In addition to
the clean-cut filter specified in Table \ref{tab:limits} and section
\ref{blast} the data were restricted to polar angles
$\theta _{cm}>20^{\circ }$ .
It is evident from the figure that the nuclear charge distribution is much
more robust to the cut than the velocity-space topologies discussed
earlier (see also \cite{kuhn93}).
This fulfills a necessary, although not sufficient, condition for the
application of a one temperature concept as used in the present
analysis.

Such clean-cut distributions formed the basis for extrapolating the data to
$4\pi$.
Specifically, a blast-model event-generator with the mass-dependent
velocity-space parameterization determined earlier, was fed with the
measured filtered Z-distribution and the missing parts were numerically
calculated as function of Z and $\theta _{cm}$.
The resulting $4\pi$ integrated yields are given in Table
\ref{tab:multiplicity} and plotted in Fig.~\ref{fig:zdist}.
As an important consistency check the $4\pi$ integrated total charge obtained
in this way should be 158 for the Au on Au reaction.
Table \ref{tab:results} shows the \% deviations we found from the adopted
WS2 scenario.
Except for the 150 A MeV data, where a 13\% deviation was obtained, the
total charge is reproduced within 5\% or better.
This shows that the isotropic blast scenario is not dramatically different
from the data and in particular that the missing high-momentum parts near
$90^{\circ}$, that will be measured in future runs with the full detector,
are unlikely to change this conclusion qualitatively.
The higher deviation for the 150 A MeV data was to be expected in view of
the somewhat prolate topologies at this energy and for the selections
chosen here.
As a zero-order correction and neglecting pionic contributions, the yields
in Table \ref{tab:multiplicity} were renormalized to 158.
The errors given in the Table are statistical only.
The uncertainty from the cut-dependence is 10\% at the most till about Z=6
and increases up to 30\% at the end of the shown distributions.
The uncertainty from the extrapolation is of the order of the normalization
deviations just discussed and concern primarily the ratio of the hydrogen
fragments to the rest, as the phase-space distributions of the former are
not described so well, especially away from mid-rapidity.
As we shall see all these uncertainties are well below the uncertainties of
theoretical models.  

Fig.~\ref{fig:zdist} shows also the exponential curves
$N\exp(-\lambda Z)$,  that were least-squares
fitted to the data excluding the light particles and the $Z=4$ point (which
is low because $^8$Be decays before it reaches the detectors).
The resulting slopes $\lambda$ were $0.625\pm 0.010$, $0.908\pm 0.012$ and
$1.170\pm 0.018$ for 150, 250 and 400 A MeV, respectively. 
The resulting average cluster masses are $3.20\pm 0.05$, $2.20\pm 0.03$ and
$1.71\pm 0.03$, respectively (the errors imply a $\chi ^{2}$ per degree of
freedom equal to one).
These fits  serve to
enhance the special role played by hydrogen and helium fragments and also
reveal some odd-even structures: the fragment yields for $Z=9$ (F) and 11
(Na) are presumably somewhat low because $^{18}$F and $^{22}$Na are less
stable isotopes.
All of this suggests the influence of evaporative decays.
As a matter of fact, if one were to speculate that
{\em all} the surplus of hydrogen
and helium nuclei above the exponential extrapolation is due to
evaporation,
then less than half of the light charged particles are 'primordial' in this
energy regime.

With the aid of the light-particle data of ref.~\cite{poggi95} and the
neutron emission estimate presented in section~\ref{blast},
 it is possible to complete
the picture with various other multiplicities given in Table
\ref{tab:varmult}. 
In using Table~\ref{tab:varmult} it should be kept in mind that the
isotopic composition of the Z=1 and 2 fragments at 400 A MeV has been
extrapolated from \cite{poggi95} (section~\ref{blast}) and that neutron
yields were estimated using eq.(~\ref{eq:Mn}).
The error in the latter was estimated by replacing the assumption $A=2Z$ by
 $A=2Z+1$ for $Z>2$.
An idea of how 'clusterized' the hot nuclear matter is at freeze out is
given in Table~\ref{tab:cluster} which shows the percentage of protons and
nucleons bound into clusters. These numbers are also plotted in
Fig.~\ref{fig:cluster} together with the deduced IMF multiplicities, an
observable that finds great attention in the multifragmentation literature.
The 'degree of clusterization' of protons is high (about 80\%) and hardly
varies with incident energy.  What is changing with energy is the \emph
{size} of the clusters.

Another point of interest are the event-by-event fluctuations of some of
these multiplicities within this tightly selected central-collisions
sample.
This is shown in Fig.~\ref{fig:fluct} together with distributions obtained
with our blast-model simulation, duly filtered to allow a direct
comparison.
It is useful to note that the simulation uses a simple Monte Carlo
procedure to generate events from the multiplicity distributions given in
Table \ref{tab:multiplicity} and with the velocity-space parameters of our
WS2 scenario.
Only events that strictly conserve mass, charge, energy and momentum are
allowed. 
No other correlation of any kind is assumed.
The simulation does extremely well in reproducing the External Wall
multiplicity distribution (panel a in Fig.~\ref{fig:fluct}), the IMF
multiplicity distribution (panel b) and the anticorrelation of the average
IMF multiplicity with the light-particle ($Z<3$) multiplicity. 
This latter plot (panel c) illustrates the sort of bias one introduces on
IMF multiplicities if one selects events by making cuts on the
light-particle multiplicity, as is sometimes done in the literature.
The major conclusion is that we find no evidence for fluctuations other
than those expected from finite number fluctuations under the constraints
of conservation laws.

Does the apparent purely statistical nature of the fluctuations, together with
the exponential behaviour of the nuclear charge distributions beyond Z=2,
suggest a thermal mechanism?
Since we have determined the 'thermal' part of the available energy (see
Table \ref{tab:results}) we can use existing statistical multifragmentation
models to check this idea.
The only  parameter left was the bulk freeze-out density  which we fixed to
the value $0.4 \rho _{0}$, where $\rho _{0}$ is the equilibrium density of
nuclear matter in its ground state.
This value is derived from the IMF-IMF correlations analysed in
\cite{kotte95}.

We have done calculations with three codes:
QSM or 'Quantum Statistical Model' \cite{hahn88,konopka94},     
WIX \cite{randrup93} an extension of an older code FREESCO \cite{fai86},
and SMM or 'Statistical Multifragmentation Model' \cite{bondorf95}. 
The results of these calculations are shown in Fig.~\ref{fig:zdisttheo}
together with our data.
As these models are well documented we shall only recall here the general
idea: following statistical principles all possible final states are sorted
out and their relative probabilities are calculated.
The probability of a certain final state of the decaying system is
proportional to its statistical weight, i.e. to the number of microscopic
states leading to this final state.
In all three implementations an initial 'fast' multifragmentation is
followed by a 'slow' evaporation from the 'primordial' fragments.
While QSM gives 'exact' ensemble averaged results, WIX and SMM generate
individual events and hence also contain fluctuations.

Coming back to our data, what we find is that all calculations
underestimate the formation of heavy clusters.
For QSM
the degree of underestimation at the oxygen level is also shown in one of
the panels. At 400 A MeV the oxygen rate exceeds the values calculated with
QSM by four orders of magnitude.
In order to give an idea on what is changed if one increases the collective
energy (and hence decreases the thermal energy) by the maximum amount that
seems tolerable from our analysis (see Table~\ref{tab:results}), we show
the corresponding results as dotted line in the same panel.
We note also that switching to the lower freeze-out densities usually used
in the literature (and contributions from the tails of the density
distribution implied by our diffuse flow density profile),
 increases the discrepancy between theory and data.  
The same is true to an even higher degree, if we were to use the lower
collective flows deduced from the light-particle data
\cite{poggi95,lisa95}.

A very old problem in applying statistical models to the calculation of
yields is the number of very high lying,
usually unresolved, levels that are included: these are
either ignored (QSM) or a level density formula with a cutoff function
(SMM, WIX) is used. In particular WIX \cite{randrup93} uses as cutoff
function
\begin{equation}
F=\exp \left [-\frac{1}{2}\left (
         \frac{\epsilon - B}{\ecut }\right )^{2}\right ]
\label{eq:cutoff}
\end{equation}
where B is the barrier against the most favourable light-particle emission
mode and the width is given by $\ecut = \ccut   (\sqrt{A-3}-1)$,
where $\ccut $ is a parameter. 
Fig.~\ref{fig:ccut} illustrates the change in the WIX prediction if one
modifies the cut-off energy.
The oxygen yield (none within a statistics of 6000 events in the case
$\ccut  =2$ rises considerably (to 16 fragments) when $\ccut  =8$ is
chosen.
The right panel in Fig.~\ref{fig:ccut} shows the distribution of the
$^{16}$O levels \cite{tilley93} and compares it to the corresponding
predicted average excitation energies $E^{*}$ of oxygen in the two cases.
For $\ccut  =8$, the 'standard' value in WIX, 
one finds $E^{*}\approx 60$ MeV ,well outside the region of
known levels. 
The $\ccut  =2$ parameterization (which we also used for
Fig.~\ref{fig:zdisttheo}) leads to $E^{*} \approx 20$ MeV.
(In ref. \cite{csernai87} QSM and FREESCO were shown to be equivalent, but
there only light-particle predictions were compared).

As an aside, we present in fig.~\ref{fig:ccut} also a standard calculation
for the 250 A MeV data that does not include evaporation (open-circle
symbols).
This approximation was used in \cite{petrovici95}.
The quality of data description achieved here  is
comparable to that obtained in ref.~\cite{petrovici95}.

Coming back to the cut-off problem, one may suspect that one is doing some
double counting if one is including many high lying broad resonances, as
these must, in virtue of Levinson's theorem \cite{levinson49} have a
negative echo in the undisturbed continuum spectrum, an effect that is not
usually taken into account.
As these uncertainties affect the predictions and prevent a definite
conclusion concerning the more fundamental issue, namely the adequacy of
the statistical approach, it will be necessary in the future to set the
treatment of, largely unknown, high lying levels on a better foundation.

One consideration that could save the statistical model (although its
practical implementation would have to be rethought) would be to postulate
a higher density at which fragment sizes are 'frozen out'.
IMF-IMF correlation radii \cite{kotte95} are not necessarily 'cluster-size
freezeout' radii, as such correlations are primarily sensitive to the last
Coulomb-interaction when sizes are fixed already.

Interpreting naively the probability of oxygen
($^{16}$O)  formation as $P_{c}^{16}$
where $P_{c}$ is the clusterization probability, one could account for the
oxygen yields by raising $P_{c}$ by a factor of 1.7-1.8.
This could be done by switching to about twice the density assumed in our
calculations, or about $0.8*\rho _{0}$.
At this density, of course, the present implementations of statistical models
\cite{gross90,bondorf95,hahn88,fai82} are not valid.
Three considerations make this speculation interesting however, and perhaps
worth pursuing:
\begin{enumerate}
\item It seems reasonable that clusters heavier than hydrogen would form at a
time where the average density is still comparable (athough somewhat lower)
than the densities of the clusters that are about to form, which presumably
are not too far from 'normal' density.
This would also provide a fast mechanism as at high density rearrangements
are more frequent.
It is interesting in this connection that new algorithms to detect clusters
in simulations are now able to discern clusters at a very early stage
\cite{dorso93,puri96}.
\item This could bring out-of-conflict the very low apparent chemical
temperatures needed to explain the data with low freeze-out densities.
At higher density clusters statistically 'survive' at higher temperature
(see also fig. 5 in ref.~\cite{kuhn93}). 
The combined high-density and temperature would put cluster formation in
these central collisions well outside the spinodal regime \cite{mueller95}.
The almost perfectly exponential yield-distribution and the absence of
non-trivial multiplicity fluctuations also favour this viewpoint.
\item The measured radial flow for heavy clusters would now be a signature of
{\em compression} as it could not have developped in the low density phase.
It is also interesting to note that the QMD model with Pauli-potential
\cite{peilert92} predicts multifragmentation only if compression is
present ('mechanically excited').
In this connection we also refer to section \ref{clusterization} where we
found, using QMD, a mean-field dependence of clusterization. 
\end{enumerate}

\section{Experimental data and molecular dynamics}
\label{qmd}
It is attractive to use transport models because they do not rely on a
priori assumptions of global or even local equilibration.
Such models, which are able to simulate collisions event-by-event,
are more ambitious, but in order to be usable they need to
introduce many approximations that must stand the test of time.
One such model is Quantum Molecular Dynamics (QMD) \cite{aichelin86} and more
specifically one of its variants the Isospin (I)QMD \cite{hartnack93,bass95}.

In the context of this work we were primarily interested in checking what
such models have to say about the general topology of central events: can
we at all expect a 'radially exploding fireball' in such collisions with a
maximum of 'nuclear material' (in earthly laboratories) and a
minimum of 'spectator' material?
In particular is the observable ERAT, eq.~(\ref{eq:erat}), anywhere close to
the reference value of two, expected from a global thermalization scenario?
Are our methods of selecting central collisions by maximizing ERAT and
minimizing side flow supported by such simulations?
Because of the claim of QMD to handle non-trivial fluctuations, what does
it have to say to the degree of clusterization (in the structureless
'liquid-drop' kind of sense)  and its possible connection to the phenomenon
of radial flow extracted in the previous sections?

The general idea of QMD \cite{aichelin86} is to use the wide spread  method
of molecular dynamics with some quantum modifications specific to the nuclear
context.
For a derivation of the theory from a variational principle we refer to
\cite{hartnack94}.
As in molecular dynamics one solves the classical
(non-relativistic)
hamiltonian equations of motion for a system of A nucleons (and in the case
of IQMD more explicitly a system of Z protons and N neutrons) that evolve
under the influence of nucleon-nucleon interactions.
These involve in the case of IQMD \cite{hartnack93}
 Skyrme-type (density dependent)
interactions, a Yukawa term with a range parameter $\mu = 0.4$ fm,
a symmetry potential between protons and neutrons corresponding to the
Bethe-Weizsaecker mass formula, a Coulomb interaction between charged
particles and, finally, a momentum dependent term.
An important decision taken within the model is to adjust the parameters of
these phenomenological interactions to properties of cold, i.e. highly
degenerate nuclei: binding energy per nucleon, asymmetry energy, saturation
density and the real part of the nucleon-nucleus optical-model potential.
One important difference to conventional molecular dynamics is the
following: for the calculation of the potential gradient that is driving
each individual nucleon, these are not treated as point particles, but are
given a finite, but fixed spread around their classical phase-space
location. This density spread is Gaussian
 with a variance $2L$ in coordinate space and
an associated uncertainty-principle-conserving spread $\hbar^{2}/2L$ in
momentum space. This finite spread introduces de facto a finite range into
the Skyrme-($\delta -$function)-like interactions and hence influences the
finite-system dynamics and in particular also the establishment of
correlations such as clusterization.
There is some freedom in the choice of $L$: in the IQMD model the canonical
value is $L=2.165$ fm$^2$ for Au on Au simulations.
The freedom in the other parameters of the phenomenological interactions
allows to tailor them in such a way that they imply a fixed
compressibility $K$ when applied to homogeneous nuclear matter. 
It has been customary to allow for a so-called 'hard' (H) version of the
equation of state (EOS) corresponding to $K\approx  400$ MeV (388 MeV to
be exact) and a 'soft' (S) version, corresponding to $K=200$ MeV.
The present status from 'low-energy' nuclear physics is a value $K=215$ MeV
from a theoretical analysis \cite{blaizot95} of the 'breathing modes' of
nuclei, i.e. the giant monopole resonances (GMR) as excited typically by
high-energy alpha scattering.
(A very recent search for still missing strength in the GMR
\cite{youngblood96} came to the conclusion that $K$ values as high as 300 MeV
could not be excluded from the GMR data).  

Another important difference to molecular dynamics, but a point in common
with BUU-type approaches \cite{bertsch88} is the treatment of so-called
'hard' collisions by a Boltzmann-like binary-collision integral, borrowed
from classical transport theory for dilute systems, and supplemented by a
Nordheim-Uehling-Uhlenbeck term to account for Pauli-blocking at least on
the local level.
In this model it acts as
 an entropy-producing non-deterministic term as in practice
collisions are dialed with Monte-Carlo methods whenever two nucleons
approach within the radius corresponding to the total 'hard' cross section.
Again, an important decision within the model is to use free scattering
nucleon-nucleon cross sections, i.e. this part of the model (in contrast to
the 'soft' collisions) is adjusted to the properties of infinitely dilute
nuclear matter.
IQMD distinguishes neutrons, n, and protons, p, i.e. it uses the relevant
cross sections $\sigma _{np}$, $\sigma_{pp}$ and $\sigma_{nn}$.
Pions and $\Delta$'s are included, but elastic scattering is dominant, still,
in the present energy range.

Finally it seems worthwhile to mention also details of the initialization
of each nucleus-nucleus collision,
 since they seem to have a surprisingly large influence
on the final outcome \cite{hartnack96}: in IQMD the centroids of the
Gaussians in a nucleus are randomly distributed in a sharp phase-space
sphere $(r \leq R$ and $p \leq p_{F})$, 
with the radius $R = 1.12\cdot A^{1/3}$ fm and the Fermi-momentum $p_{F}=268$
MeV/c.
One of our prime encouragements to use IQMD was its reasonable stability
against the variation of the initial distance between the two nuclear
surfaces \cite{hartnack96}.
For the present calculations, starting from a zero initialization distance,
we followed the evolution for 200 fm/c (i.e. a width of 1 MeV in terms of
final levels) at which time clusters were formed with the minimum spanning
tree method using a clusterization radius of 3 fm \cite{aichelin86}.
 Decay past this time was
not taken into account.
For reasons of economy we took a randomly sampled unweighted impact
parameter $b$ distribution from $b=0$ to $b_{max}$, where $b_{max}$ was
mostly 14 fm and sometimes 7 fm. Experimental observables were
then constructed  weighting them eventwise with the factor $b$.
In general both the hard and soft variants of the EOS were explored and
some 300-1000 events per fm interval were generated.
  
\subsection{Transverse energy (ERAT)}
\label{erat}
We shall start with the observable ERAT i.e. eq.(\ref{eq:erat}) since it played
a key role in our event selection.
IQMD in its standard implementation just described, predicts average values
that are close to two in the energy range 150-400 A MeV and for central
collisions ($b=0-1$ fm).  
The results are shown in figure \ref{fig:eratiqmd} for both the hard and
soft equations of state with momentum dependence included (HM and SM).
No apparatus filtering was done here, except that neutrons were excluded.
We have already mentioned the fluctuations to which ERAT is subjected (fig.
\ref{fig:eratfluct}).
The filtering effect in the Phase I apparatus is considerable: the ERAT
values are lowered and the relative fluctuations are increased.
For unfiltered
ERAT values in the range $1.5-2.5$ we found however that the filtered value was
lowered to a relatively constant fraction of $46\pm 2\%$, so that
differences between different parameterizations are preserved.
Therefore we can roughly say that a modification of the unfiltered average
ERAT value by 0.2 units will give rise, after filtering, to a shift 
along the abscissa of
about 0.1 units in the tail of the distribution (at the 20-60 mb level in fig.
\ref{fig:daniel}) .
Also, we observed in all model simulations that parameterizations that lead
to higher values of ERAT for $b<1$  showed increased 'transversality' at other
impact parameters as well.
As the value for the hard EOS happens to be almost exactly two 
for central collisions at 250 A MeV,
we compare the ERAT distribution ($b=0-7$ fm), after filtering with
the corresponding FOPI data in fig. \ref{fig:daniel} and find that it
follows the data rather well.
As a curiosity we have also plotted in the same figure the (filtered)
blast-model prediction using the chemical composition and momentum space
distribution as fitted to the data (but not to the ERAT distribution).
Of course the unfiltered average value for the blast-model is exactly
two.
We have already shown in section \ref{centrality} that the blast model
fluctuations which are purely statistical, but obey constraints of mass,
charge, energy and momentum conservation, come out identical to those of
IQMD (fig. \ref{fig:eratfluct}).
Since the blast-model is conceived only for central collisions, its
normalisation is dependent on what is defined to be 'central'.
We see from the figure that a central component of the innermost 1 fm (or
30 mb) seems to hug the tail of the measured distribution.

The top panel of fig.~\ref{fig:daniel} shows the $b$-dependence of
(unfiltered) ERAT values predicted by the HM and SM versions of IQMD.
In the same figure we have also plotted the values predicted by a
Vlasov-Uehling-Uhlenbeck (VUU) calculation using a different code
\cite{daniel92}.
This model was used earlier in our Collaboration \cite{poggi95} for
comparison with light-particle data since it includes formation of clusters
up to mass three.
This feature seems to lead, in the model, to very high ERAT values.
Since VUU by principle does not allow to study fluctuations, we have
constructed by randomly mixing the 'test' particles \cite{bertsch88}
'events' which have the correct total mass and charge.
These 'events' then have the same relative, finite number, fluctuations as
both the blast model and IQMD (fig.~\ref{fig:eratfluct}).
With these trivial (but unavoidable) fluctuations added on top of the
model, we can then again compare with the data.
As can be seen in fig.~\ref{fig:daniel} (this calculation is limited to
$b=0-5$fm), models with such copious transverse momentum generation are
incompatible with the data.
This particular model predicts substantially lower ERAT values however, if
composite particle formation is not included \cite{daniel96}.
 
There are other, more subtle deviations between simulation and data.
From fig.~\ref{fig:eratiqmd} we see that IQMD predicts for central
collisions a rise of ERAT of approximately $30 \%$ (0.4-0.5 units) within the
analysed energy range which, after filtering, should lead to a positive
shift by about 0.2 units of the tail of the distribution when going from
150 to 400 A MeV.
In contrast, the data, fig.~\ref{fig:erat}, reveal, if at all, a trend to
{\em smaller} ERAT values at the higher energy.
(The measured ERAT distribution for 150 A MeV lies almost exactly on top of
the distribution for 400 A MeV if it is shifted {\em down} by 0.1 units).
Despite the estimated uncertainty of the data (maximally 0.1 units) one can
therefore say that ERAT, in the experiment, does {\em not} increase with
the incident energy.
 
If the naive concept of global thermalization were to be fulfilled in
nature, ERAT would be 'dynamically inert' in the sense that its value for
central collisions would not depend sensitively on details of the
calculation, but always have the value of two.
Instead, however, all the simulations that we have done ourselves, or
became aware of, show that ERAT seems to 'react' to many parameter
variations.
Besides the disturbing dependence on the more 'technical' parameters
\cite{hartnack96} such as the spread parameter $L$ described earlier, and
even the details of the initialization (sharp spheres versus Woods-Saxon
profiles, choice of Fermi-energy \cite{hartnack96}), there are the more
fundamental dependences on the EOS (according to IQMD a $20-30\%$ effect,
see fig.~\ref{fig:eratiqmd})  and above all on the implemented
nucleon-nucleon cross sections that enter the 'hard' collision term.
 
From a recent paper by Botvina et al. \cite{botvina95} it was shown that
the predicted final event shape in momentum space could vary from strongly
oblate shapes (high ERAT values), obtained if constant cross sections
$\sigma_{np}=\sigma_{pp}=40$ mb were assumed (see also \cite{konopka94a,roy96}),
to slightly prolate shapes if the 'in-medium' nucleon-nucleon cross
sections were lowered.
ERAT values higher than two for highly central collisions ($b<2$fm) 
should lead to
polar angular distributions peaked at $90^{\circ}$ in the c.o.m.
\cite{daniel92,konopka94a}.
This important cross connection and the possibility to draw conclusions on
the magnitude of in-medium cross sections will be discussed more
quantitatively in a separate paper \cite{roy96}.

Our general conclusion is that measured ERAT distributions are an important
constraint to models.
  
\subsection{Sideflow and centrality} 
\label{sideflow}
In the lower panel of fig.~\ref{fig:sideflow} we use IQMD to study the
evolution of sideflow with the true impact parameter $b$ (solid curve).
Sideflow is quantized in terms of $\pxdir $, eq.(\ref{eq:pxdir}), in
scaled units.
The soft (momentum dependent) EOS was used and the apparatus filter
applied.
Notice the enlarged abscissa scale: of the approximately 14 fm range of
impact parameters only the innermost half is shown.
When using ERAT binning instead of the true impact parameter we find that
the curve follows the original $b$-dependence amazingly well down to the 1
fm level (where the 'true' $\pxdir $ becomes negative).
When using multiplicity binning there is a lack of $b$-resolution below 3
fm resulting in an almost flat dependence, much the same as in the data,
fig.~\ref{fig:dirpx}.
This supports our method of picking central events: low side-flow plus high
transverse momenta.
The efficiency of this method should depend on the scaled strength of the
sideflow at intermediate ($3-5$ fm) impact parameters where it is peaking.

How does this strength evolve with incident energy?
In the figure~\ref{fig:erat} we have seen that the ERAT distributions vary
rather modestly with incident energy.
One of our main results from the blast-model analysis,
table~\ref{tab:results}, was that the radial flow takes up a nearly
constant amount of the available energy.
A more detailed look back at the varying topologies, even in the most
central collisions, see Fig.~\ref{fig:ptye},
shows that 'scaling' is approximate at best, however.
Scaling of observables in heavy ion collisions has been considered in refs.
\cite{balazs84,bonasera87}. 
In analogy to Reynold's famous laws governing viscous flow, one can expect
\cite{balazs84} in the hydrodynamics framework that scaling might exist
under certain conditions.
Scaling would reduce many systems of different size and incident energy
into one.
Deviations from scaling could then isolate interesting phenomena known to
violate scaling: the most obvious ones are the influence of the EOS, and
connected with it, possible phase transitions.
  
IQMD predicts that sideflow {\em does not} scale in this energy range, see
fig.~\ref{fig:sideflow}.
As shown in the upper two panels, this agrees almost quantitatively with
experiment.
The conclusion does not depend on the specific choice of the sideflow
observable: using $\fdo $, eq.(\ref{eq:fdo}), which does not involve the
so called reaction plane explicitly, leads to very similar results.
There is relatively less sideflow at the lower energy and hence the method
of using 'absence of sideflow' to increase centrality will be less
efficient as the energy is lowered even further.
We note in this context that the observables $\pxdir $ and $\fdo $ are
expected to be 'coalescence-invariant' and hence better suited for scaling
studies than the slope of $dp_{x}/dy$ taken at midrapidity which is steeper
for clusters \cite{doss87,partlan95,dona96} than for nucleons.

Is the EOS responsible for this lack of scaling?
We found only very modest differences in the calculations using the hard
instead of the soft version of the EOS.
(Momentum dependence in this model seems to affect primarily the high
impact parameter side of the peaks in fig.~\ref{fig:sideflow} leading to a
broadening of the peak and a slight shift to higher $b$).
'Disappearence of sideflow' \cite{krofcheck89,ogilvie90,soff95,zhang90}
 in the energy
range $50-150$ A MeV has been primarily associated with the transition from
predominantly attractive to repulsive forces, i.e. to the EOS.
In view of our IQMD results and our observations of the evolution of phase
space with energy (fig.~\ref{fig:ptye}) another cause for non-scaling is
suggested:
the variation of the time-scale with incident energy versus the fixed
time-scale typical of Fermi-motion.
Sideflow, quantized with the observables $\pxdir $ or $\fdo $,
is primarily a 'spectator' phenomenon (a more detailed study of
relevant FOPI data will be published separately \cite{dona96}):
it develops into 'full bloom' only when there is a clearcut
spectator-participant separation.
This separation, as we have qualitatively observed, is less pronounced at
lower energy (see for example fig.~\ref{fig:ptypm}).
In this picture the 'spectators' are seen to be recoiling under the
explosive power of the fireball.
Although this push from the fireball scales in ultracentral collisions,
there is, at energies still comparable with the Fermi-energy, sufficient
time for some 'cross-talk' possibly via a different, one-body mechanism
known from the theory of deep-inelastic heavy-ion collisions
\cite{swiatecki81,feldmeier87} to mitigate the recoil.
The 'spectators' are cooling the fireball and the fireball is heating the
spectators, i.e. there is some start towards a more global equilibration
that weakens sideflows at freeze-out.
In this picture, at some high energy, when the above globalization seizes
to work because of a lack of time, sideflow at sufficiently small $b$ might
become just a reflection of (scale with) the radial flow.
At very high energy (CERN SPS energies?) another time scale might come in:
the spectators pass by with essentially light velocity, but see an
increasingly contracted participant.
If the explosion time scale is too long, again, there will be little scaled
sideflow seen.
  
Another aspect of participant-spectator geometries, enhanced flow out of
the 'reaction plane' or 'squeeze-out' \cite{stoecker82} might also be
governed by such time scale comparisons.
Eventually, this may turn out to be an important clue to setting scales for
the fireball explosion (see also \cite{daniel95}).
{\em Quantitative} assessments will definitely require the use of highly
refined transport theories.
It will be interesting to establish the low as well as the high energy
connection by using the same type of analyses and ideas.

\subsection{Clusterization}          
\label{clusterization}
We were only semi-successful in understanding the degree of clusterization
using statistical concepts.
Is a full-blown transport theory more successful?
Presumably a theory of clusterization in heavy-ion collisions will have to
contain 'the right amount of quantum mechanics'.
Antisymmetrization of fermionic matter, which has {\em nonlocal}
consequences, is a key ingredient in understanding 'the making of nuclei',
one of the fascinating subjects of 'hot nuclear matter'.
So far, fully antisymmetrized approaches to heavy-ion collisions
\cite{feldmeier90,ono92} could not be extended to heavy systems such as Au
on Au owing to exorbitant needs for central processor time.
{\em Locally}, some requirements of antisymmetrization are taken into
account in QMD as well as BUU/VUU by the introduction of a Pauli blocking
term into the collision integral.
Some QMD approaches have also introduced Pauli-potentials \cite{peilert92}
to prevent Pauli-forbidden local overlap in phase space.
It is still an open question whether this sufficiently mimics the needs for
antisymmetrization.
The momentum dependence introduced by Pauli potentials has a profound
influence on the local kinetics \cite{donangelo94}.
Basically, the clusters in QMD remain classical objects and 
for example do not have the
correct heat capacity at low temperatures \cite{peilert92,donangelo94}.
Of course shell effects that might come in at freeze out time, namely
preferred alpha formation \cite{ono92,ono93} are completely missing in (I)QMD.
At best one can hope to obtain 'macroscopic' kinds of clusterizations in a
spirit similar to macroscopic mass formulas (which are known to be
successful in explaining global trends).

We have added in Fig.~\ref{fig:zdisttheo} the charge yield predictions of
IQMD to be compared to the statistical model calculations.
IQMD is doing well on a logarithmic scale, but does not include decays
after 200 fm/c.
As IQMD is a dynamic model, it is worthwhile looking at clusterization in
more detail.
There are different ways of quantifying the probability of clusterization.
In order to allow a direct comparison with $4\pi$ extrapolated data, we
have plotted in fig.~\ref{fig:cluster} for the momentum dependent versions
(SM and HM) the IQMD predictions for the multiplicities of the IMF's and
for the degree of proton clusterization (in a cluster of any size).
It is obvious that the probability to clusterize is underestimated by the
present implementation of IQMD.
We recall that no decay after 200 fm/c is included: such sequential decays
(see section \ref{chemical}) would increase the discrepancy.
 
The failure of the model is most dramatic at the highest energy and, there,
is specifically associated to {\em central} collisions.
This is illustrated in fig.~\ref{fig:mimfvsb} where a combined ERAT (low
$b$) PM (intermediate $b$) binning was used to take advantage of the better
'resolution' of $b$ (suggested by the simulation) when a combined method is
used. 
Experimental 'impact parameter' dependences of IMF multiplicities in Au on
Au collisions have been shown earlier by the MSU/ALADIN collaboration
\cite{tsang93}, where however charged particle multiplicity was used for
the entire range of cross sections, which may represent a bias for such
observables in the tails of the distributions (see our
fig.~\ref{fig:fluct}).
A direct comparison with \cite{tsang93}
 is not straightforward as both apparatus have different
filtering properties.
It seems however that there is a real discrepancy for the most central
collisions, namely at 400 A MeV, as our $4\pi$ extrapolated IMF
multiplicities (see table~\ref{tab:multiplicity}) are higher by at least a
factor two.

The failure to predict sufficient clusterization was also reported in
\cite{tsang93}.
It is known \cite{hartnack96} that the clusterization probability is
influenced by the width parameter $L$.   
Varying $L$ does however have also complex influences on other observables
\cite{hartnack96}.
We stress that the failure to clusterize with sufficient probability is not
limited to large clusters (IMF's) but holds also, and even more so, for the
so-called light charged particles (LCP's).
Using the soft EOS, which seems to be more favourable to clusterization
than the hard EOS (see later),  the He/H ratio is underpredicted by a
factor 5 at 150 A MeV and of 6 at 400 A MeV.

Despite this problem, we show a  very interesting effect: in
IQMD the potential interaction seems to influence
 {\em dynamically} the probability of
clusterization,see Fig.~\ref{fig:mimfvseos}.
In general, more clusters are made using the soft EOS, and at high impact
parameters the momentum dependence seems to favour clusterization as well.
In a different context the breaking apart in explosions has been connected
to the speed of the expansion \cite{grady82} :
 loosely speaking, the faster the explosive
moves apart the smaller the pieces.
If fragments are not formed by density fluctuations in a quasi-equilibrium
situation but in an out-of-equilibrium regime, what is at stake is an
energy balance between the kinetic energy of expansion and the potential
energy of broken surface bonds 
\cite{grady82,remaud96}.
In the nuclear context the time scale could be partially set by the EOS.
A harder EOS, intuitively, could drive a faster expansion, leading to
smaller pieces.
Under certain conditions, following \cite{grady82}, it is predicted that
the average cluster size should vary linearly with the inverse flow energy
per nucleon. Among the restrictions are a fixed freeze-out density and the
assumption that the later evaporation does not modify the linear
relationship.
For the average cluster masses determined at three incident energies
(see section~\ref{chemical} and Fig.~\ref{fig:zdist}) and the collective
flows extracted from the spectra (see Table~\ref{tab:results}) we find that
the linear relationship mentioned above is fulfilled with an accuracy
better than $1\%$. 
It will be highly interesting to see if these speculations stand the test
of time as the current transport codes are further refined.
  
\subsection{Radial flow}             
\label{radialflow}    
The collective flow energies per nucleon listed in Table~\ref{tab:results}
($\ecoll /A$, row 6) are in amazing agreement with the theoretical
simulations in \cite{daniel95} where values of 21, 35 and 55 A MeV were
predicted for the three energies.
In \cite{daniel95} it was argued that these values vary only slowly with
impact parameter.
However the splitting into longitudinal and transverse degrees of freedom
{\em does} vary with the impact parameter.
Our discussion in section~\ref{erat} already suggested that the simulation
in \cite{daniel95} predicts too much 'transversality'.

Our method of extracting the radial flow, which does not need many events,
was also applied to IQMD events.
The main results of our analysis are reflected in a series of numbers for
the fraction of the final total kinetic energy caught in collective flow
that can be found in Table \ref{tab:iqmdflow}.
Rows 1 and 2 in the Table show the incident energy and EOS (HM vs SM)
dependence of the radial flow for impact parameters $b < 1$ fm and for
charged particles emitted under large polar angles
$(45^{\circ}-135^{\circ})$.
The method used, called {\em slope} in the Table, is based on the relation
(\ref{eq:avekin}) (see also Fig. \ref{fig:avekinqmd}) which implies a linear
dependence of the average kinetic energy as a function of mass.
The flow determined in this way (using the spectra for masses $A=1-10$)
 is not necessarily identical to the
'theoretical' flow as determined from a careful cell-analysis using the 
full information of momentum and position as a function of time.
This method is not available to experiment however.
 
Judging from the numbers (row 1 and 2) in the table, there is a weak
increase of the flow fraction with energy which is more pronounced at the
low energy end, and there is a small bracket between the hard and the soft
EOS which gets larger as the energy is lowered.
This higher sensitivity to the {\em cold} EOS at lower energies was
suggested in ref. \cite{stoecker81} where it was actually recommended at
that time (1981) that the 'best' incident energy for this purpose was
around 100 A MeV!
Intuitively, however, it is rather puzzling that the hard version of the
cold EOS should give {\em less} radial flow in the IQMD simulation.
The work of ref. \cite{stoecker81} suggested that a better than $10 \%$
accuracy would be needed to solve the question of the stiffness of the EOS.
The argumentation stemmed from solving the one-dimensional Rankine-Hugoniot
shock equations (implying validity of hydrodynamics) to obtain the maximum
density and pressure in the shocked zone.
Such considerations usually indicate that the maximum pressure (which is a
sum of two terms from the cold and the thermal EOS, respectively, the
latter being smaller than the former only at sufficiently low incident
energies) rises relatively modestly with the stiffness of the cold EOS due
to a compensatory effect: the soft EOS leads to larger densities.  
However there is no inversion as to which EOS leads to higher pressure as
implied in Table \ref{tab:iqmdflow}!

Rows 3-8 in Table \ref{tab:iqmdflow} serve the purpose to test and defend
the likelihood method as applied to the (less complete and less perfect)
experimental data.
Most importantly, we have checked that all the numbers listed in the table,
which were obtained without experimental filtering, could be recovered
with a typical deviation of $2\%$ of the available total kinetic energy 
if we applied the threshold and angular cuts of the Phase I FOPI apparatus.

It is seen by comparing rows 2 and 3 in the table that the likelihood
method with our preferred profile WS2 
gives virtually
identical results to the 'slope' method if applied to the same data with
the same restrictions on impact parameter and polar angles.
Enlarging the angular range to include more forward (backward) angles tends
to increase the average flow, a faint remnant of the initial purely
longitudinal 'flow'. In the IQMD simulation the effect is significant only
at the lowest energy (row 3 versus row 4 in the table).
The likelihood method, in contrast to the slope method, also tests the
shapes of the spectra; rows 4, 5 and 6
compare the results for the Woods-Saxon (WS2), the 'Box'
and the 'Shell' scenario's.
As in the experimental analysis (Fig.\ref{fig:collectivity}) the Shell
scenario underestimates the flow, while differences between the 'Box' and
the WS2 profiles are less significant than in the experiment.
We also checked what was the influence of mixing-in higher impact
parameters (rows 7 and 8 to be compared with rows 3 and 4, respectively).
For reasons of simplicity a flat impact parameter population between 0 and
3 fm was taken to approximate the 'experimental' distribution
\cite{reisdorf92}.
While one finds for the larger angular range a general rise of the apparent
flow as compared to the values for the more central impact parameters
(rows 8 and 4), there is virtually no change if the angular range is
restricted to angles closer to $90^{\circ}$.

This observation is our main argument for concluding that our extracted
experimental flow values (Table \ref{tab:results}) , which hold for both
angular ranges, should be representative of very central collisions and are
significantly larger than the theoretical values.

It is tempting to relate this smaller radial flow in the model to the
failure to predict sufficient clusterization.
Again, future investigations, will have to clarify if the  problem is
technical or if it indicates a need for a refinement of the physics of the
model.
On the experimental side, there is a need to fill the gap visible in the
topology plots figs.~\ref{fig:ptyc} and \ref{fig:ptye}.
  
Finally, a note on the flow of very heavy clusters ($A>10$ or $Z>5$):
comparing the two panels in Fig. \ref{fig:avekinqmd} it is seen that IQMD
predicts a saturation of the mass dependence of average kinetic energies
that can however be seen clearly only at the lower incident energies for
obvious statistical reasons.
Such a trend is weakly present also in the experimental data at more
forward angles (Fig.\ref{fig:avekin}) and was seen more clearly at lower
energies \cite{hsi94,kunde95}.
A possible interpretation of this effect might be of geometric origin if
one assumes that there is a positive position-momentum correlation at freeze-out.
a) Since heavy clusters have central densities close to saturation density,
they must coagulate further inside the expanding fireball 
 rather than at the less
dense surface.
b) Because of the larger sizes the flow velocity gradients 'inside' the
potential cluster may disrupt cells in the surface region
\cite{kunde95,grady82}.
c) A more subtle effect might be that already in the high compression phase
the innermost cells of the fireball had smaller entropies per nucleon which
then were conserved in the subsequent adiabatic expansion 
\cite{petrovici95}.

\section{Summary and outlook}
\label{summary}
We have presented a rather complete account of the momentum space
distributions in central collisions of Au on Au at three different incident
energies, 150, 250 and 400 A MeV located around or well above the Fermi
energy.
These data should represent a challenge to future theoretical
analyses.

The strong sensitivity to flow of heavy clusters emitted copiously in these
collisions allowed us to select very central collisions by combining high
total transverse energies with the absence of sideflow which could be shown
to be maximal for impact parameters of 3-5 fm.
This method is supported by IQMD simulations which showed, in excellent
agreement with the data, that the scaled sideflow was increasing with
incident energy in the present regime.
The total transverse energy in the data scales however better with energy
than predicted by IQMD and its magnitude excludes models that predict
strongly oblate event shapes for impact parameters below 2 fm.

Autocorrelation effects which are severe in these rare events ($1\%$ of the
reaction cross section), were avoided by always excluding the particle of
interest from the selection condition.
For these central collisions we could show that invariant cross section
topologies for the IMF's in the twodimensional space of scaled transverse
velocity vs rapidity varied in a subtle way when raising the energy.
All topologies are basically centered around midrapidity, but tend to be
somewhat prolate, becoming more compact however along the rapidity axis at
the higher energy.
Isotropy is reached only approximately.
From somewhat less central events selected with high multiplicity window,
we were able to expose the increasing participant-spectator separation with
rising energy; some faintly dissipative features, known from lower energies
were still evident at the lowest energy measured by us.

In a phenomenological data analysis we explored how well a thermal model
that includes radial flow could reproduce the data.
An isotropic blast scenario that allowed for a complex flow profile was
adjusted to the data under the constraint of energy and charge
conservation, yielding the subdivision of the available energy into a
collective energy, found to take up $62\pm 8\%$, and a rest energy
interpreted as thermal and used later as input into statistical model
calculations.
It could be shown that the heavier clusters were sensitive to the assumed
flow profile and a Woods-Saxon profile with a large diffuseness was able to
describe the measured rapidity distributions (applying a low
transverse-velocity cut) and the transverse-velocity distributions
(applying a mid-rapidity cut).
In contrast, the velocity distributions of single nucleons were found to be
insensitive to the assumed profiles.

The model, fitted to the full measured phase space, reproduces the size
dependence of the average kinetic energies of the fragments around
$35^{\circ}$, however there is an offset indicating
that the fragments are somewhat more energetic at these angles than at more
backward angles, another indication of deviations from perfect isotropy.
The spectral shapes of the IMF kinetic energies are rather well rendered by
the model.
A close comparison with the light particle data suggest that their spectra
and average energies are influenced by late decays and reorganization
effects accompanying the local amalgamations.

The polar angular distributions in these highly central collisions do not
support a dramatic $90^{\circ}$ squeezeout.
It seems that the presence of spectator matter is necessary for the
squeezeout pattern.    

The blast-model fits were used to extrapolate the fragment yields to
$4\pi$.
Although the average size of the clusters decreases with increasing
incident energy we found that only about $20\%$ of the protons were emitted
as single nucleons even at the highest energy (400 A MeV).
The thermal energy determined from the flow analysis was used as input to
statistical calculations.
Using a freeze-out density of $0.4\rho _{0}$ suggested by the IMF-IMF
correlation studies, it was found that such calculations underestimate the
oxygen yields by up to 4 orders of magnitude in two of the models used
\cite{bondorf95,hahn88}.
Large uncertainties in these calculations subsist however, because of the
difficulty to assess the contributions from unknown very high lying levels
avoiding double-counting.

We have suggested that higher cluster yields could still be explained in
the quasi-statistical framework if size-freeze-out densities around
$0.8\rho_{0}$ were postulated.
Such a possibility would also suggest that most of the observed radial flow
actually already developped in the high density phase making this kind of
flow a potentially very interesting observable.
The relatively high apparent temperatures concurrent with this density
would put cluster formation in these central collisions well outside the
spinodal region.
The almost perfect 'primordial' exponential yield curves that are suggested
after consideration of late decay processes and the lack of nontrivial
fluctuations in multiplicities and other observables, like ERAT, also
favour a fast cluster formation mechanism that is only possible at high
densities.

The multifragmentation could be favoured by the explosive high velocity
expansion \cite{grady82} and the related 'mechanical instability'
\cite{peilert92}.
This is also in line with QMD calculations  which suggested that cluster
formation was sensitive to the EOS.
The present implementation (IQMD) of the QMD model however underestimates the
degree of clusterization and does not seem to generate sufficient radial
flow.

Looking forward to new experiments, besides the obvious extension of this
kind of data and analysis both to higher {\em and} lower energy, it is
clear that one should aim at resolving some of the ambiguities left in the
present understanding of the data.
The role of losses on the way to equilibrium could be elucidated by
studying the system-size dependence of yields and momentum distributions.
The freedom about the correct fragment-size freezeout density should be
further constrained by determining and comparing correlation radii also
with other particles than IMF's, for instance $p-p$, $^3$He-$^3$He (that
seem to be more primordial) and, perhaps starting from 400 A MeV, $\pi -
\pi$ correlations.
Close comparisons of the momentum space distributions of equal-mass
particle pairs, $^3$He/t, $\pi^{+}/\pi^{-}$, would also give useful clues.
In general correlation studies could give the information necessary to try
to reconstruct primordial yields allowing to extract primordial
clusterization probabilities, one of the clues to setting up the EOS. 
The degree to which complete mixing of projectile matter and target matter
takes place (a necessary condition for the application of statistical
considerations) could be checked by studying collisions between nuclei with
different neutron to proton ratio.

On the theoretical side a {\em simultaneous} account of momentum space
distributions {\em and} yields of all the emitted particles and clusters
remains a challenge, a convincing mechanism accounting for the large
observed radial flow must be found.
Statistical fragmentation models appropriate for weakly sub-saturation
densities ought to be developped and the role of high lying continuum
levels explored.
Refined transport models, in addition to statistical models, will be needed as
supporting tools to try to extract the time scales of the explosions,
perhaps by studying the scaling properties of the participant/spectator
interplay in the form of sideflow and squeezeout phenomena.
The future 'microscopic' theory will probably have to include the momentum
dependent mean field on a relativistic basis 
\cite{sorge89,cassing90,liwuko89}, even at low
energies, and have to treat the 'soft' collisions (mean field) and the
'hard' collisions (nucleon-nucleon scatterings) on a more consistent basis
\cite{khoa92,fuchs96}. 

\begin{ack} 
This work was supported in part by the Bundesministerium
f\"{u}r Forschung und Technologie under contract 06~HD~525~I(3) and
Gesellschaft f\"{u}r Schwerionenforschung under contract HD Pel K.
\end{ack}

\newpage


\newpage
\begin{table}[p]
\caption{Summary of event samples}              
\label{tab:eventsample}
\begin{tabular}{cccccc}
\hline
energy & name &  PM cut & ERAT cut & D cut & $\sigma $ \\
(MeV/A) &     &         &         &        & (mb) \\
\hline
150 & PM200     & PM $>$ 36    &               &            & 215 \\
    & ERAT200   &              & ERAT $>$ 0.69 &            & 217 \\
    & ERAT200D  &              & ERAT $>$ 0.69 & D $<$ 0.24 &  82 \\
    & ERAT50    &              & ERAT $>$ 0.92 &            &  54 \\
\hline
250 & PM200     & PM $>$ 44    &               &            & 186 \\
    & ERAT200   &              & ERAT $>$ 0.68 &            & 197 \\
    & ERAT200D  &              & ERAT $>$ 0.68 & D $<$ 0.20 &  47 \\
    & ERAT50    &              & ERAT $>$ 0.93 &            &  49 \\
\hline
400 & PM200     & PM $>$ 53    &               &            & 191 \\
    & PM200D    & PM $>$ 53    &               & D $<$ 0.19 &  22 \\
    & ERAT200   &              & ERAT $>$ 0.66 &            & 188 \\
    & ERAT200D  &              & ERAT $>$ 0.66 & D $<$ 0.19 &  42 \\
    & ERAT50    &              & ERAT $>$ 0.88 &            &  47 \\
\hline
\end{tabular}
\end{table}
\begin{table}[p]
\caption{Limits imposed on the likelihood fits}
\label{tab:limits}
\begin{tabular}{ccccc}
\hline
Z & $\ylabiw $   & $\tcmew $ & $\tcmew $ & $\tcmew $ \\
& & 150 A MeV & 250 A MeV & 400 A MeV \\
\hline
1  & 0.238 &  90 & 120 & 130 \\
2  & 0.238 &  90 & 110 & 120 \\
3  & 0.252 & 145 & 150 & 150 \\
4  & 0.280 & 140 & 150 & 150 \\
5  & 0.300 & 130 & 140 & 150 \\
6  & 0.322 & 125 & 140 & 150 \\
7  & 0.331 & 115 & 140 & 150 \\
8  & 0.342 & 110 & 130 & 140 \\
9  & 0.350 & 100 & 130 & 140 \\
10 & 0.359 &  90 & 120 & 140 \\
\hline
\end{tabular}
\end{table}
\clearpage
\begin{table}[p]
\caption{Results from blast model fit}
\label{tab:results}
\begin{tabular}{ccccc}
\hline
E/A & 150  & 250 & 400 & MeV/A \hspace{5mm} (incident) \\
\hline \hline
$\ecoll$ & $61.3\pm 7.0$ & $61.3\pm 7.0$ & $ 63.4 \pm 7.0$ & \% of TKE \\
\hline
$\ecm /A$ & 36.8 & 60.5 & 95.1 & MeV/A \\
$\qval /A$ & -4.3 & -5.0 & -5.5 & MeV/A \\
$<\beta _{f}>$ & $0.204\pm 0.011$ & $0.263\pm 0.014$ & $0.334\pm 0.017$
 & units of c \\
$\ecoll /A$ & $19.9\pm 2.3$ & $34.0\pm 3.9$ & $56.8\pm 6.3$ & MeV \\
$\eth /A$ & $12.6\pm 2.3$ & $21.5\pm 3.9$ & $32.8\pm 6.3$ & MeV \\
T & $17.2\pm 3.4$ & $26.2\pm 5.1$ & $36.7\pm 7.5$ & MeV \\
$M$      & $191\pm 11$  & $208\pm 9$  & $223\pm 6$  & includes neutrons \\
$\zsum $ error & -14\% & -2\% & +5\% & extrapol. to $4\pi$ \\
\hline
\end{tabular}

\end{table}
\begin{table}[p]
\caption{Multiplicities as function of the nuclear charge}
\label{tab:multiplicity}
\begin{tabular}{rr@{.}lcr@{.}lr@{.}lcr@{.}lr@{.}lcr@{.}l}
\hline
Z & \multicolumn{5}{c}{multiplicity} & \multicolumn{5}{c}{multiplicity}
& \multicolumn{5}{c}{multiplicity}\\
& \multicolumn{5}{c}{150 A MeV} & \multicolumn{5}{c}{250 A MeV}
& \multicolumn{5}{c}{400 A MeV}\\
\hline
   1 & 61&84   & $\pm$ & 0&58   & 75&82   &$\pm$ & 0&62  
                                                & 92&04   &$\pm$ & 0&62\\
   2 & 26&76   & $\pm$ & 0&36   & 27&27   &$\pm$ & 0&36 
                                                & 24&16   &$\pm$ & 0&30\\
   3 &  5&39   & $\pm$ & 0&041  &  4&89   &$\pm$ & 0&034
                                                &  3&75   &$\pm$ & 0&03\\
   4 &  1&789  & $\pm$ & 0&024  &  1&358  &$\pm$ & 0&178
                                                &  0&909  &$\pm$ & 0&014\\
   5 &  1&438  & $\pm$ & 0&022  &  0&772  &$\pm$ & 0&014 
                                                &  0&335  &$\pm$ & 0&009\\
   6 &  0&868  & $\pm$ & 0&017  &  0&355  &$\pm$ & 0&009 
                                                &  0&110  &$\pm$ & 0&005\\
   7 &  0&447  & $\pm$ & 0&012  &  0&127  &$\pm$ & 0&006 
                                                &  0&0397 &$\pm$ & 0&0029\\
   8 &  0&223  & $\pm$ & 0&0087 &  0&0497 &$\pm$ & 0&0035
                                                &  0&0156 &$\pm$ & 0&0018\\
   9 &  0&0898 & $\pm$ & 0&0057 &  0&0132 &$\pm$ & 0&0018
                                                &  0&0029 &$\pm$ & 0&0008\\
  10 &  0&0644 & $\pm$ & 0&0049 &  0&0097 &$\pm$ & 0&0015\\
  11 &  0&0231 & $\pm$ & 0&0031 &  0&0025 &$\pm$ & 0&0008\\
  12 &  0&0245 & $\pm$ & 0&0033 &  0&0012 &$\pm$ & 0&0006 \\
\hline
\end{tabular}
\end{table}
\clearpage
\begin{table}[p]
\caption{Various multiplicities}                           
\label{tab:varmult}      
\begin{tabular}{cr@{.}lcr@{.}lr@{.}lcr@{.}lr@{.}lcr@{.}l}
\hline
& \multicolumn{5}{c}{150 A MeV} & \multicolumn{5}{c}{250 A MeV}
& \multicolumn{5}{c}{400 A MeV}\\
\hline
proton  & 26&1   & $\pm$ & 1&4   & 31&9   &$\pm$ & 1&6   
                                                & 38&7   &$\pm$ & 2&0\\
deuteron& 18&6   & $\pm$ & 1&0   & 23&0   &$\pm$ & 1&2 
                                                & 27&9   &$\pm$ & 1&4\\
triton  & 17&2   & $\pm$ & 0&9   & 21&0   &$\pm$ & 1&1 
                                                & 25&5   &$\pm$ & 1&3\\
$^{3}$He&  5&7   & $\pm$ & 0&3   &  9&1   &$\pm$ & 0&5 
                                                & 10&6   &$\pm$ & 0&6 \\
$^{4}$He& 21&0   & $\pm$ & 1&1   & 18&2   &$\pm$ & 1&0  
                                                & 13&6   &$\pm$ & 0&7 \\
neutron & 92&6   & $\pm$ &10&8   & 97&9   &$\pm$ & 8&2  
                                                &101&7   &$\pm$ & 6&2 \\
charged p.& 99&0   & $\pm$ & 0&7   &110&7   &$\pm$ & 0&8  
                                                &121&4   &$\pm$ & 0&7  \\
IMF     & 10&4   & $\pm$ & 0&1   &  7&6   &$\pm$ & 0&1  
                                                &  5&2   &$\pm$ & 0&1  \\
\hline
\end{tabular}
\end{table}
\begin{table}[p]
\caption{Degrees of clusterization (\%)}                   
\label{tab:cluster}     
\begin{tabular}{cr@{.}lcr@{.}lr@{.}lcr@{.}lr@{.}lcr@{.}l}
\hline
& \multicolumn{5}{c}{150 A MeV} & \multicolumn{5}{c}{250 A MeV}
& \multicolumn{5}{c}{400 A MeV}\\
\hline
protons in IMF& 27&0   & $\pm$ & 0&2   & 17&5   &$\pm$ & 0&1  
                                                & 11&2   &$\pm$ & 0&1\\
protons in clusters & 83&5   & $\pm$ & 1&5   & 79&8   &$\pm$ & 1&8 
                                                & 75&5   &$\pm$ & 2&0\\
nucleons in IMF & 21&6   & $\pm$ & 2&7   & 14&0   &$\pm$ & 2&0 
                                                &  9&0   &$\pm$ & 1&4\\
nucleons in clusters  & 69&9   & $\pm$ &11&8   & 67&1   &$\pm$ & 9&5 
                                                & 64&4   &$\pm$ & 7&8 \\
\hline
\end{tabular}
\end{table}
\begin{table}[p]
\caption{Collective energy from the IQMD model in percent of the total
kinetic energy of emitted nucleons and clusters. } 
\label{tab:iqmdflow}
\begin{tabular}{ccccc}
\hline
Row & specification & 150 A MeV & 250 A MeV & 400 A MeV \\
\hline 
1 & $b<1$ \ \   45$^o$-135$^o$ \ \ HM \ \  slope &
                   $21.6\pm 2.8$ & $33.9\pm 1.7$ & $42.1\pm 1.1$ \\
2 & $b<1$ \ \   45$^o$-135$^o$ \ \ SM \ \  slope &
                   $32.4\pm 2.6$ & $38.9\pm 2.4$ & $41.2\pm 1.7$ \\
3 & $b<1$ \ \   45$^o$-135$^o$ \ \ SM \ \  WS2    &
                   $31.5\pm 0.9$ & $38.3\pm 0.9$ & $41.2\pm 0.7$ \\
4 & $b<1$ \ \   20$^o$-160$^o$ \ \ SM \ \  WS2    &
                   $37.6\pm 0.6$ & $40.1\pm 0.7$ & $41.9\pm 0.4$ \\
5 & $b<1$ \ \   20$^o$-160$^o$ \ \ SM \ \  Box&
                   $36.9\pm 0.7$ & $40.3\pm 0.8$ & $43.8\pm 0.8$ \\
6 & $b<1$ \ \   20$^o$-160$^o$ \ \ SM \ \  Shell&
                   $30.2\pm 0.5$ & $32.5\pm 0.5$ & $35.5\pm 0.7$ \\
7 & $b<3$ \ \   45$^o$-135$^o$ \ \ SM \ \  WS2    &
                   $31.1\pm 1.1$ & $38.8\pm 1.2$ & $42.4\pm 0.9$ \\
8 & $b<3$ \ \   20$^o$-160$^o$ \ \ SM \ \  WS2    &
                   $40.9\pm 0.8$ & $44.7\pm 0.6$ & $46.1\pm 0.8$ \\
\hline
\end{tabular}

\end{table}
\clearpage


\newpage
\newpage

\clearpage

\begin{figure}[p]
\caption{\small Invariant cross section $d^{2}\sigma/u_{t}du_{t}dy$ contour
plots (scaled units) for Li (Z=3) fragments observed in the system Au+Au at
150, 250 and 400 A MeV for the selection ERAT200D.
 See text and the caption
fig. \ref{fig:ptyc} for further details.} 
\label{fig:ptye}
\end{figure}

\begin{figure}[p]
\caption{\small Invariant cross section $d^{2}\sigma/u_{t}du_{t}dy$ contour
plots (scaled units) for Li (Z=3) fragments observed in the system Au+Au at
150, 250 and 400 A MeV for the selection PM200.   
 See text and the caption
fig. \ref{fig:ptyc} for further details.} 
\label{fig:ptypm}
\end{figure}

\begin{figure}[p]
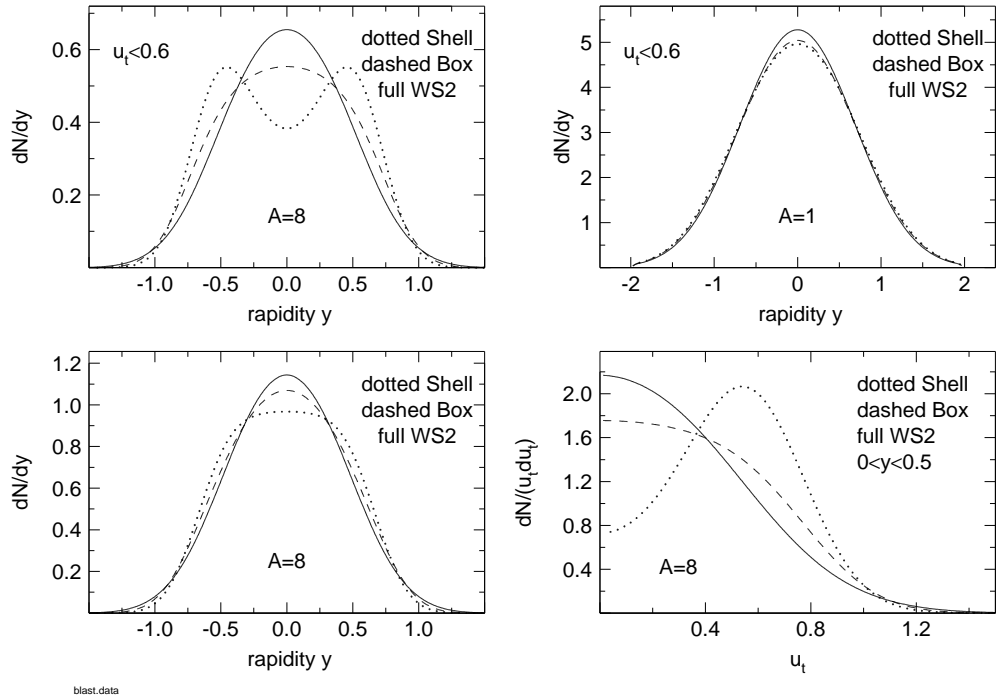

\caption{\small Blast model spectra for a scenario with 50\% collectivity.
The spectra for three different flow-velocity profiles 
Shell, Box, WS2, explained
in the text, are shown.
Lower left panel: rapidity distribution for mass $A=8$ fragments.
Upper left panel: Same, but with a transverse 4-velocity cut $u_{t} 
< 0.6$. 
Upper right panel: same cut but for mass $A=1$.
Lower right panel: transverse 4-velocity spectrum for mass $A=8$ with
a rapidity cut $0<y<0.5$.}
\label{fig:blast}
\end{figure}

\begin{figure}[p]
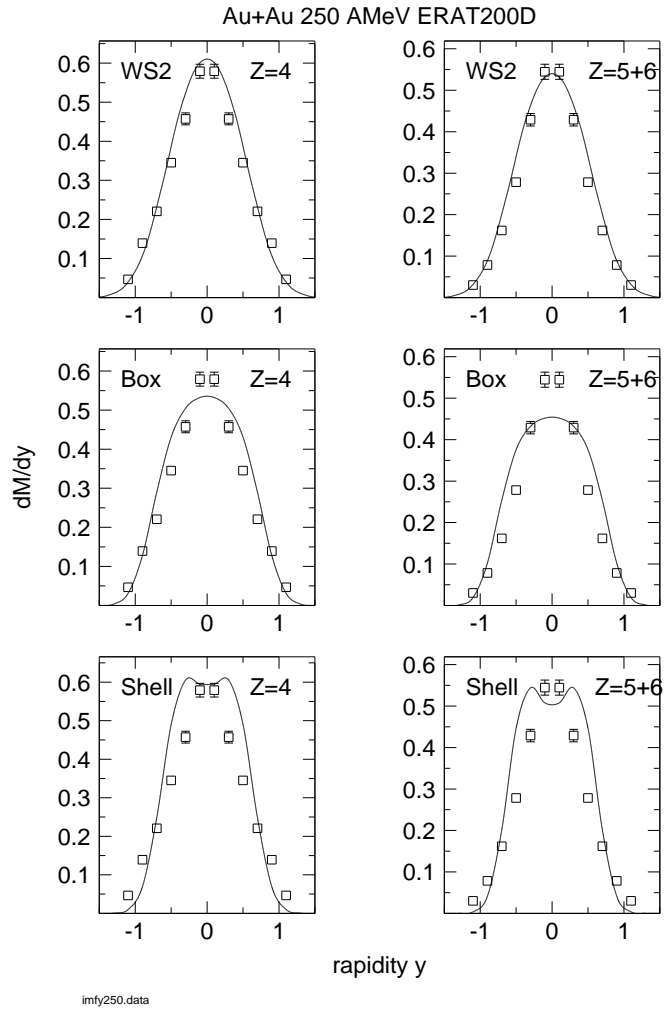

\caption{\small Measured rapidity distributions of fragments with nuclear
charge Z=4 (left) and Z=5 or 6 (right) at 250 A MeV incident energy.
The data (symbols) are compared with the global fits using the blast model
with flow profiles $Shell$ (bottom), $Box$ (middle) and $WS2$ (top).
A transverse 4-velocity cut $u_{t}<0.6$ was done.
Only the forward-hemisphere contribution was measured, the backward
contribution is obtained by reflection.}
\label{fig:rapidity250}
\end{figure}

\clearpage
\begin{figure}[p]
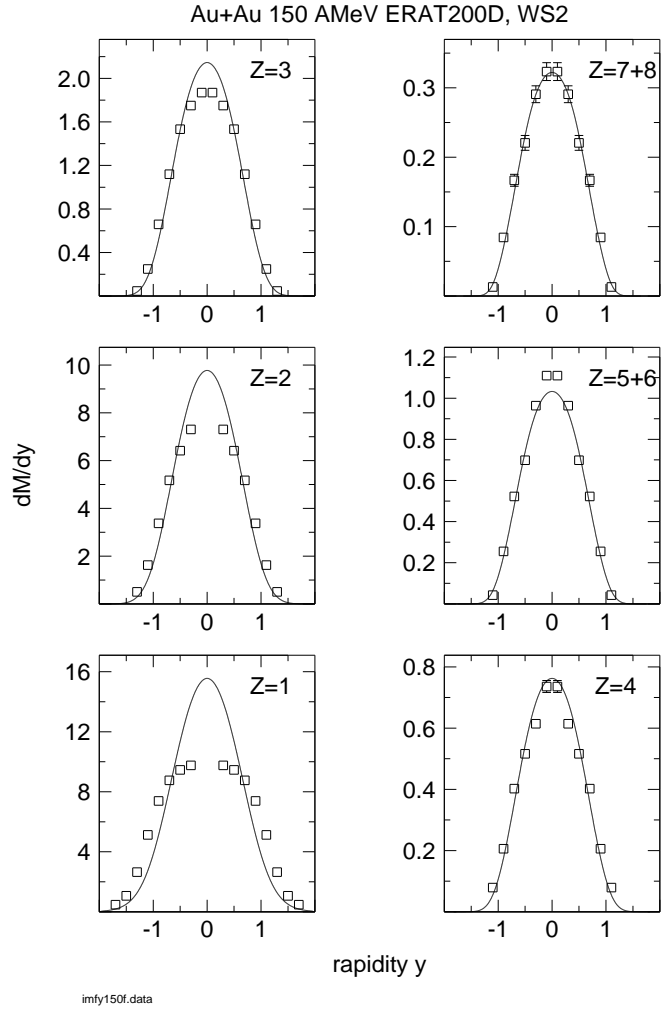

\caption{\small Measured charge-separated rapidity distributions for Au+Au
at 150 A MeV. 
The various nuclear charges are given in the figure. The smooth curves are
blast model fits with the profile WS2.}
\label{fig:rapidity150}
\end{figure}

\begin{figure}[p]
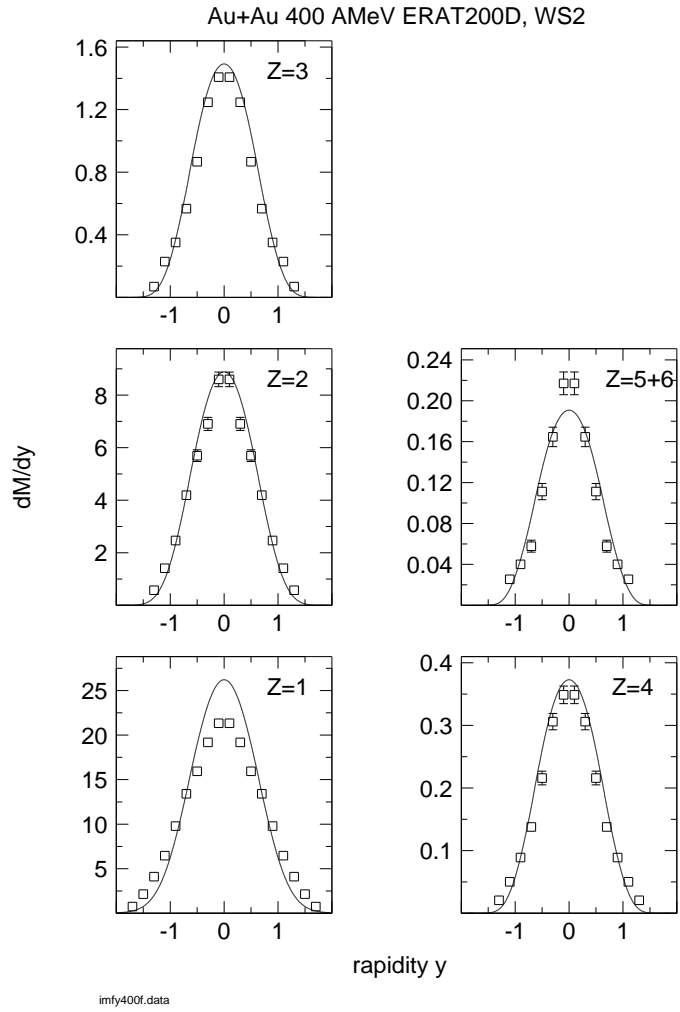

\caption{\small Same as Fig.~\ref{fig:rapidity150}, but at 400 A MeV
incident energy.}
\label{fig:rapidity400}
\end{figure}

\begin{figure}[p]
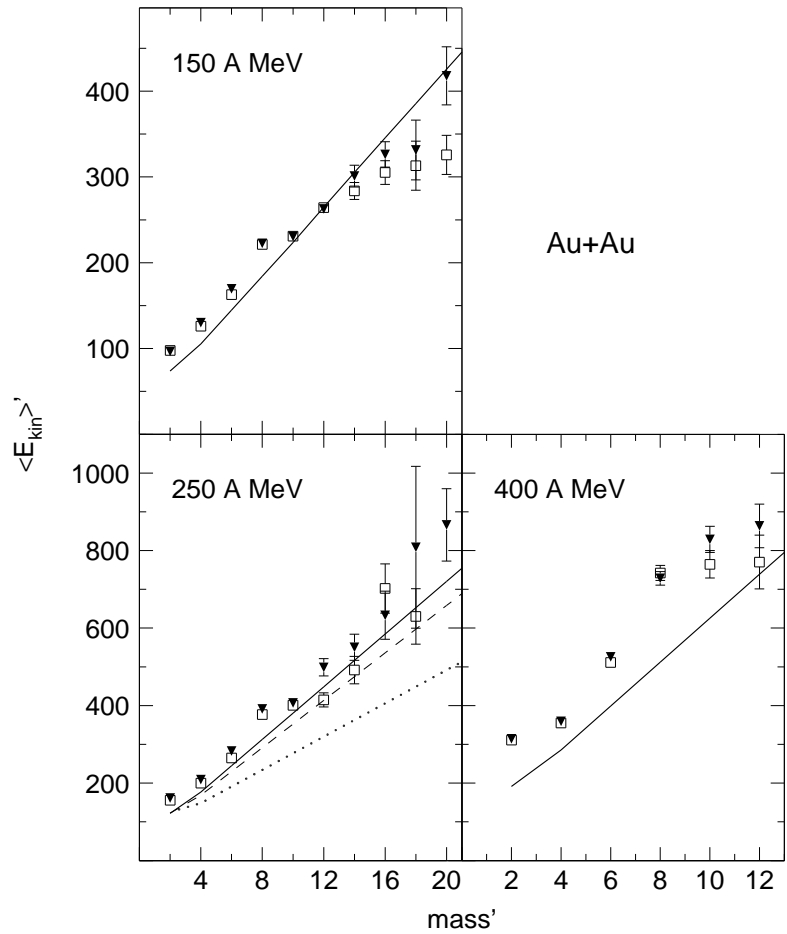

\caption{\small Average kinetic energy as function of mass in ther polar
angle range $25-45^{\circ}$ for the three indicated energies.
Open symbols: selection ERAT200D;
full triangles: selection ERAT50.
The solid lines are blast model fits with the 
velocity profile $WS2$. The dashed line
holds for profile $Box$, the dotted line for profile $Shell$, both shown only
for 250 A MeV.}
\label{fig:avekin}
\end{figure}

\begin{figure}[p]
\caption{\small Variation of the deduced collectivity with increasingly
dilute flow profiles.}
\label{fig:collectivity}
\end{figure}

\begin{figure}[p]
\caption{\small Average kinetic energies for hydrogen and helium isotopes
at polar angles      $60-90^{\circ}$.
The data are from  ref.~\cite{poggi95}. 
The solid line represents the mass dependence inferred from the global fit
of the blast model with velocity profile $WS2$.}
\label{fig:avekinlp}
\end{figure}

\begin{figure}[p]
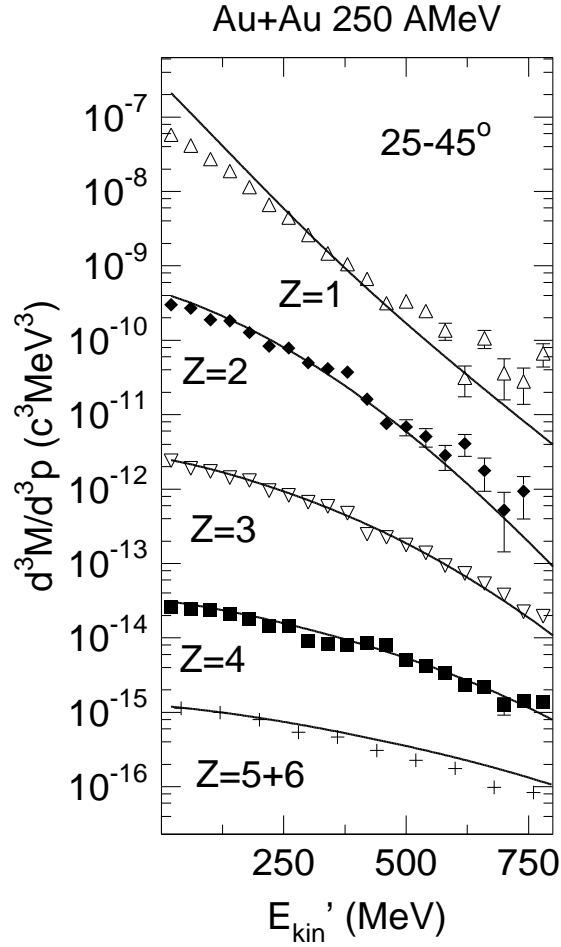

\caption{\small Kinetic energy spectra at center-of-mass polar angles
$(25-45)^{\circ}$ for fragments with nuclear charge Z=1, 2($\times$0.1),
3($\times$0.01), 4($\times$0.001),and 5+6($\times$0.0001).
The system is Au+Au at 250 A MeV, the event sample is ERAT200D.
The smooth curves are blast model descriptions.}
\label{fig:ekin250}
\end{figure}

\clearpage
\begin{figure}[p]
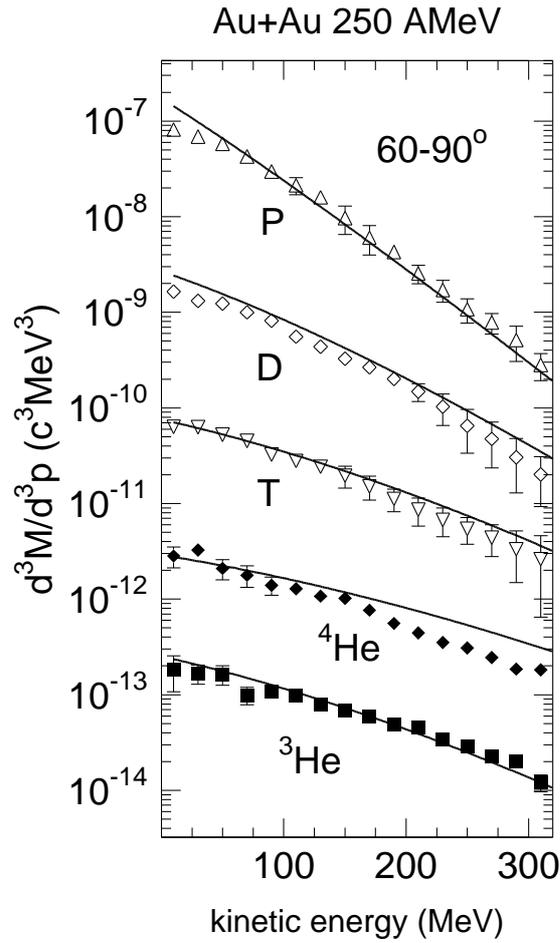

\caption{\small Kinetic energy spectra of light charged particles
(indicated in the figure)  at polar angles
$(60-90)^{\circ}$ for Au on Au at 250 A MeV.
The data (symbols) are from reference \protect \cite{poggi95}, the smooth curves are
predictions of the blast model. 
The event sample is approximately ERAT200.}
\label{fig:ekin250lp}
\end{figure}

\begin{figure}[p]
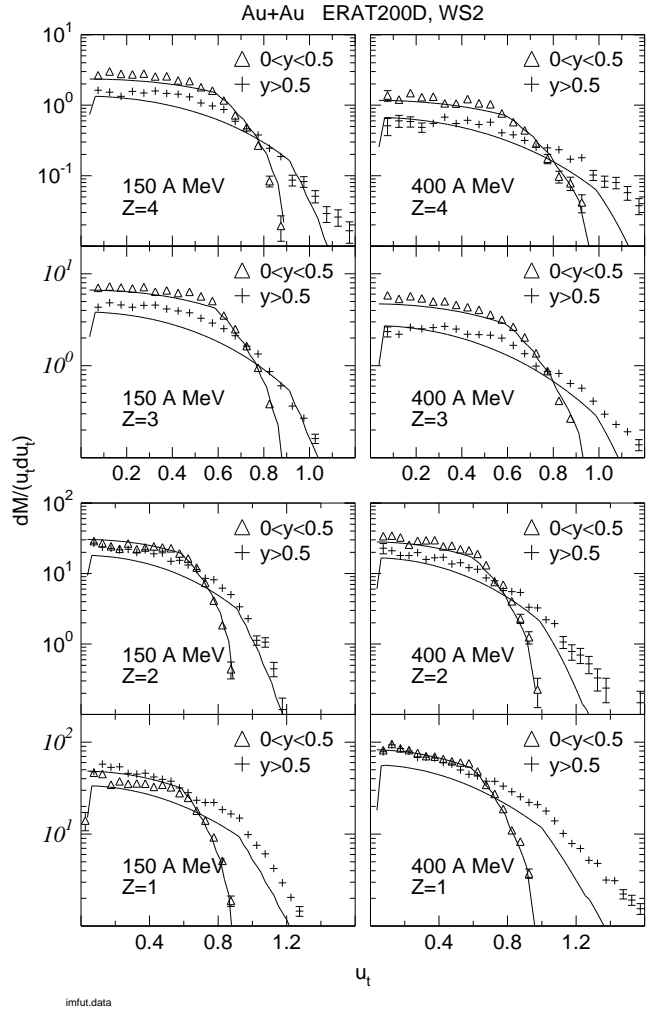

\caption{\small Invariant transverse four-velocity spectra for Au on Au at
150 (left panels) and 400 A MeV for fragments with nuclear charges Z=1, 2,
3 and 4 (bottom to top). The spectra are shown for two indicated rapidity
intervals. 
The data are represented by symbols, smooth curves were calculated with the
blast model. 
The event sample is ERAT200D.}
\label{fig:pt}
\end{figure}

\begin{figure}[p]
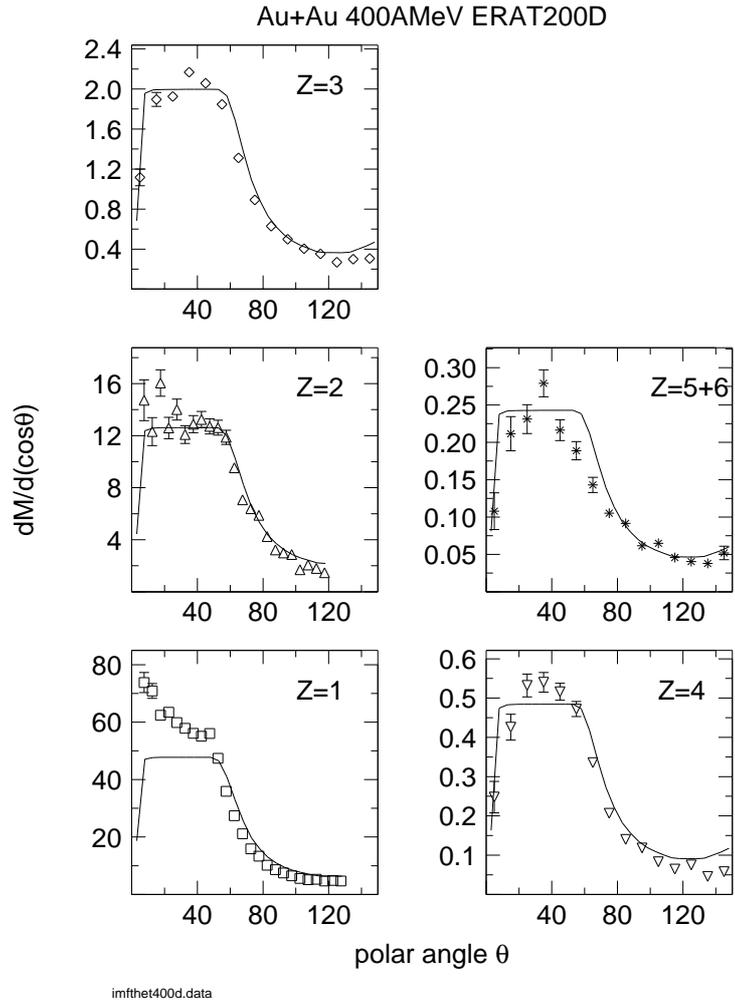

\caption{\small Polar angular distributions for Au on Au at 400 A MeV for
fragments with nuclear charges Z=1 to 6.
The smooth curves are blast model calculations.
The deviations from a flat distribution in the model are due to the
apparatus filter.
The event sample is ERAT200D.}
\label{fig:polar1}
\end{figure}

\begin{figure}[p]
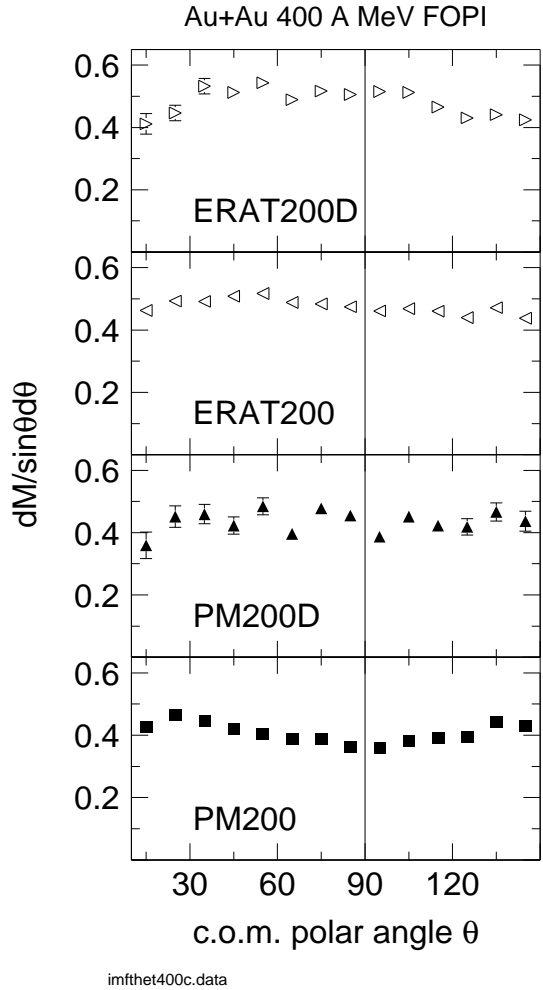

\caption{\small Polar angular distributions of intermediate mass fragments
for Au on Au at 400 A MeV. 
A scaled four-velocity cut $u < 0.5$ is applied.     
The distributions are shown for various indicated event samples.}
\label{fig:polar2}
\end{figure}

\begin{figure}[p]
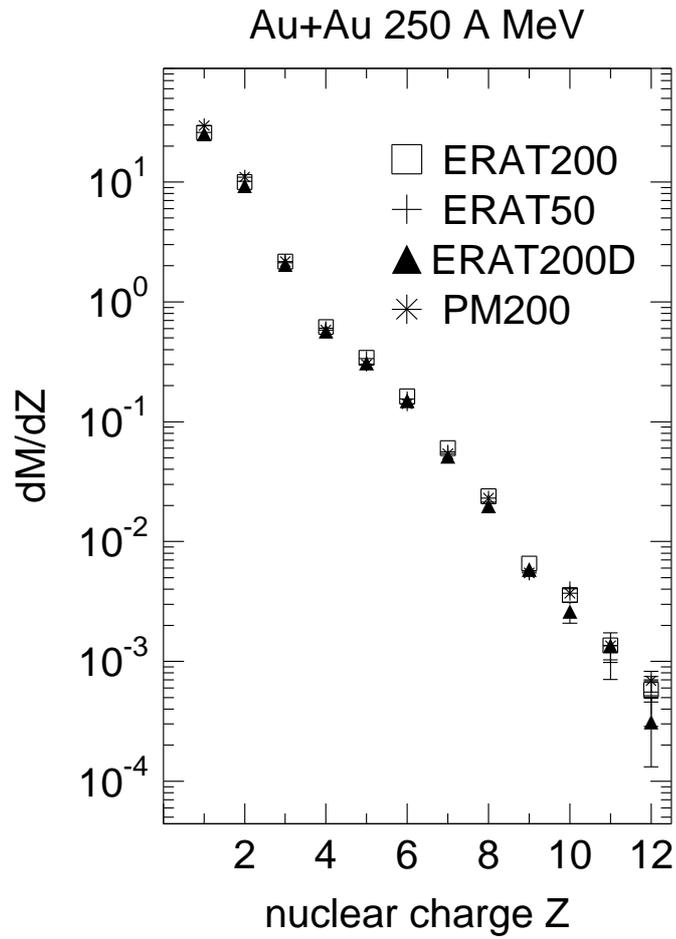

\caption{\small Measured multiplicities for Au on Au at 250 A MeV
as function of nuclear charge for various event selections described in the
text.}
\label{fig:zrobust}
\end{figure}

\begin{figure}[p]
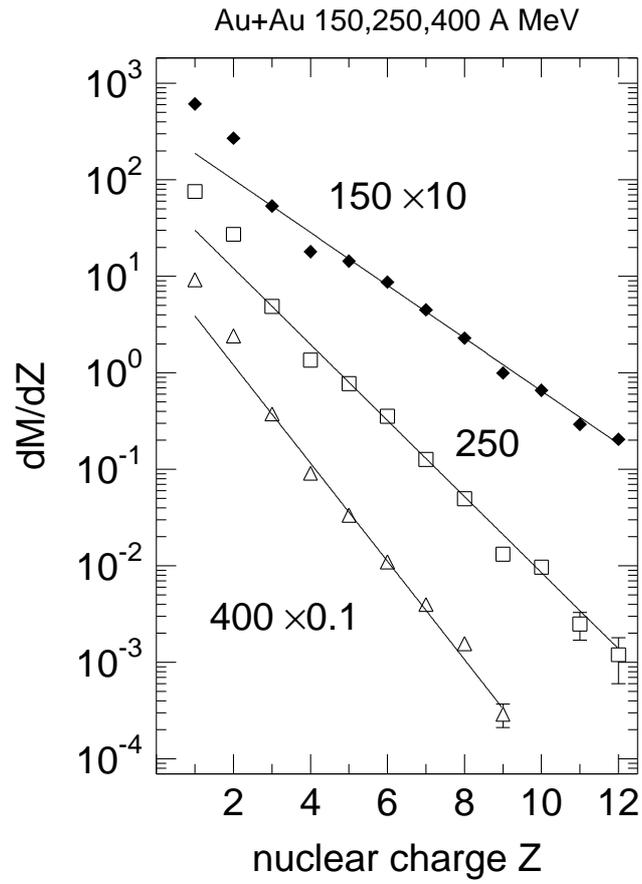

\caption{\small Measured $4\pi$-integrated multiplicities for Au on Au at
150, 250 and 400 A MeV as function of nuclear charge. The straight lines
are exponential fits excluding Z=1, 2 and 4 fragments.}
\label{fig:zdist}
\end{figure}

\clearpage
\begin{figure}[p]
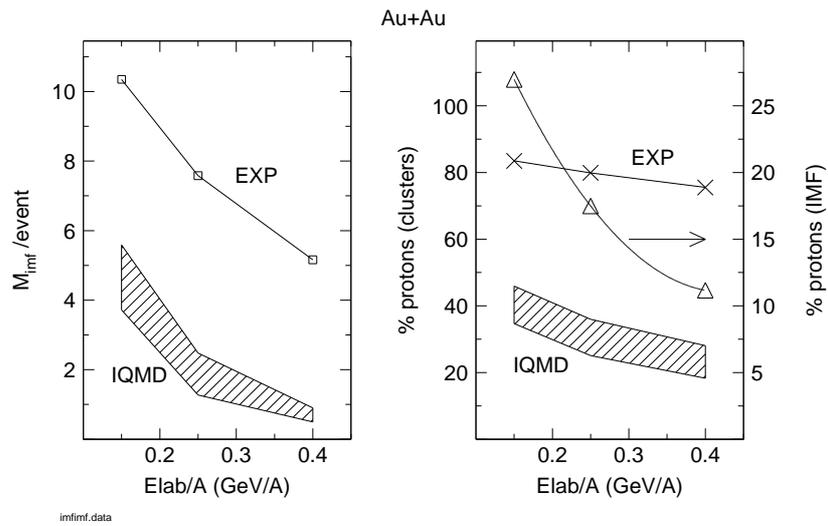

\caption{\small Clusterization in experiment (EXP) and theory (IQMD) versus
incident energy.
Left panel: IMF multiplicity in $4\pi$ geometry.
Right panel: percent of protons bound in any cluster (left ordinate, EXP
and IQMD) and protons bound in IMF's (right ordinate, triangles,EXP only). 
In both panels the height of the hatched area indicates the span when
switching from soft (higher values) to hard EOS.}
\label{fig:cluster}
\end{figure}

\begin{figure}[p]
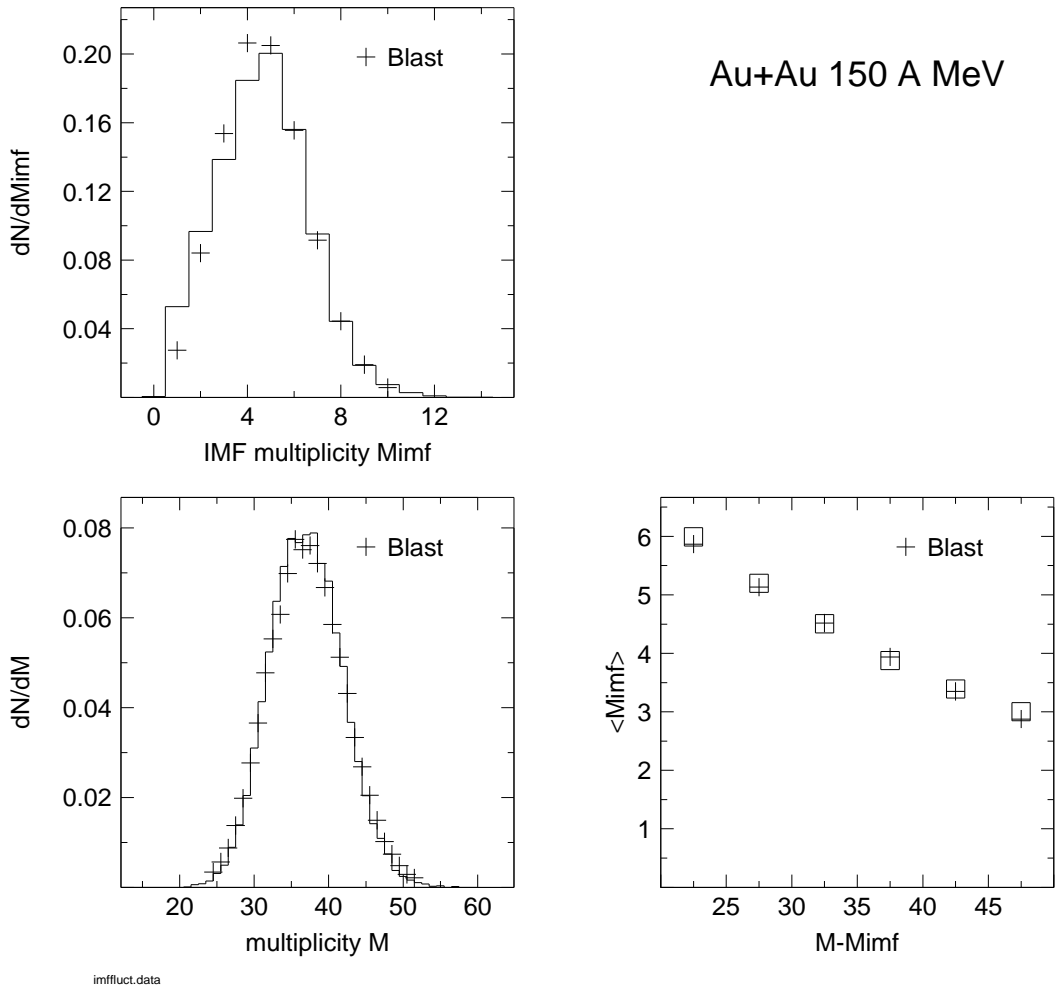

\caption{\small Fluctuations and correlations of multiplicities.
The histograms are measurements, the crosses are blast model simulations.
Lower left panel: External Wall multiplicity (M) distribution.     
Upper left panel: IMF multiplicity distribution. Lower right
panel: average IMF multiplicity versus light-particle multiplicity.}
\label{fig:fluct}
\end{figure}

\begin{figure}[p]
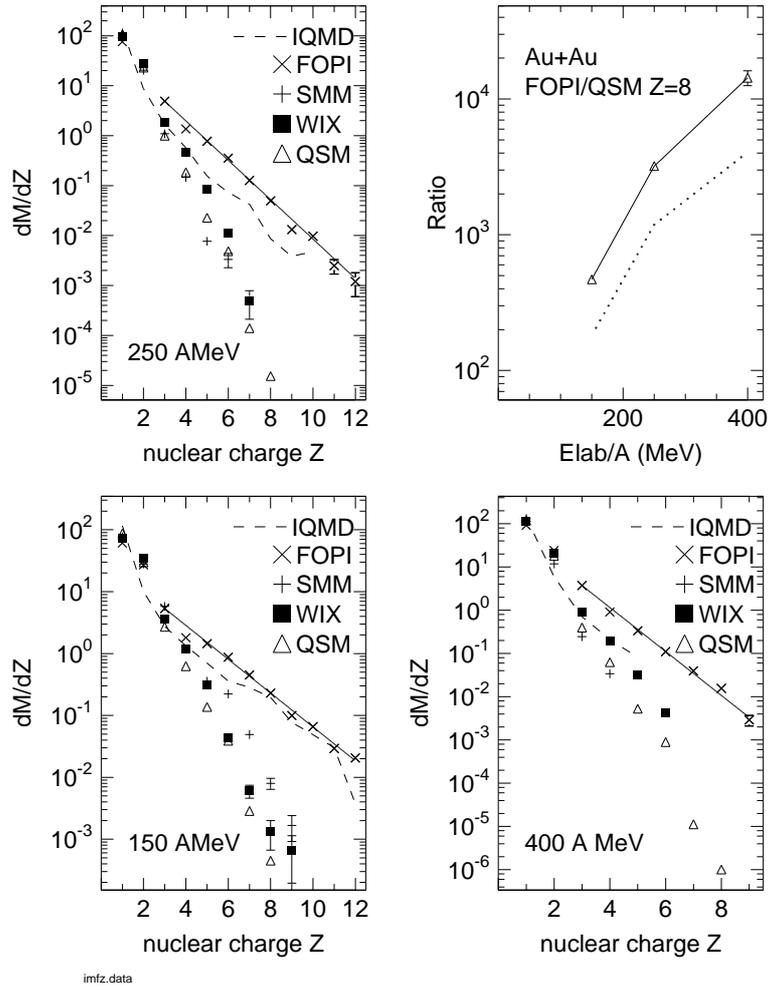

\caption{\small Comparison of measured (FOPI) charged particle
multiplicities with calculations using three different statistical models,
QSM, WIX, SMMM for three indicated incident energies
The predictions of IQMD (SM) are indicated as dashed lines.
 The upper right panel gives the ratio of measured oxygen
yields to predictions of the QSM (triangles joined by solid lines). The
dotted line in the panel is a calculation for the maximum collective energy
allowed by the blast model fits.}
\label{fig:zdisttheo}
\end{figure}

\begin{figure}[p]
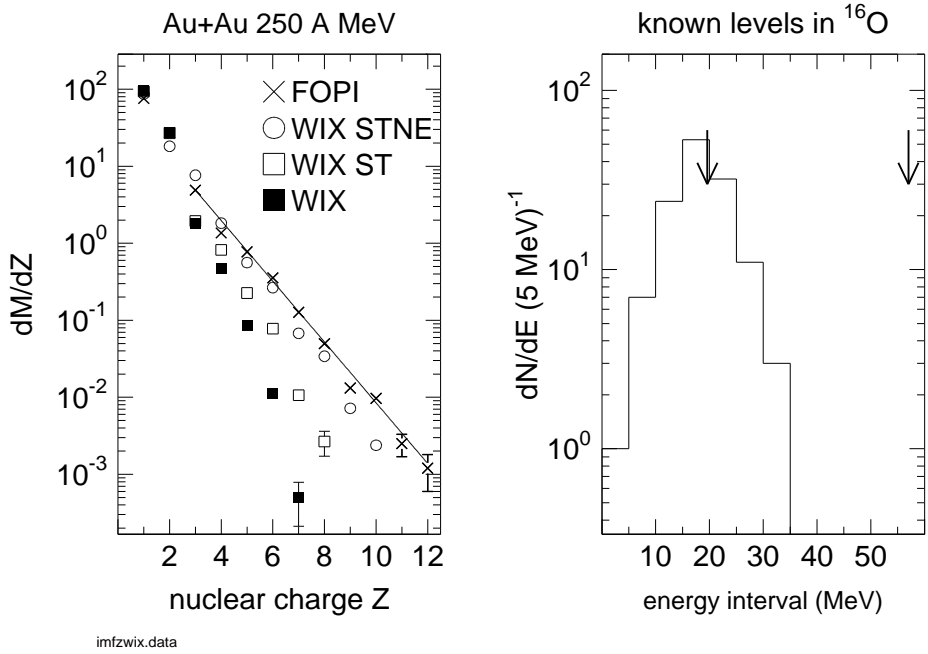

\caption{\small Left panel: measured (crosses) charged particle
multiplicities at 250 A MeV (together with the exponential fit to guide the
eye) and various calculations with the code WIX. Full squares: $\ccut =2$,
open squares: $\ccut =8$, open circles: $\ccut =8$ and no evaporation.
Right panel: known level distribution for $^{16}$O (histogram), left
(right) arrow: average excitation energy of $^{16}$O predicted by WIX for
$\ccut =2$ (8).}
\label{fig:ccut}
\end{figure}

\begin{figure}[p]
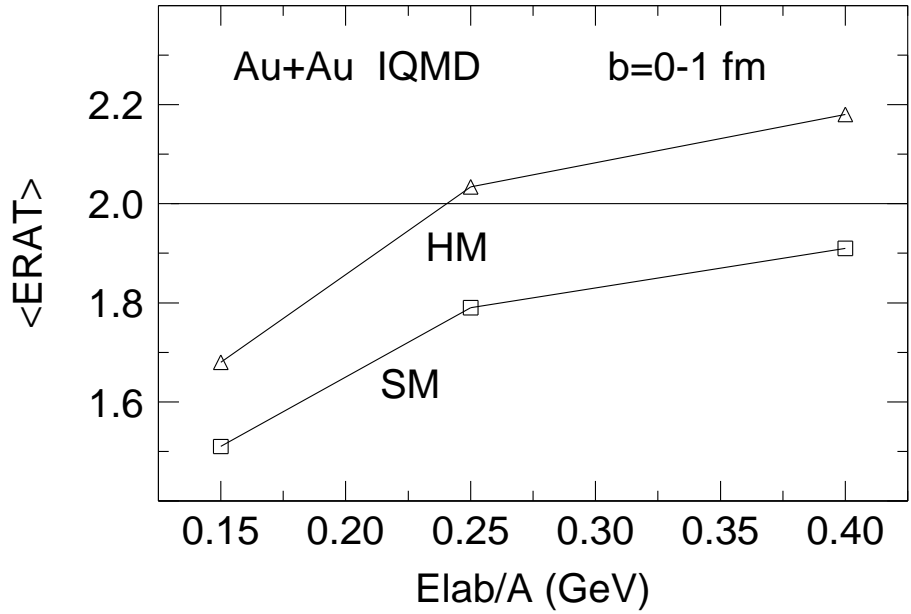

\caption{\small Variation of the average value of ERAT with incident energy
in the IQMD model.
The dependences are shown for the hard ($HM$) and the soft ($SM$) momentum
dependent EOS. The Au on Au collisions are restricted to impact parameters
$<1$ fm.
ERAT is calculated from charged particles only, but using full $4\pi$
geometry.}
\label{fig:eratiqmd}
\end{figure}

\begin{figure}[p]
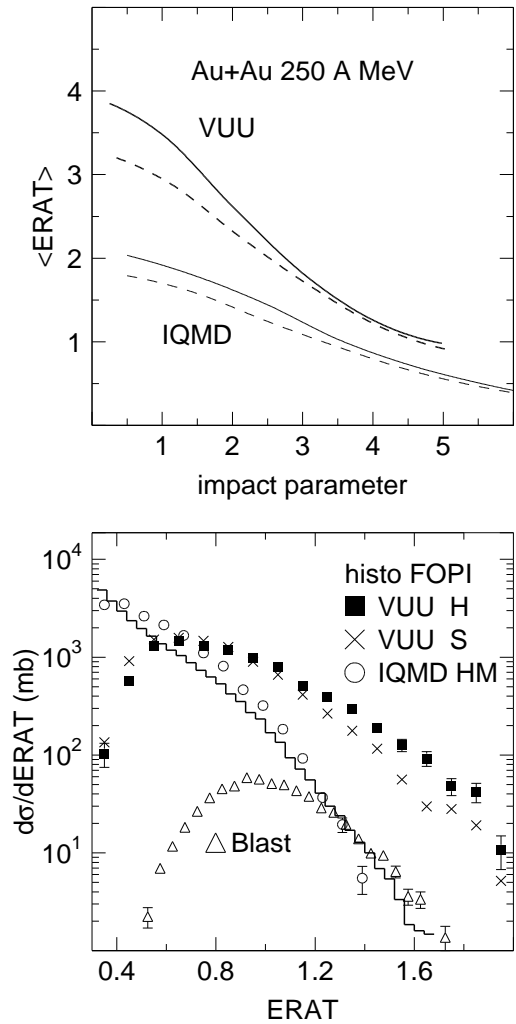

\caption{\small ERAT for Au on Au at 250 AMeV.
Top panel: predicted dependence on the impact parameter in $4\pi$ geometry,
using charged particles only.
Two different codes, VUU \protect \cite{daniel92} and IQMD \protect
\cite{hartnack93} were used. The solid (dashed) curves are for the hard
(soft) EOS.
Bottom panel: measured ERAT distribution (histogram) and filtered
theoretical simulations.
Triangles: Blast Model normalized to 31 mb ($b\leq 1$fm).
Open circles: IQMD hard momentum dependent EOS.
The other symbols are for the VUU calculation with hard (H) or soft (S)
EOS.}
\label{fig:daniel}
\end{figure}

\clearpage
\begin{figure}[p]
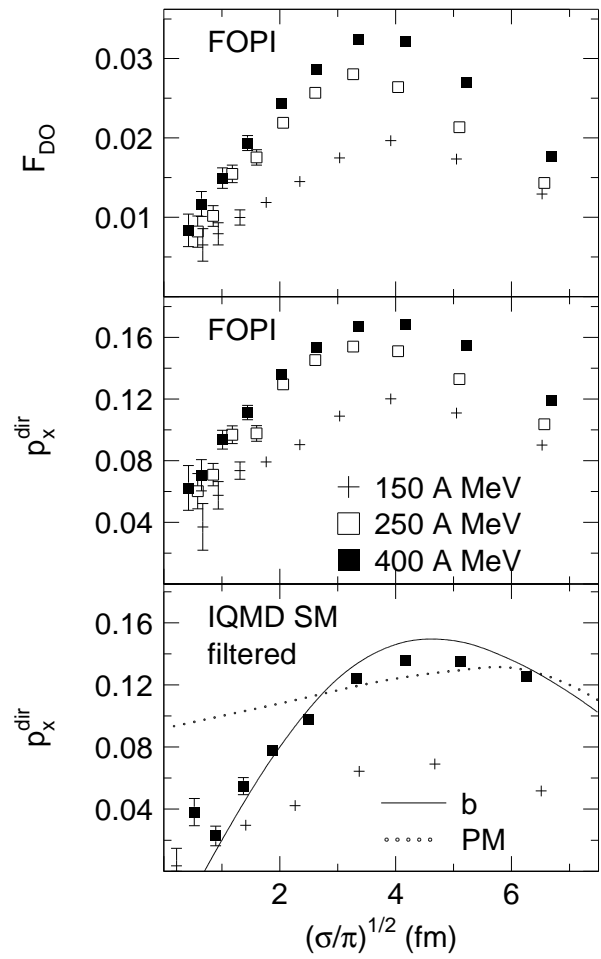

\caption{\small Scaled sideflow in semicentral Au on Au collisions.
Lower panel: filtered IQMD calculations for $\pxdir $ with the soft
momentum-dependent EOS (SM) and for 400 A MeV incident energy.
Solid curve: binning with the true impact parameter.
Squares: binning with ERAT.
Dotted: binning with PM.
Crosses: binning with ERAT, but for 150 A MeV incident energy.
Middle panel: FOPI data for $\pxdir $ binned with ERAT and for the
three indicated energies.
Upper panel: same as middle panel, but for the alternate sideflow observable
$\fdo $.}
\label{fig:sideflow}
\end{figure}

\begin{figure}[p]
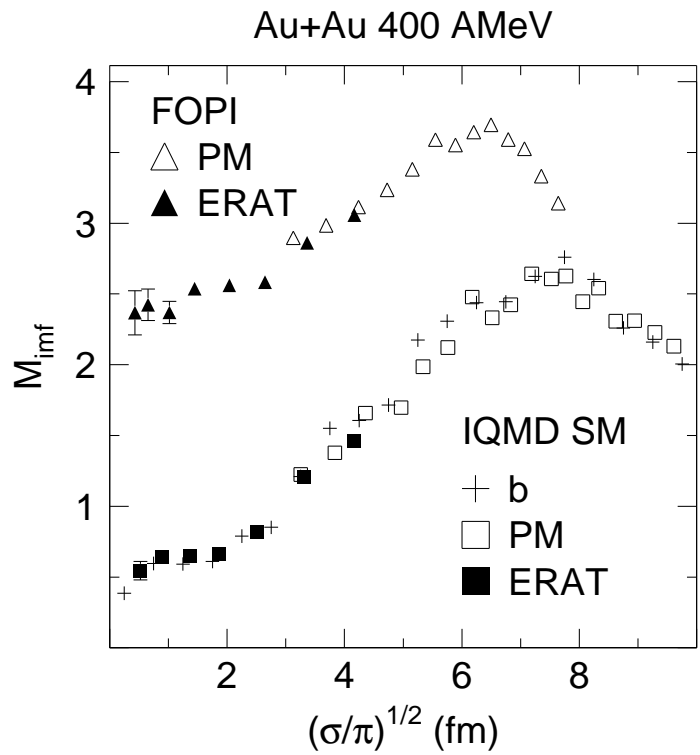

\caption{\small Comparison of the centrality dependence of the IMF
multiplicity in experiment (triangles) and theory.                    
Open symbols: multiplicity binning, closed symbols: ERAT binning, crosses:
impact parameter binning.}
\label{fig:mimfvsb}
\end{figure}

\begin{figure}[p]
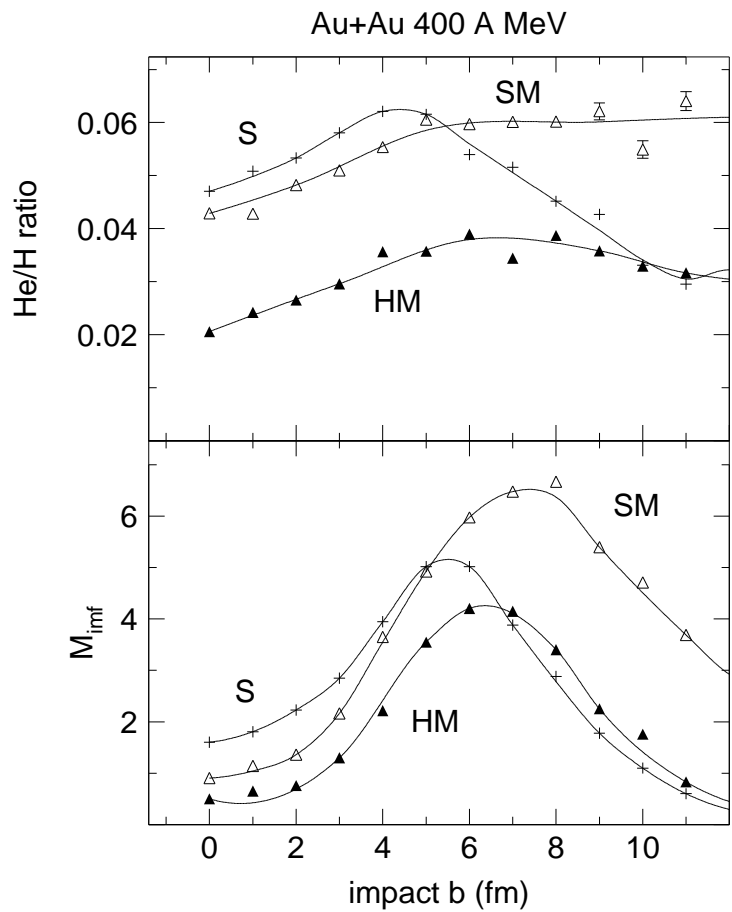

\caption{\small Equation of state and impact parameter dependence of the
degree of clusterization in the IQMD simulation.
Lower panel: IMF multiplicity, upper panel: He/H ratio.
$S$ ($H$) stands for soft (hard) EOS, $M$ indicates inclusion of momentum
dependence.
The curves are to guide the eye.
For comparison, the experimental He/H ratio for central collisions is 0.26.
All data are for Au on Au at 400 A MeV.}
\label{fig:mimfvseos}
\end{figure}

\begin{figure}[p]
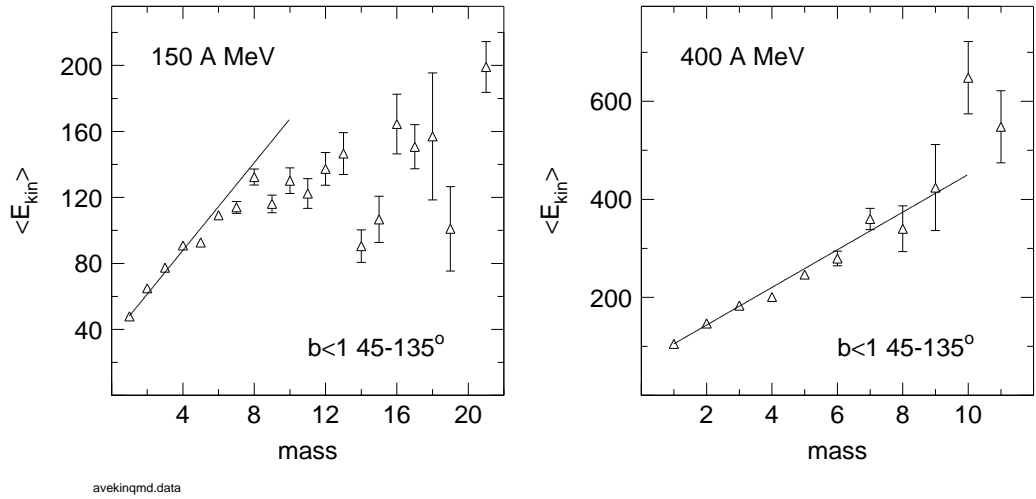

\caption{\small IQMD predictions (soft momentum dependent EOS) for average
kinetic energies as function of mass. The straight lines are linear least
squares fits in the mass range $A=1-10$.
The simulations are for Au on Au collisions 
at 150 (left) and 400 (right) A MeV with impact parameters up to 1
fm and for particles emitted with polar angles $45-135^{\circ}$.}
\label{fig:avekinqmd}
\end{figure}

\clearpage


\end{document}